\documentclass[letterpaper,11pt]{article}
\pdfoutput=1
\usepackage{jheppub}
\usepackage{mathtools,amssymb,braket,mathrsfs,tikz,subfig,xspace,xcolor,float,color,siunitx,comment,afterpage,multirow,array,hhline}
\usepackage[utf8]{inputenc}
\usepackage[countmax]{subfloat}

\providecommand{\href}[2]{#2}
\newcommand{\pd}[2]{\frac{\partial #1}{\partial #2}}
\DeclareMathOperator{\tr}{tr}

\DeclareRobustCommand{\Sec}[1]{Sec.~\ref{#1}}
\DeclareRobustCommand{\Secs}[2]{Secs.~\ref{#1} and \ref{#2}}

\DeclareRobustCommand{\App}[1]{App.~\ref{#1}}
\DeclareRobustCommand{\Tab}[1]{Table~\ref{#1}}
\DeclareRobustCommand{\Tabs}[2]{Tables~\ref{#1} and \ref{#2}}
\DeclareRobustCommand{\Fig}[1]{Fig.~\ref{#1}}

\DeclareRobustCommand{\Eq}[1]{Eq.~(\ref{#1})}
\DeclareRobustCommand{\Eqs}[2]{Eqs.~(\ref{#1}) and (\ref{#2})}
\DeclareRobustCommand{\Eqss}[3]{Eqs.~(\ref{#1}), (\ref{#2}), and (\ref{#3})}
\DeclareRobustCommand{\Ref}[1]{Ref.~\cite{#1}}
\DeclareRobustCommand{\Refs}[1]{Refs.~\cite{#1}}

\newcommand{\B}{\text{EFP}\xspace}
\newcommand{\Bs}{\text{EFPs}\xspace}

\definecolor{jdtcolor}{rgb}{0.8,0,0}
\definecolor{emmcolor}{rgb}{0,0.8,0}
\definecolor{ptkcolor}{rgb}{0,0,0.8}

\usepackage{amsthm}
\newtheorem*{EFPMultiCorre}{Multigraph/EFP Correspondence}

\title{Energy flow polynomials:\\A complete linear basis for jet substructure}

\preprint{MIT--CTP 4965}

\author{Patrick T. Komiske,}
\author{Eric M. Metodiev,}
\author{Jesse Thaler}

\affiliation{Center for Theoretical Physics, Massachusetts Institute of Technology, Cambridge, MA 02139, USA}

\emailAdd{pkomiske@mit.edu}
\emailAdd{metodiev@mit.edu}
\emailAdd{jthaler@mit.edu}

\abstract{
We introduce the energy flow polynomials: a complete set of jet substructure observables which form a discrete linear basis for all infrared- and collinear-safe observables.
Energy flow polynomials are multiparticle energy correlators with specific angular structures that are a direct consequence of infrared and collinear safety.
We establish a powerful graph-theoretic representation of the energy flow polynomials which allows us to design efficient algorithms for their computation.
Many common jet observables are exact linear combinations of energy flow polynomials, and we demonstrate the linear spanning nature of the energy flow basis by performing regression for several common jet observables.
Using linear classification with energy flow polynomials, we achieve excellent performance on three representative jet tagging problems:  quark/gluon discrimination, boosted $W$ tagging, and boosted top tagging.
The energy flow basis provides a systematic framework for complete investigations of jet substructure using linear methods.
}

\begin{document} 
\flushbottom
\maketitle

\addtocontents{toc}{\protect\enlargethispage{9mm}}

\section{Introduction}
\label{sec:intro}

Jet substructure is the analysis of radiation patterns and particle distributions within the collimated sprays of particles (jets) emerging from high-energy collisions~\cite{Seymour:1991cb,Seymour:1993mx,Butterworth:2002tt,Butterworth:2007ke,Butterworth:2008iy}.
Jet substructure is central to many analyses at the Large Hadron Collider (LHC), finding applications in both Standard Model measurements~\cite{Chatrchyan:2012sn,CMS:2013cda,Aad:2015lxa,Aad:2015cua,TheATLAScollaboration:2015ynv,Aad:2015hna,Aad:2015rpa,ATLAS:2016dpc,ATLAS:2016wlr,ATLAS:2016wzt,CMS:2016rtp,Rauco:2017xzb,ATLAS:2017jiz,Sirunyan:2017dgc} and in searches for physics beyond the Standard Model~\cite{CMS:2011bqa,Chatrchyan:2012ku,Fleischmann:2013woa,Pilot:2013bla,TheATLAScollaboration:2013qia,CMS:2014afa,CMS:2014aka,Khachatryan:2015axa,TheATLAScollaboration:2015elo,TheATLAScollaboration:2015xqi,Khachatryan:2015bma,CMS:2016qwm,CMS:2016bja,CMS:2016ehh,CMS:2016rfr,Aaboud:2016okv,Khachatryan:2016mdm,CMS:2016flr,CMS:2016pod,Aaboud:2016qgg,Sirunyan:2016wqt,Sirunyan:2017usq,Sirunyan:2017acf,Sirunyan:2017nvi}. 
An enormous catalog of jet substructure observables has been developed to tackle specific collider physics tasks~\cite{Abdesselam:2010pt,Altheimer:2012mn,Altheimer:2013yza,Adams:2015hiv,Larkoski:2017jix}, such as the identification of boosted heavy particles or the discrimination of quark- from gluon-initiated jets.

The space of possible jet substructure observables is formidable, with few known complete and systematic organizations.
Previous efforts to define classes of observables around organizing principles include:
the jet energy moments and related Zernike polynomials to classify energy flow observables~\cite{GurAri:2011vx};
a pixelated jet image~\cite{Cogan:2014oua} to represent energy deposits in a calorimeter;
the energy correlation functions (ECFs)~\cite{Larkoski:2013eya} to highlight the $N$-prong substructure of jets;
the generalized energy correlation functions (ECFGs)~\cite{Moult:2016cvt} based around soft-collinear power counting~\cite{Larkoski:2014gra};
and a set of $N$-subjettiness observables~\cite{Thaler:2010tr,Thaler:2011gf,Stewart:2010tn} to capture $N$-body phase space information~\cite{Datta:2017rhs}.
With any of these representations, there is no simple method to combine individual observables, so one typically uses sophisticated multivariate techniques such as neural networks to fully access the information contained in several observables~\cite{Gallicchio:2010dq,Gallicchio:2011xq,Gallicchio:2012ez,Baldi:2014kfa,Baldi:2014pta,deOliveira:2015xxd,Barnard:2016qma,Komiske:2016rsd,Kasieczka:2017nvn,Almeida:2015jua,Baldi:2016fql,Guest:2016iqz,Louppe:2017ipp,Pearkes:2017hku,Butter:2017cot,Aguilar-Saavedra:2017rzt,Datta:2017rhs}.
Furthermore, the sense in which these sets ``span'' the space of jet substructure is often unclear, sometimes relying on the existence of complicated nonlinear functions to map observables to kinematic phase space.

In this paper, we introduce a powerful set of jet substructure observables organized directly around the principle of infrared and collinear (IRC) safety.
These observables are multiparticle energy correlators with specific angular structures which directly result from IRC safety.
Since they trace their lineage to the hadronic energy flow analysis of \Ref{Tkachov:1995kk}, we call these observables the \emph{energy flow polynomials} (\Bs) and we refer to the set of \Bs as the \emph{energy flow basis}.
In the language of \Ref{Tkachov:1995kk}, the \Bs can be viewed as a discrete set of $C$-correlators, though our analysis is independent from the original $C$-correlator logic.
Crucially, the EFPs form a \emph{linear} basis of all IRC-safe observables, making them suitable for a wide variety of jet substructure contexts where linear methods are applicable.

There is a one-to-one correspondence between \Bs and loopless multigraphs, which helps to visualize and calculate the \Bs.
A multigraph is a graph where any two vertices can be connected by multiple edges; in this context, a \emph{loop} is an edge from a vertex to itself, while a closed chain of edges is instead referred to as a \emph{cycle}.
For a multigraph $G$ with $N$ vertices and edges $(k,\ell)\in G$, the corresponding \B takes the form:
\begin{equation}\label{eq:introefp}
\B_{G} = \sum_{i_1=1}^M\cdots\sum_{i_N=1}^M z_{i_1}\cdots z_{i_N}\prod_{(k,\ell) \in G} \theta_{i_k i_\ell},
\end{equation}
where the jet consists of $M$ particles, $z_i \equiv E_i/\sum_{j=1}^M E_j$ is the energy fraction carried by particle $i$, and $\theta_{ij}$ is the angular distance between particles $i$ and $j$.
The precise definitions of $E_i$ and $\theta_{ij}$ will depend on the collider context, with energy and spherical ($\theta,\phi$) coordinates typically used for $e^+e^-$ collisions, and transverse momentum $p_T$ and rapidity-azimuth $(y,\phi)$ coordinates for hadronic collisions.
For brevity, we often use the multigraph $G$ to represent the formula for $\B_G$ in \Eq{eq:introefp}, e.g.:
\begin{equation}\label{eq:wedgegraphex}
\begin{gathered}
\includegraphics[scale=.3]{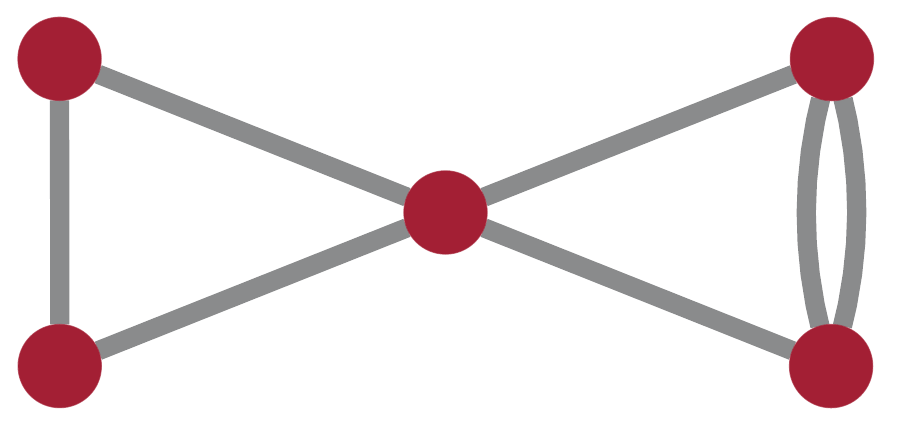}
\end{gathered}
= \sum_{i_1=1}^M\sum_{i_2 = 1}^M \sum_{i_3 = 1}^M \sum_{i_4=1}^M\sum_{i_5=1}^M  z_{i_1} z_{i_2} z_{i_3}z_{i_4}z_{i_5}\theta_{i_1i_2} \theta_{i_2i_3}\theta_{i_1i_3}\theta_{i_1i_4}\theta_{i_1i_5}\theta_{i_4i_5}^2.
\end{equation}

This paper is a self-contained introduction to the energy flow basis, with the following organization.
\Sec{sec:efps} contains a general overview of the \Bs, with more detailed descriptions of \Eq{eq:introefp} and the correspondence to multigraphs.
We also discuss a few different choices of measure for $z_i$ and $\theta_{ij}$.
As already mentioned, EFPs are a special case of $C$-correlators~\cite{Tkachov:1995kk}, so not surprisingly, we find a close relationship between \Bs and other classes of observables that are themselves $C$-correlators, including jet mass, ECFs~\cite{Larkoski:2013eya}, certain generalized angularities~\cite{Larkoski:2014pca}, and energy distribution moments~\cite{GurAri:2011vx}.
We also highlight features of the \Bs which are less well-known in the $C$-correlator-based literature.
 
In \Sec{sec:basis}, we give a detailed derivation of the \Bs as an (over)complete linear basis of all IRC-safe observables in the case of massless particles.
Because this section is rather technical, it can be omitted on a first reading, though the logic just amounts to systematically imposing the constraints of IRC safety.
In \Sec{sec:Eexpansion}, we use an independent (and arguably more transparent) logic from \Ref{Tkachov:1995kk} to show that any IRC-safe observable can be written as a linear combination of $C$-correlators:
\begin{equation}\label{eq:genccorr}
\mathcal C_N^{f_N}=\sum_{i_1=1}^M \cdots\sum_{i_N=1}^M E_{i_1}\cdots E_{i_N} \, f_N(\hat p_{i_1}^\mu,\ldots,\hat p_{i_N}^\mu),
\end{equation}
where $f_N$ is an angular weighting function that is only a function of the particle directions $\hat{p}_i^\mu=p_i^\mu/E_i$ (and not their energies $E_i$).
To derive \Eq{eq:genccorr}, we use the Stone-Weierstrass theorem~\cite{stone1948generalized} to expand an arbitrary IRC-safe observable in polynomials of particle energies, and then directly impose IRC safety and particle relabeling invariance.
In \Sec{sec:angleexpansion}, we determine the angular structures of the \Bs by expanding $f_N$ in terms of a discrete set of polynomials in pairwise angular distances.
Remarkably, the discrete set of polynomials appearing in this expansion is in one-to-one correspondence with the set of non-isomorphic multigraphs, which facilitates indexing the \Bs by multigraphs to encode the geometric structure in \Eq{eq:introefp}.

In \Sec{sec:complexity}, we investigate the complexity of computing \Bs.
Naively, \Eq{eq:introefp} has complexity $\mathcal O(M^N)$ due to the $N$ nested sums over $M$ particles.
However, the rich analytic structure of \Eq{eq:introefp} and the graph representations of \Bs allow for numerous algorithmic speedups.
Any \B with a disconnected graph can be computed as the product of the \Bs corresponding to its connected components.
Furthermore, we find that the Variable Elimination (VE) algorithm~\cite{zhang1996exploiting} can be used to vastly speed up the computation of many \Bs compared to the naive $\mathcal O(M^N)$ algorithm.
VE uses the factorability of the summand to systematically determine a more efficient order for performing nested sums.
For instance, all tree graphs can be computed in $\mathcal O(M^2)$ using VE. 
As an explicit example, consider an EFP with naive $\mathcal O(M^6)$ scaling:
\begin{equation}\label{eq:extree}
\begin{gathered}
\includegraphics[scale=.25]{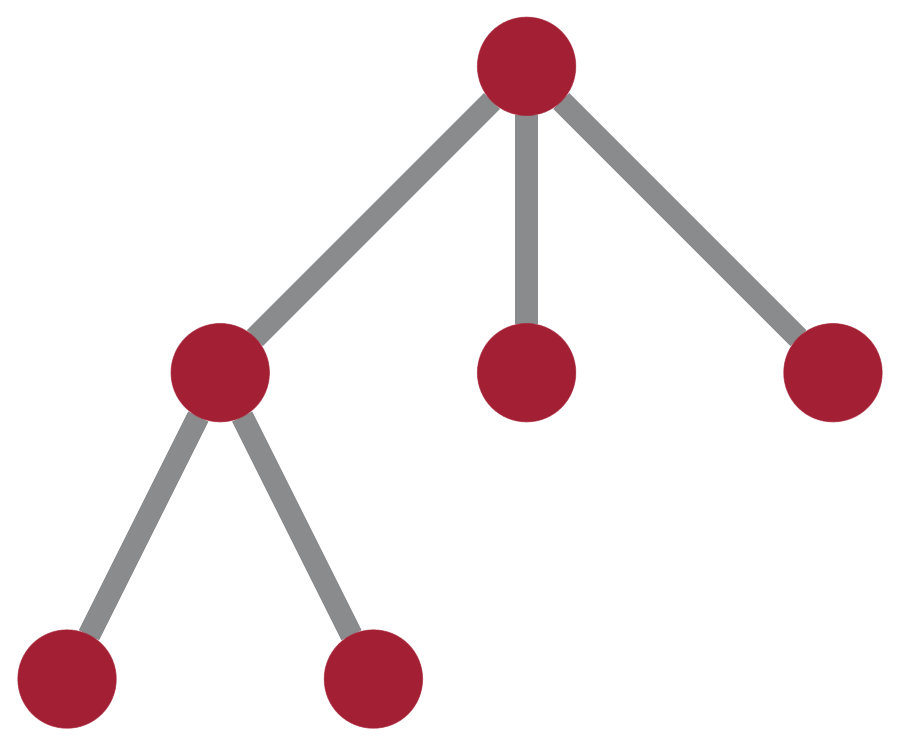}
\end{gathered}
\,=\sum_{i_1=1}^M\sum_{i_2=1}^Mz_{i_1}z_{i_2}\theta_{i_1i_2}\left(\sum_{i_3=1}^Mz_{i_3}\theta_{i_1i_3}\right)^2\left(\sum_{i_4=1}^Mz_{i_4}\theta_{i_2i_4}\right)^2.
\end{equation}
The quantities in parentheses are computable in $\mathcal O(M^2)$, since they are length $M$ lists with each element a sum over $M$ objects, making the overall expression in \Eq{eq:extree} computable in $\mathcal O(M^2)$.
The efficient computation of the \Bs overcomes one of the main previous challenges in using higher-$N$ multiparticle correlators in collider physics applications.\footnote{Sadly, fully-connected graphs, which correspond to the original ECFs~\cite{Larkoski:2013eya}, cannot be simplified using VE.} 

In \Sec{sec:linreg}, we perform numerical linear regression with \Bs for various jet observables.
The linear spanning nature of the energy flow basis means that any IRC-safe observable $\mathcal S$ can be linearly approximated by \Bs, which we write as:
\begin{equation}\label{eq:introefpsspan}
\mathcal S\simeq\sum_{G\in\mathcal G}s_G \, \B_G,
\end{equation}
for some finite set of multigraphs $\mathcal G$ and some real coefficients $s_G$.  
One might worry that the number of \Bs needed to achieve convergence could be intractably large.
In practice, though, we find that the required set of $\mathcal G$ needed for convergence is rather reasonable in a variety of jet contexts.
While we find excellent convergence for IRC-safe observables, regressing with IRC-unsafe observables does not work as well, demonstrating the importance of IRC safety for the energy flow basis.

In \Sec{sec:linclass}, we perform another test of \Eq{eq:introefpsspan} by using linear classification with \Bs to distinguish signal from background jets.
We consider three representative jet tagging problems: quark/gluon discrimination, boosted $W$ tagging, and boosted top tagging.
In this study, the observable appearing on the left-hand side of \Eq{eq:introefpsspan} is the optimal IRC-safe discriminant for the two classes of jets.
Remarkably, linear classification with \Bs performs comparably to multivariate machine learning techniques, such as jet images with convolutional neural networks (CNNs)~\cite{Cogan:2014oua,deOliveira:2015xxd,Barnard:2016qma,Komiske:2016rsd,Kasieczka:2017nvn} or dense neural networks (DNNs) with a complete set of $N$-subjettiness observables~\cite{Datta:2017rhs}.
Both the linear regression and classification models have few or no hyperparameters, illustrating the power and simplicity of linear learning methods combined with our fully general linear basis for IRC-safe jet substructure.

Our conclusions are presented in \Sec{sec:conclusion}, where we highlight the relevance of the energy flow basis to machine learning and discuss potential future applications and developments.
A review of $C$-correlators and additional tagging plots are left to the appendices.

\section{Energy flow polynomials}
\label{sec:efps}

IRC-safe observables have long been of theoretical and experimental interest because observables which lack IRC safety are not well defined~\cite{Kinoshita:1962ur,Lee:1964is,Weinberg:1995mt,sterman1995handbook}, or require additional care to calculate~\cite{Larkoski:2013paa,Larkoski:2015lea,Waalewijn:2012sv,Chang:2013rca,Elder:2017bkd}, in perturbative quantum chromodynamics (pQCD).
More broadly, though, IRC safety is a simple and natural organizing principle for high-energy physics observables, since IRC-safe observables probe the high-energy structure of an event while being insensitive to low-energy and collinear modifications.
IRC safety is also an important property experimentally as IRC-safe observables are more robust to noise and finite detector granularity.

As argued in \Refs{Tkachov:1995kk,Sveshnikov:1995vi,Cherzor:1997ak,Tkachov:1999py}, the $C$-correlators in \Eq{eq:genccorr} are a generic way to capture the IRC-safe structure of a jet, as long as one chooses an appropriate angular weighting function $f_N$.
Later in \Sec{sec:basis}, we give an alternative proof that $C$-correlators span the space of IRC-safe observables and go on to give a systematic expansion for $f_N$.
This expansion results in the \Bs, which yield an (over)complete linear basis for IRC-safe observables.
In this section, we highlight the basic features of the \Bs and their relationship to previous jet substructure observables.

\subsection{The energy flow basis}
\label{sec:efbasis}

One can think of the \Bs as $C$-correlators that make specific, discrete choices for the angular weighting function $f_N$ in \Eq{eq:genccorr}.  
True to their name, \Bs have angular weighting functions that are polynomial in pairwise angular distances $\theta_{ij}$.
The energy flow basis is therefore all $C$-correlators with angular structures that are unique monomials in $\theta_{ij}$, meaning monomials that give algebraically different expressions once the sums in \Eq{eq:genccorr} are performed.
Since we intend to apply the energy flow basis for jet substructure, we remove the dependence on the overall jet kinematics by normalizing the particle energies by the total jet energy, $E_J\equiv\sum_{i=1}^M E_i$, leading to the \Bs written in terms of the energy fractions $z_i\equiv E_i/E_{J}$ as in \Eq{eq:introefp}. 

The uniqueness requirement on angular monomials can be better understood by developing a correspondence between monomials in $\theta_{ij}$ and multigraphs:
\begin{EFPMultiCorre}
The set of loopless multigraphs on $N$ vertices corresponds exactly to the set of angular monomials in $\{\theta_{i_k i_\ell}\}_{k<\ell\in\{1,\cdots,N\}}$. Each edge $(k,\ell)$ in a multigraph is in one-to-one correspondence with a term $\theta_{i_ki_\ell}$ in an angular monomial; each vertex $j$ in the multigraph corresponds to a factor of $z_{i_j}$ and summation over $i_j$ in the \B:
\begin{align}\label{eq:correspondence}
&\begin{gathered}
\includegraphics[scale=0.04]{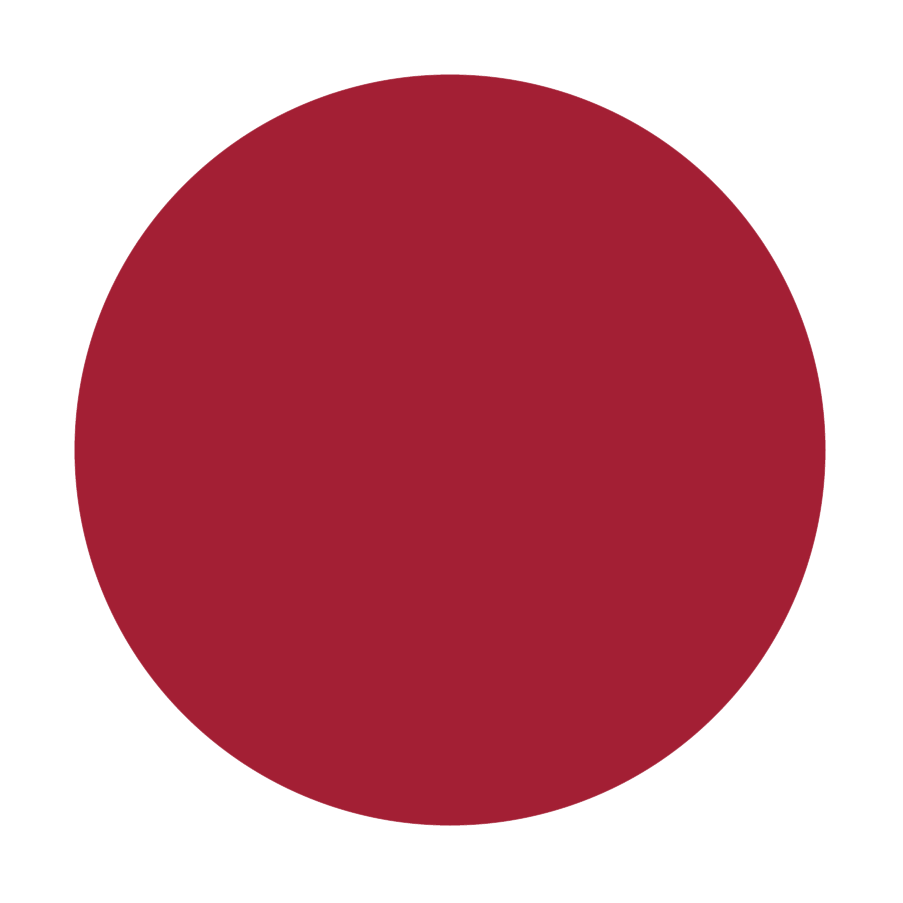}
\end{gathered}_j
 \,\,\Longleftrightarrow\,\, \sum_{i_j=1}^Mz_{i_j},
& k \begin{gathered}
\includegraphics[scale=0.3]{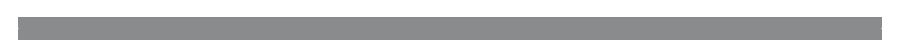}
\end{gathered} \,\ell
 \,\,\Longleftrightarrow\,\, \theta_{i_ki_\ell}.
\end{align}
\end{EFPMultiCorre}

Using \Eq{eq:correspondence}, the \Bs can be directly encoded by their corresponding multigraphs. For instance:
\begin{equation}\label{eq:flyswatter}
\begin{gathered}
\includegraphics[scale=0.32]{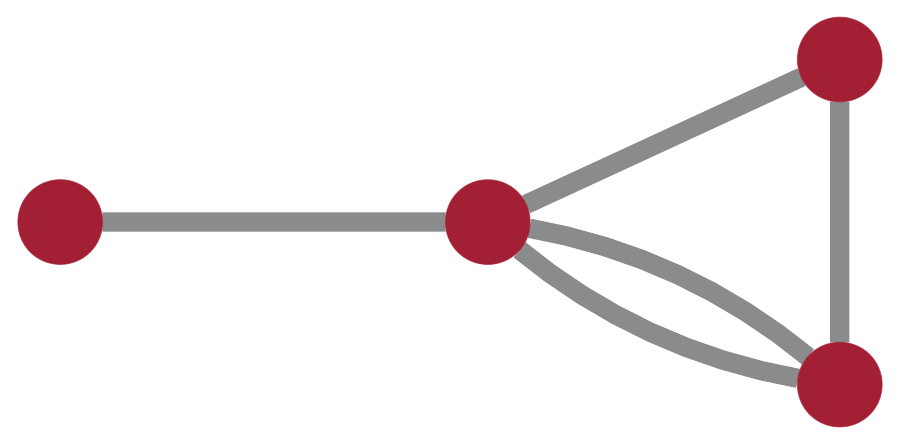}
\end{gathered}
= \sum_{i_1=1}^M\sum_{i_2 = 1}^M \sum_{i_3 = 1}^M \sum_{i_4=1}^M  z_{i_1} z_{i_2} z_{i_3}z_{i_4}\theta_{i_1i_2}\theta_{i_2i_3}\theta_{i_2i_4}^2\theta_{i_3i_4}.
\end{equation}
Since any labeling of the vertices gives an equivalent algebraic expression, we represent the graphs as unlabeled.
The specification that the \Bs are unique monomials translates into the requirement that the corresponding multigraphs are non-isomorphic.
Versions of these multigraphs have previously appeared in the physics literature in the context of many-body configurations~\cite{dallot1992reduced, mandal2017determination}, encoding all local scalar operators of a free theory~\cite{Hogervorst:2014rta}, and in graphically depicting ECFs for jets~\cite{Moult:2016cvt,Larkoski:2017iuy}.

\begin{table}[t]
\centering
\begin{tabular}{|rrcl|}
\hline
& \textbf{Multigraph}  & & \textbf{Energy Flow Polynomial}  \\ \hline \hline
$N$ : & Number of vertices & $\Longleftrightarrow$ &  $N$-particle correlator \\ 
$d$ : & Number of edges &$\Longleftrightarrow$ & Degree of angular monomial \\ 
$\chi$ : & Treewidth $+\,1$  &$\Longleftrightarrow$ & Optimal VE complexity $\mathcal O(M^\chi)$  \\
& Chromatic number &$\Longleftrightarrow$ & Minimum number of prongs to not vanish\\
\hline
& Connected& $\Longleftrightarrow$ &Prime  \\
& Disconnected& $\Longleftrightarrow$ &Composite  \\
\hline
\end{tabular}
\caption{Corresponding properties of multigraphs and \Bs.}
\label{tab:correspondence}
\end{table}

\Tab{tab:correspondence} contains a summary of the correspondence between the properties of \Bs and multigraphs.
The number of graph vertices $N$ corresponds to the number of particle sums in the EFP, and the number of graph edges $d$ corresponds to the \emph{degree} of the \B (i.e.\ the degree of the underlying angular monomial).
The number of separated prongs for which an individual \B is first non-vanishing is the \emph{chromatic number} of the graph: the smallest number of colors needed to color the vertices of the graph with no two adjacent vertices sharing a color.
For computational reasons discussed further in \Sec{sec:complexity}, we also care about the treewidth of the graph, which is related to the computational complexity $\chi$ of an EFP.
Also for computational reasons, we make a distinction between connected or \emph{prime} multigraphs and disconnected or \emph{composite} multigraphs; the value of a composite \B is simply the product of the prime \Bs corresponding to its connected components.

\begin{table}[t]
\centering
\subfloat[]{\label{tab:efpcounts:a}
\begin{tabular}{|rc||*{11}{r}|}\hline
\multicolumn{2}{|c}{Maximum degree $d$}&\multicolumn{1}{c}{\bf0}&\multicolumn{1}{c}{\bf1}&\multicolumn{1}{c}{\bf2}&\multicolumn{1}{r}{\bf3}&\multicolumn{1}{c}{\bf4}&\multicolumn{1}{c}{\,\,\bf5}&\multicolumn{1}{c}{\bf6}&\multicolumn{1}{c}{\bf7}&\multicolumn{1}{c}{\bf8}&\multicolumn{1}{c}{\bf9}&\multicolumn{1}{c|}{\bf10}\\\hhline{:==:t:*{11}{=}:}
\multirow{2}{*}{{\bf Prime \Bs}}
    & \href{https://oeis.org/A076864}{A076864} & 1 & 1 & 2 & 5 & 12 & 33 & 103 & 333 & 1\,183   & 4\,442   & 17\,576   \\ 
    & Cumul.    & 1 & 2 & 4 & 9 & 21 & 54 & 157 & 490 & 1\,673 & 6\,115 & 23\,691 \\ \hhline{|--||*{11}{-}}
\multirow{2}{*}{{\bf All \Bs}}
    & \href{https://oeis.org/A050535}{A050535} & 1 & 1 & 3 & 8   & 23 & 66   & 212 & 686     & 2\,389   & 8\,682     & 33\,160  \\
    & Cumul.    & 1 & 2 & 5 & 13 & 36 & 102 & 314 & 1\,000 & 3\,389 & 12\,071 & 45\,231 \\ \hhline{|--||*{11}{-}}
\end{tabular}}
\\\vspace{.25in}
\subfloat[]{\label{tab:efpcounts:b}
\begin{tabular}{|cc||rrrrrrrrrr|}\hline
\multicolumn{2}{|c}{$d$}&\bf1&\bf2&\bf3&\bf4&\bf5&\bf6&\bf7&\bf8&\bf9&\bf10 \\ \hhline{:==:t:*{10}{=}:}
\multirow{10}{*}{$N$} 
        & \bf2 & 1 & 1 & 1 & 1 & 1   & 1   & 1     & 1     & 1         & 1         \\
        & \bf3 &    & 1 & 2 & 3 & 4   & 6   & 7     & 9     & 11       & 13       \\
        & \bf4 &    &    & 2 & 5 & 11 & 22 & 37   & 61   & 95       & 141     \\
        & \bf5 &    &    &    & 3 & 11 & 34 & 85   & 193 & 396     & 771     \\
        & \bf6 &    &    &    &    & 6   & 29 & 110 & 348 & 969     & 2\,445 \\
        & \bf7 &    &    &    &    &      & 11 & 70   & 339 & 1\,318 & 4\,457 \\
        & \bf8 &    &    &    &    &      &      & 23   & 185 & 1\,067 & 4\,940 \\
        & \bf9 &    &    &    &    &      &      &        & 47   & 479     & 3\,294 \\
        & \bf10 &  &    &    &    &      &      &        &        & 106     & 1\,279 \\
        & \bf11 &  &    &     &    &     &       &       &        &            & 235     \\ \hhline{|--||*{10}{-}|} 
\end{tabular}}
\caption{(a) The number of \Bs (prime and all) organized by degree $d$, for $d$ up to 10. The cumulative rows tally the number of \Bs with degree at most $d$, i.e.\ the number of basis elements truncated at that $d$. While these sequences grow quickly, the total number of all basis elements is at most 1000 for $d\le7$, which is computationally tractable. (b) The number of prime \Bs broken down by number of vertices $N$ and number of edges $d$ in the multigraph.  All connected graphs (prime \Bs) for $d$ up to 5 are shown explicitly in \Tab{tab:graphs}.}
\label{tab:efpcounts}
\end{table}

Because the \B basis is infinite, a suitable organization and truncation scheme is necessary to use the basis in practice.
In this paper, we usually truncate by restricting to the set of all multigraphs with at most $d$ edges.
This is a natural choice because it corresponds to truncating the approximation of the angular function $f_N$ at degree $d$ polynomials.
Furthermore, this truncation results in a \emph{finite} number of \Bs at each order of truncation, which is not true for truncation by the number of vertices.
The number of multigraphs with exactly $d$ edges is Sequence A050535 in the On-Line Encyclopedia of Integer Sequences (OEIS)~\cite{sloane2007line,harary2014graphical}; the number of connected multigraphs with exactly $d$ edges is Sequence A076864 in the OEIS~\cite{sloane2007line}.
The numbers of \Bs in our truncation of the energy flow basis are the partial sums of these sequences, which are listed in \Tab{tab:efpcounts:a} up to $d=10$.
\Tab{tab:efpcounts:b} tabulates the number of prime \Bs of degree $d$ binned by $N$ up to $d=10$.
\Tab{tab:graphs} illustrates all connected multigraphs with $d\le 5$ edges.

\begin{table}[t]
\centering
\begin{tabular}{| >{\centering}m{.5in} | >{\centering}m{5in} |}\hline
\textbf{Degree} & \textbf{Connected Multigraphs} \tabularnewline\hline \hline
$d=0$ & \includegraphics[scale=0.02]{graphs/dot}\tabularnewline\hline
$d=1$ & \vspace{.08in}\includegraphics[scale=0.15]{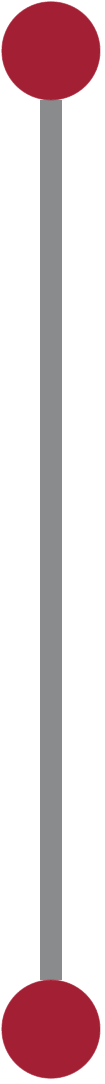}
\tabularnewline\hline
$d=2$ & \vspace{.08in}\includegraphics[scale=0.15]{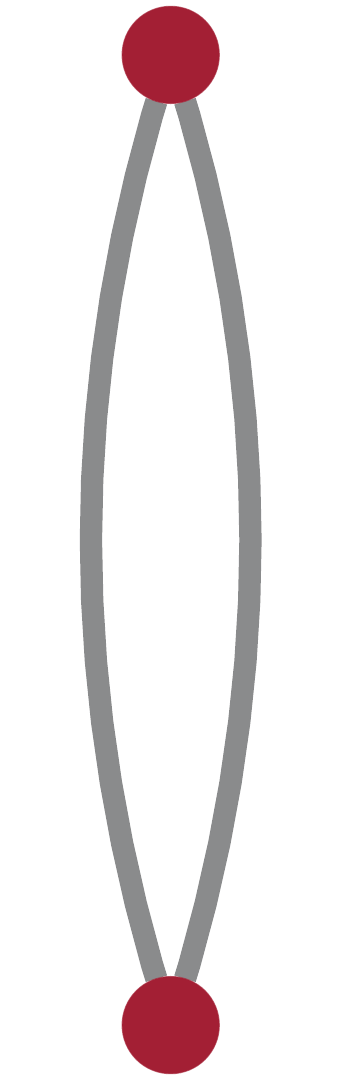}
                                     \includegraphics[scale=0.15]{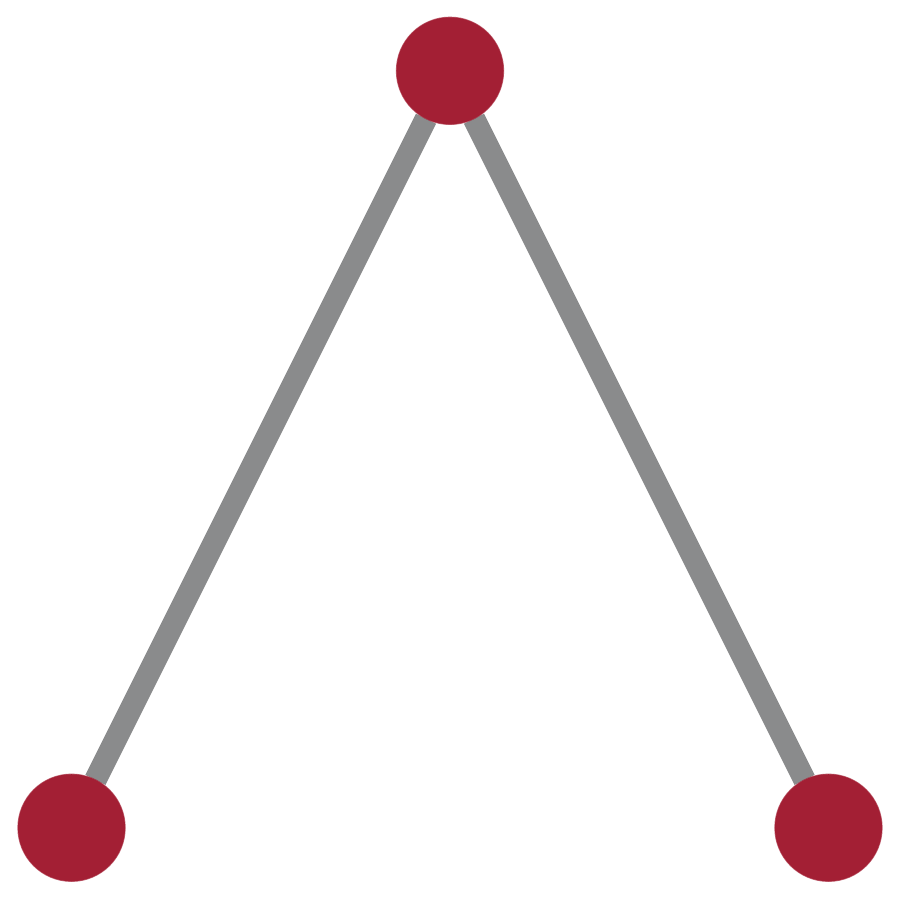} 
\tabularnewline\hline
$d=3$ & \vspace{.08in}\includegraphics[scale=0.15]{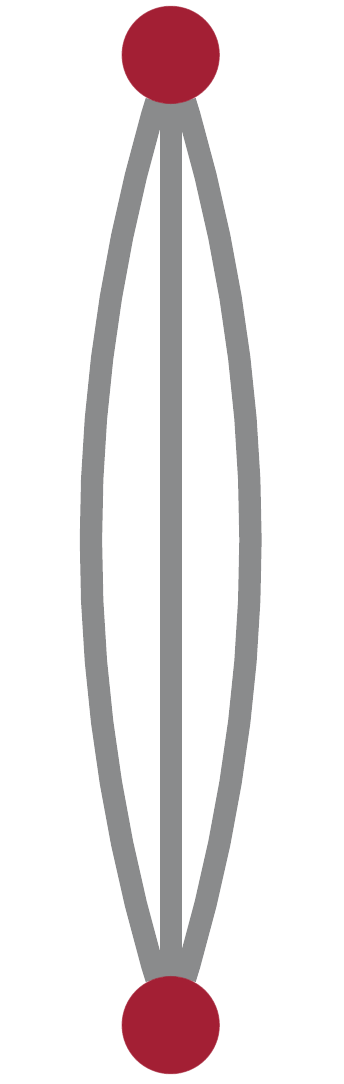}
                                     \includegraphics[scale=0.15]{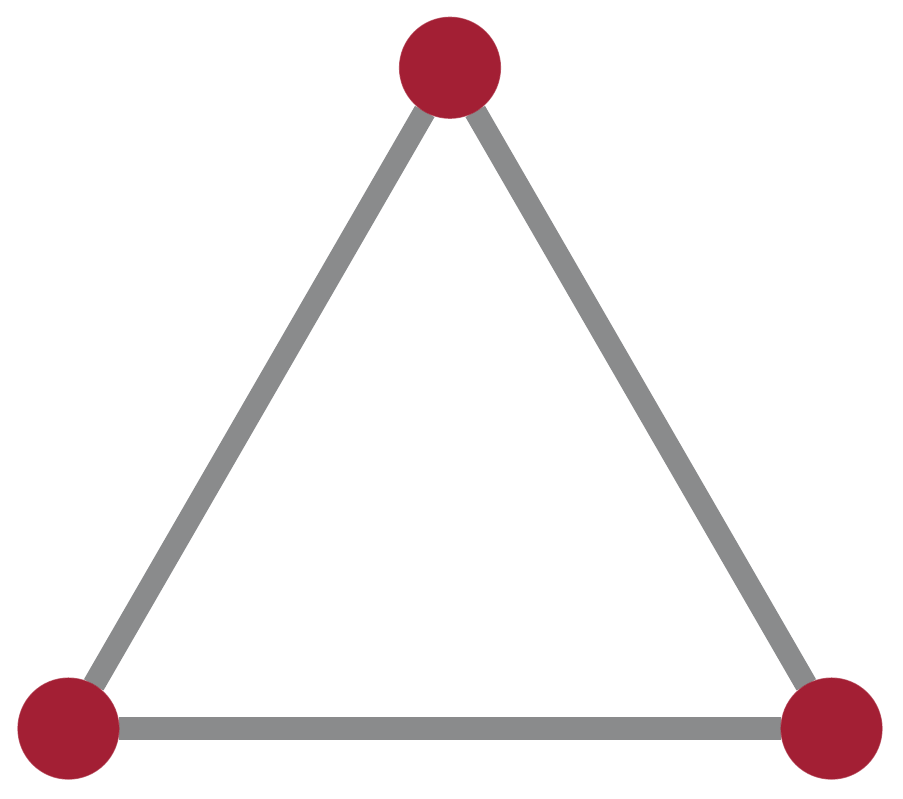}
                                     \includegraphics[scale=0.15]{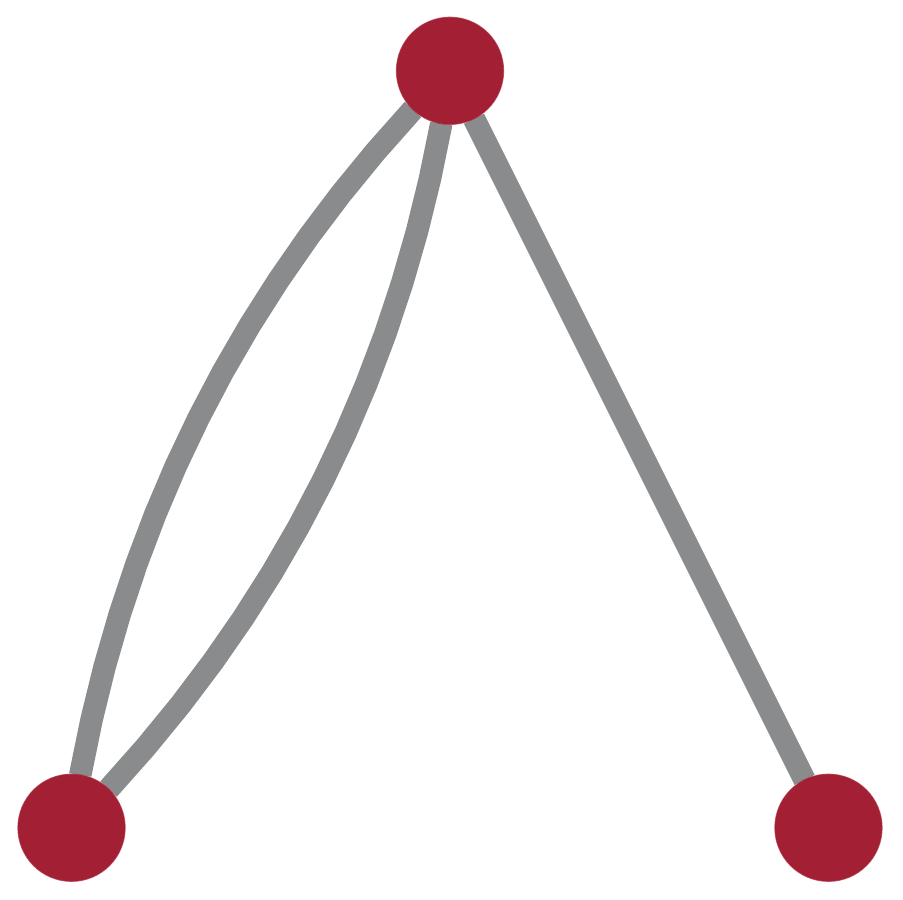}
                                     \includegraphics[scale=0.2]{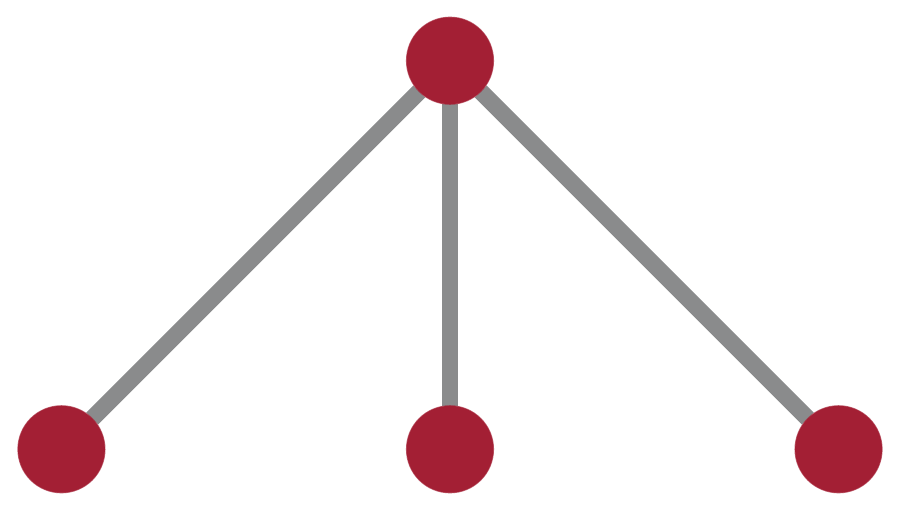}
                                     \includegraphics[scale=0.15]{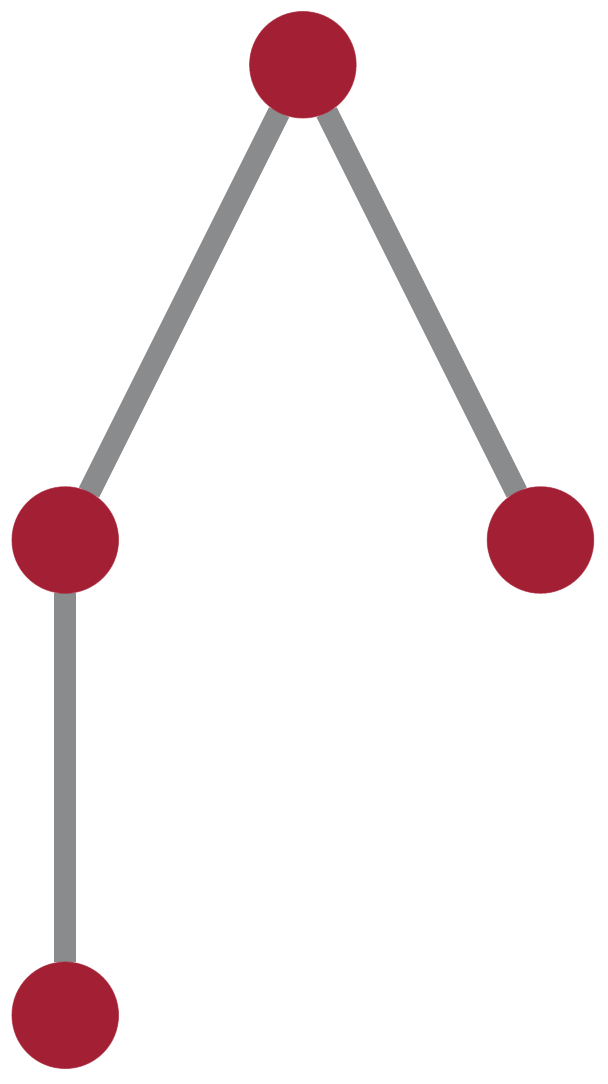}
\tabularnewline\hline
\multirow{5}{*}{$d=4$} & \vspace{.08in}
                                     \includegraphics[scale=0.15]{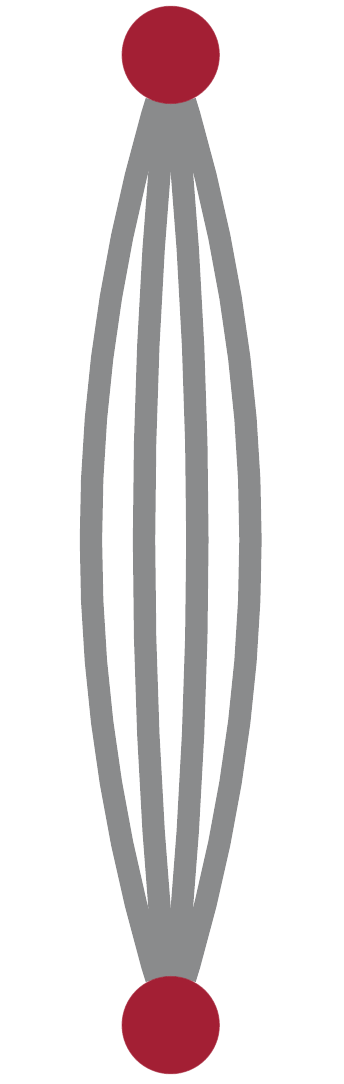}
                                     \includegraphics[scale=0.15]{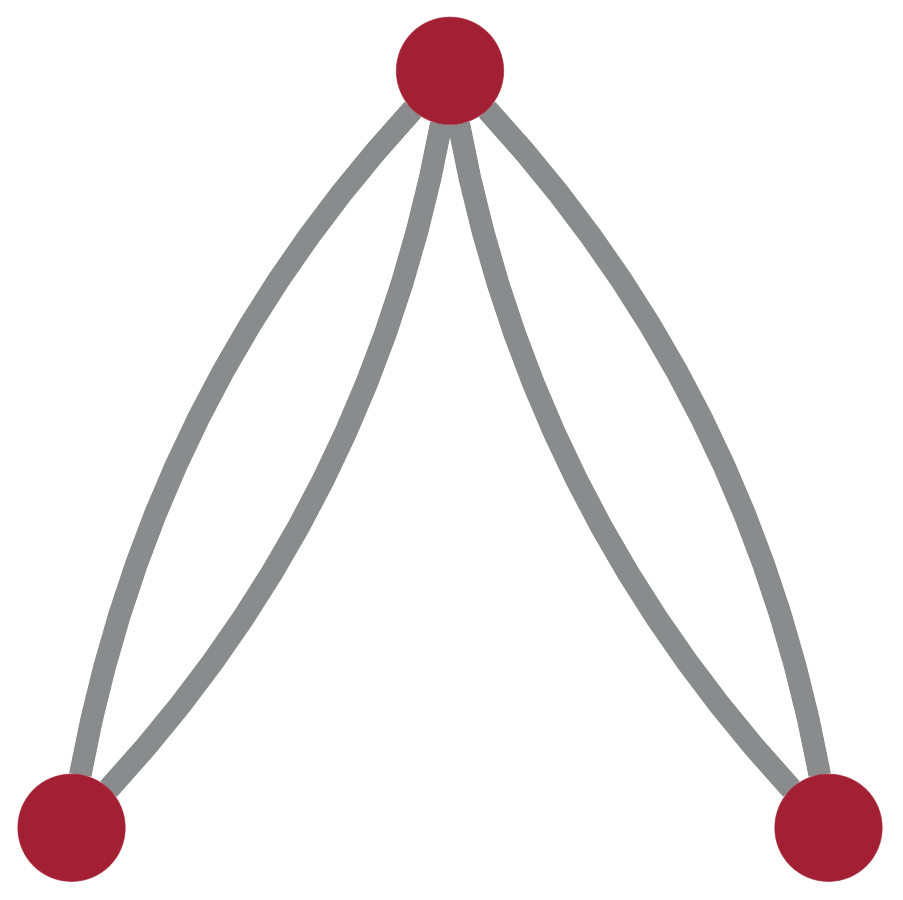}
                                     \includegraphics[scale=0.15]{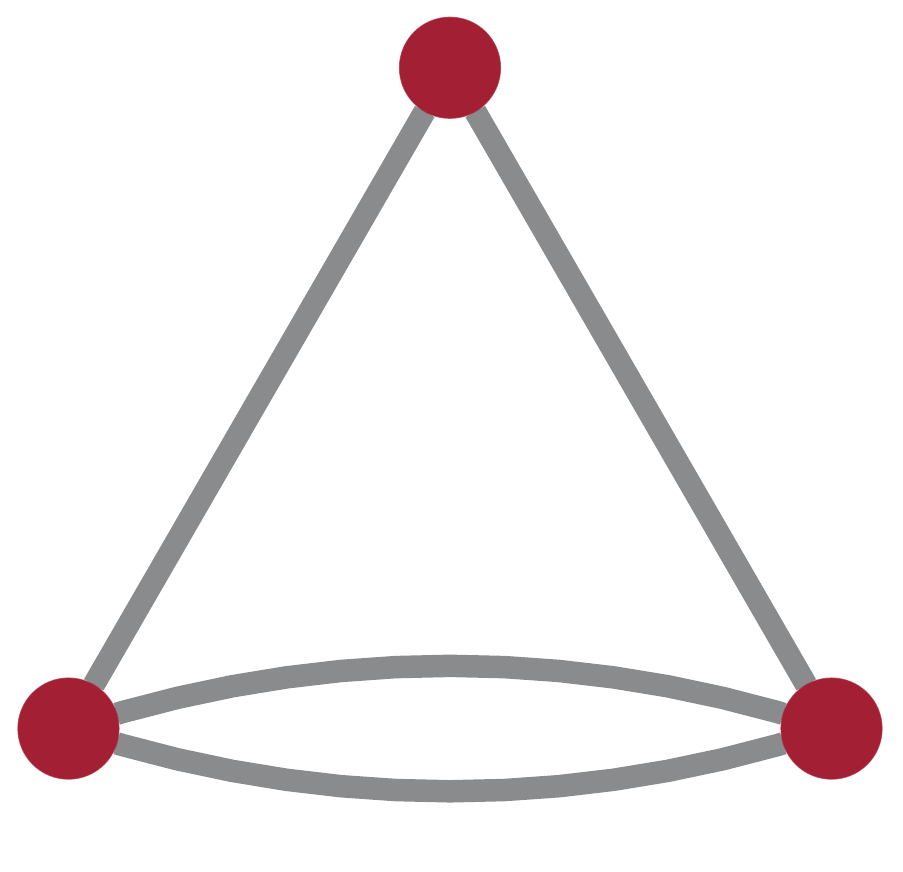}
                                     \includegraphics[scale=0.15]{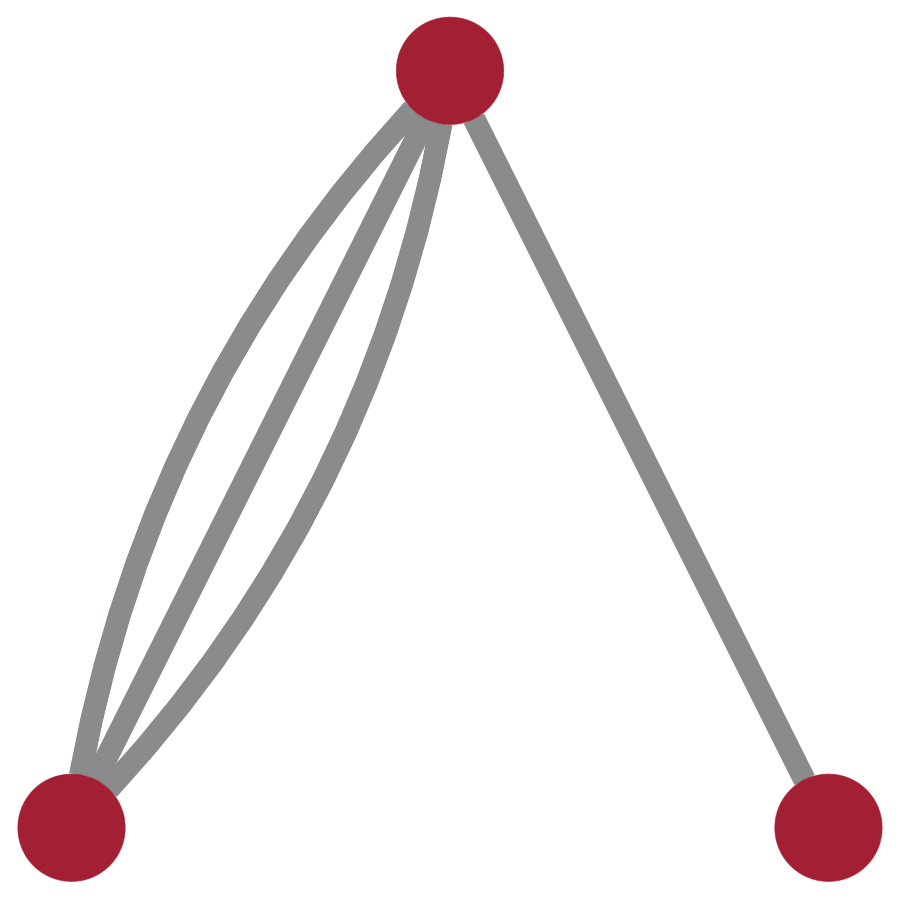}
                                     \includegraphics[scale=0.15]{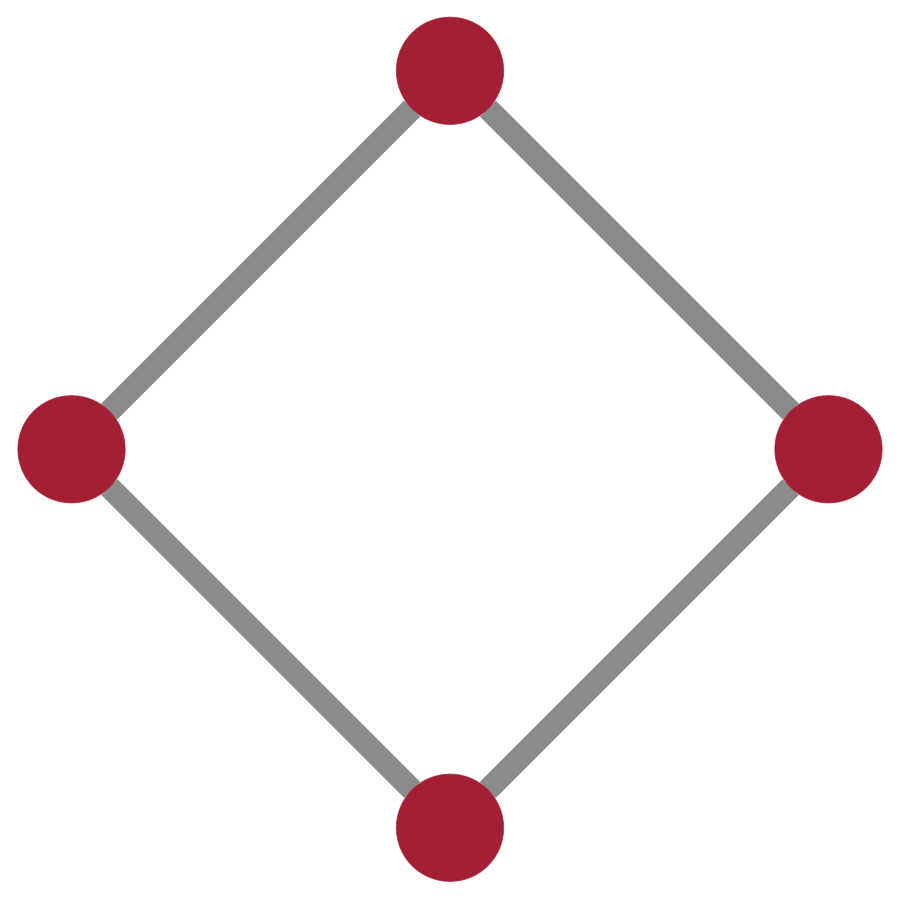}
                                     \includegraphics[scale=0.2]{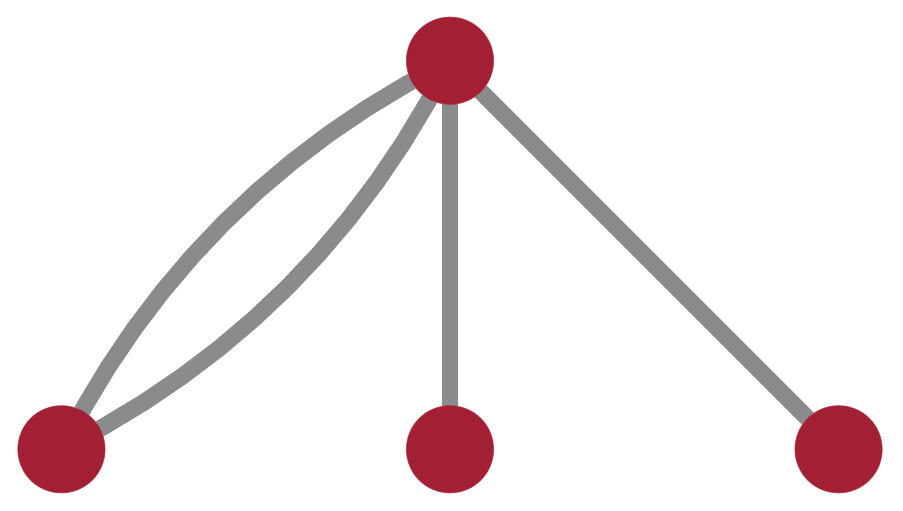}
                                     \tabularnewline &
                                     \includegraphics[scale=0.15]{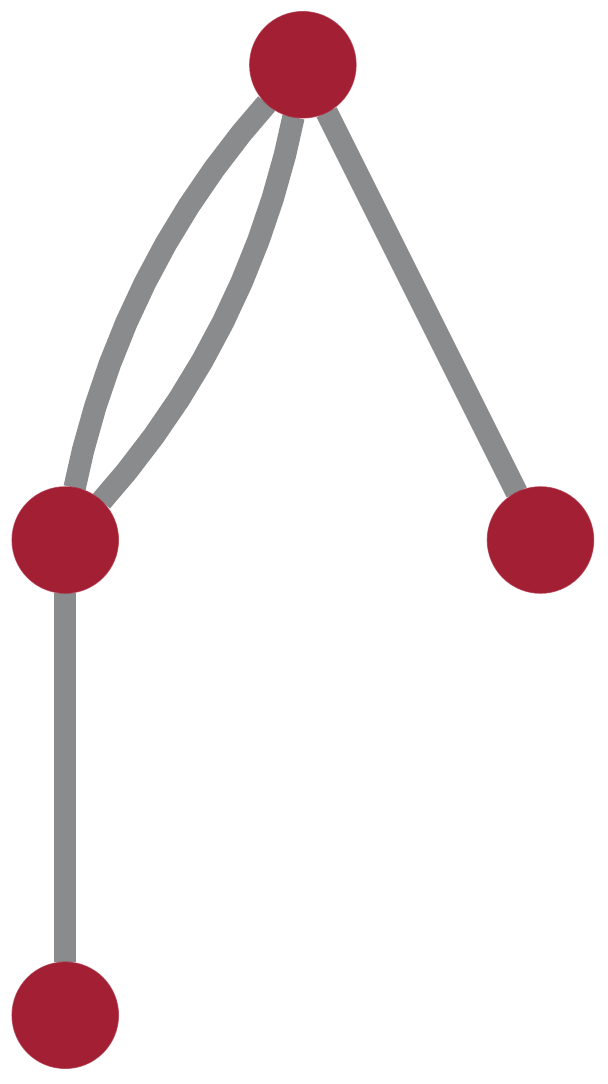}
                                     \rotatebox[origin=t]{270}{\includegraphics[scale=0.2]{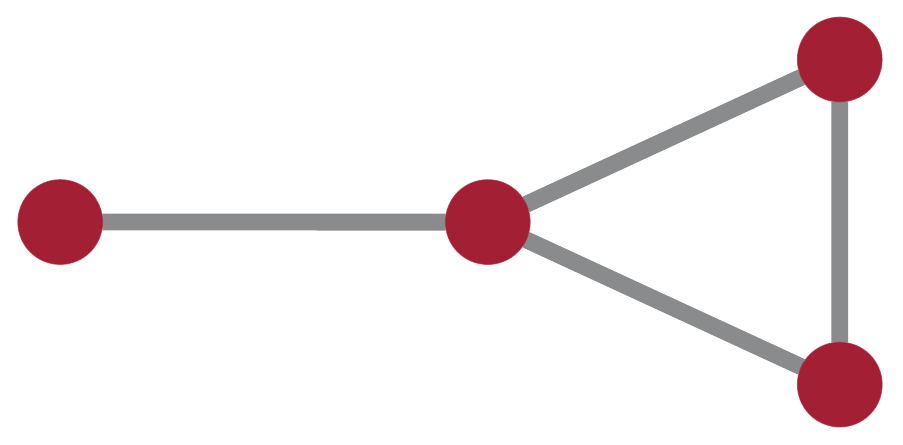}}
                                     \includegraphics[scale=0.15]{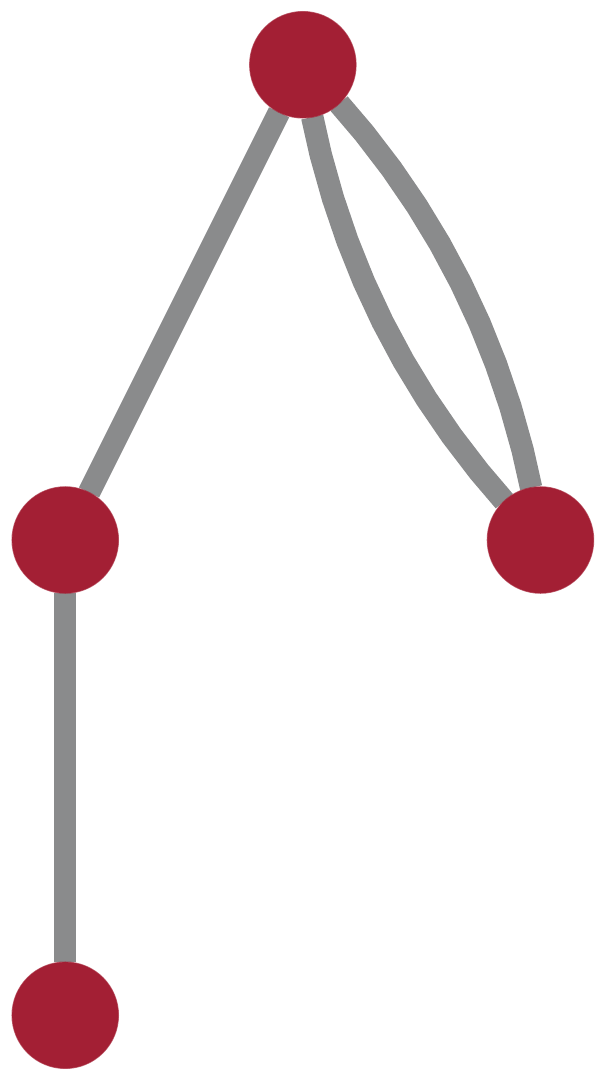}
                                     \includegraphics[scale=0.2]{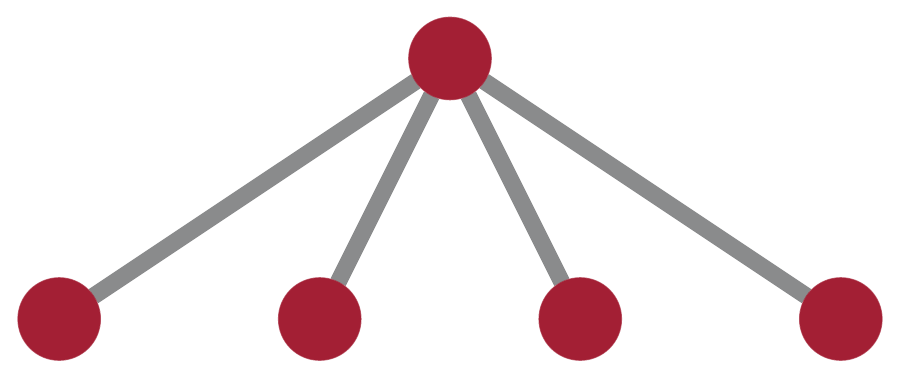}
                                     \includegraphics[scale=0.15]{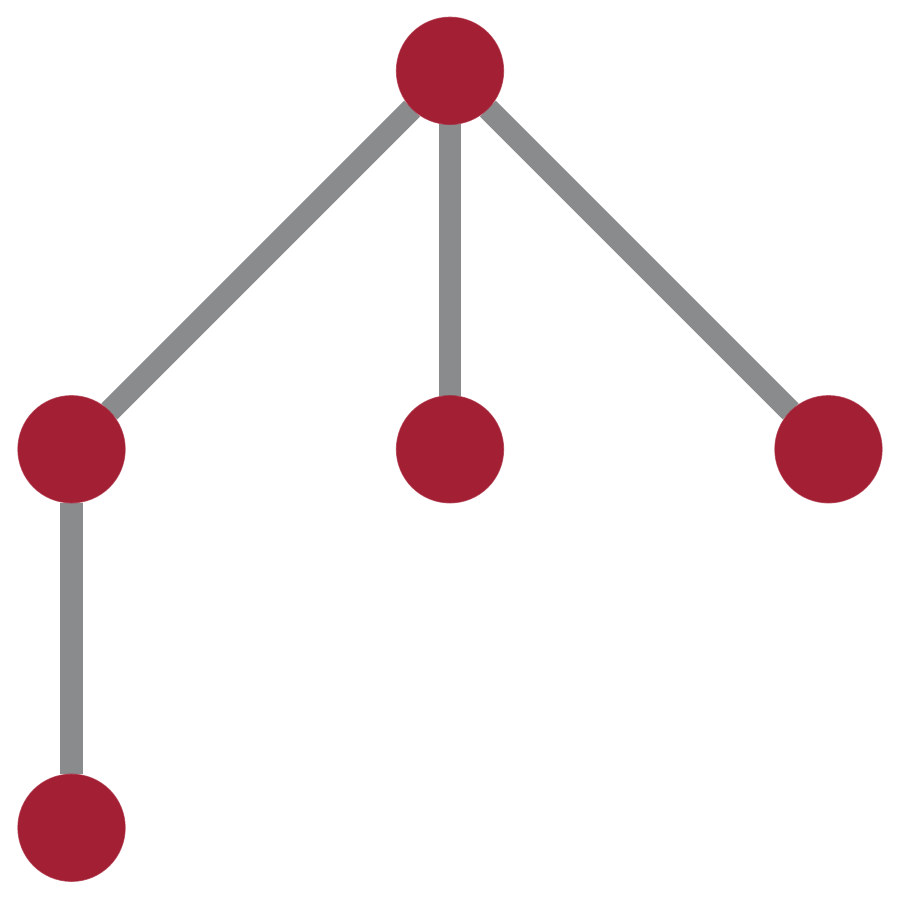}
                                     \includegraphics[scale=0.15]{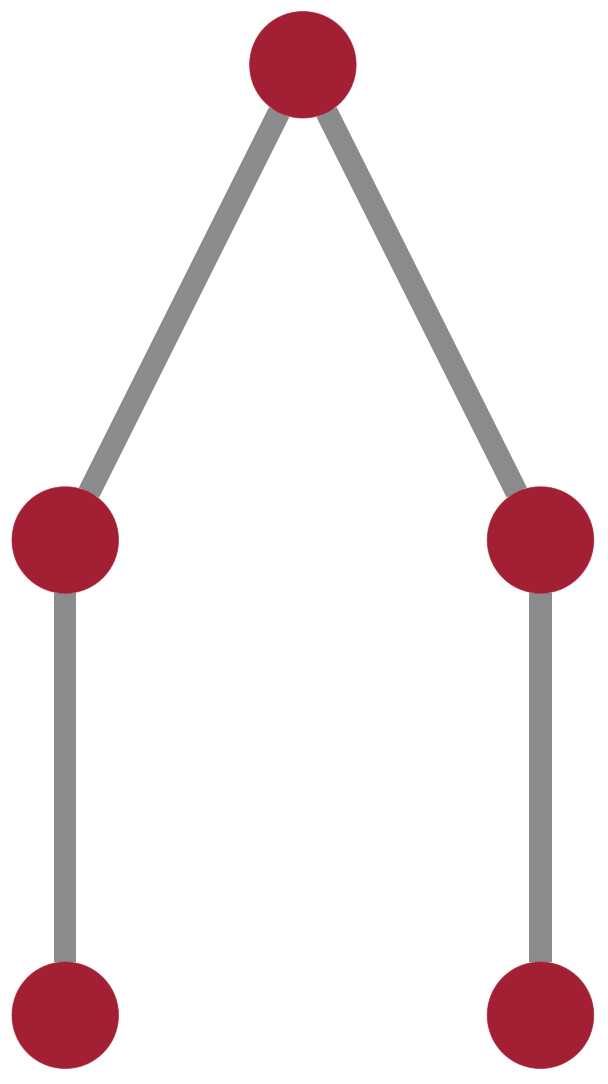}
\tabularnewline\hline
\multirow{12}{*}{$d=5$} & \vspace{.08in}
                                     \includegraphics[scale=0.15]{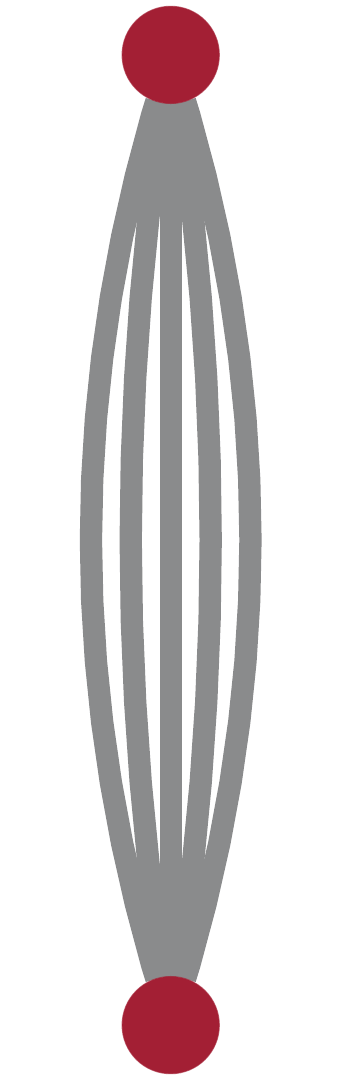}
                                     \includegraphics[scale=0.15]{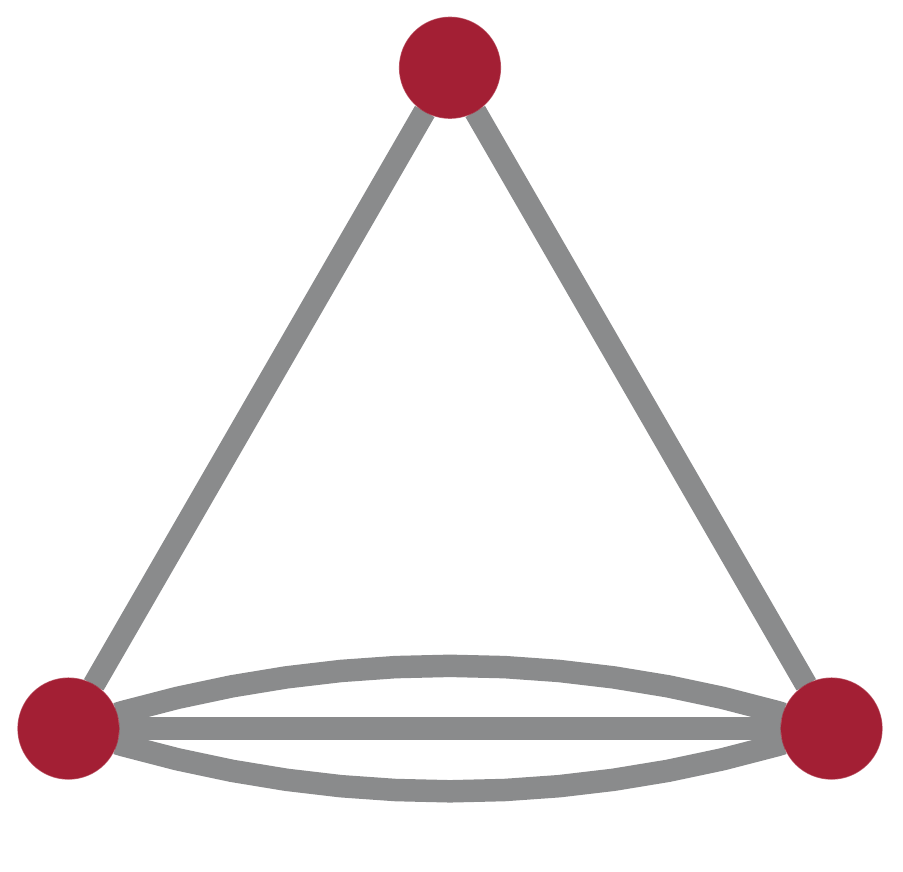}
                                     \includegraphics[scale=0.15]{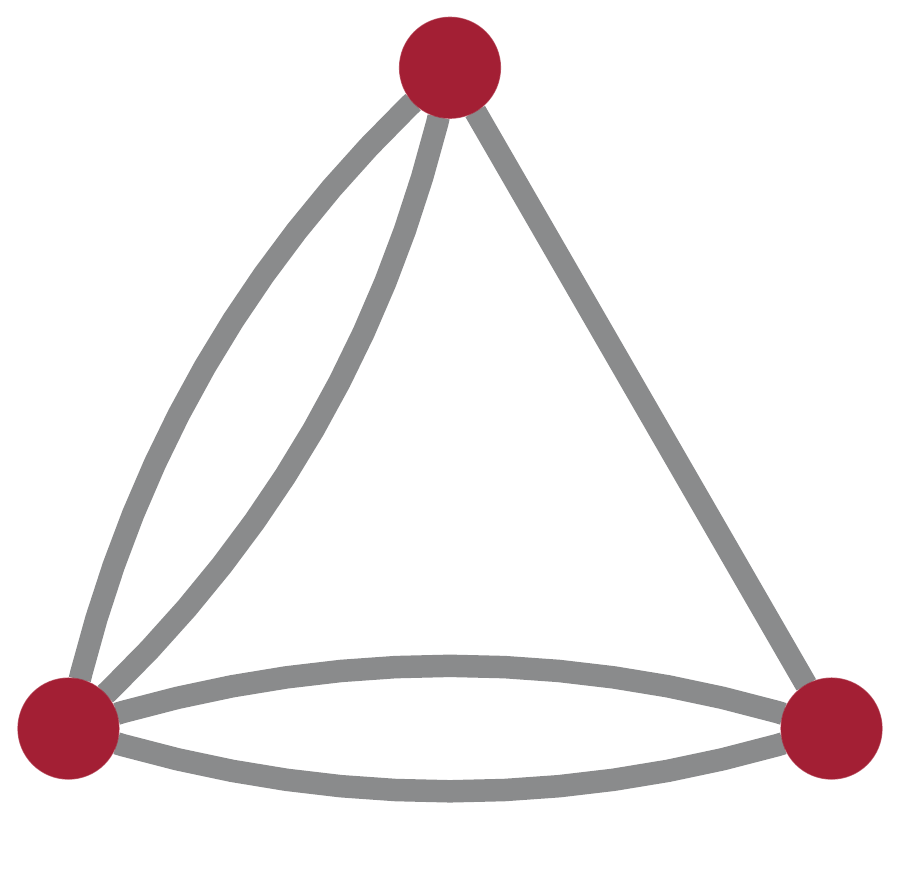}
                                     \includegraphics[scale=0.15]{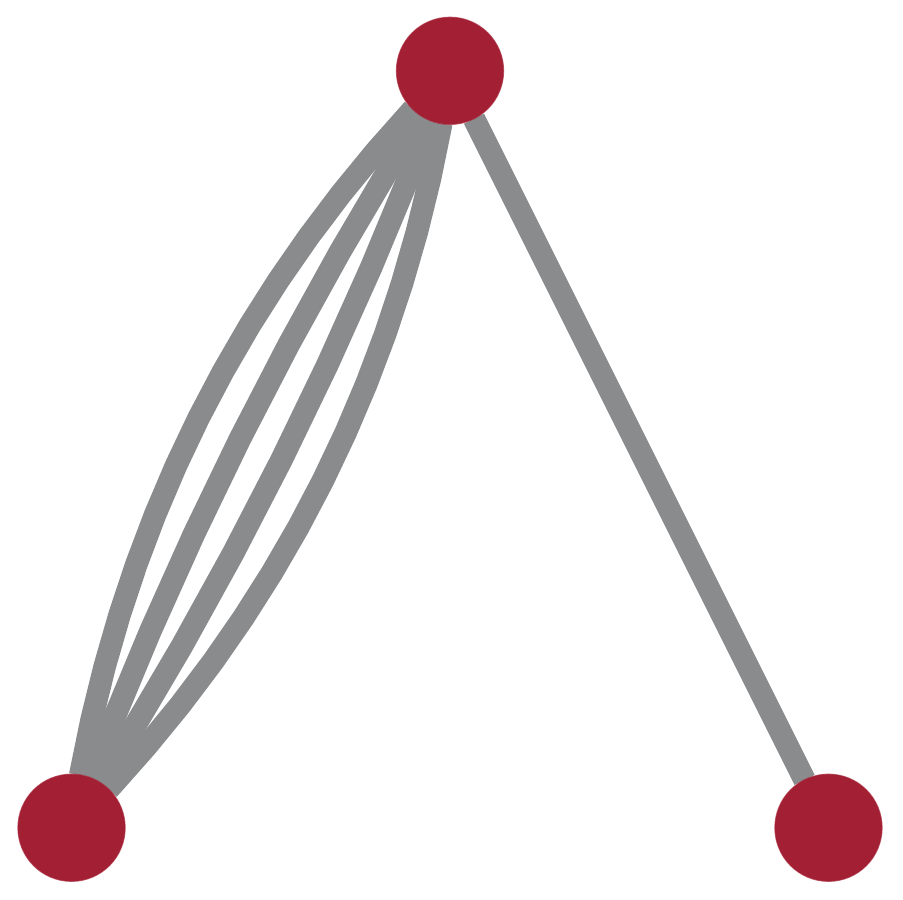}
                                     \includegraphics[scale=0.15]{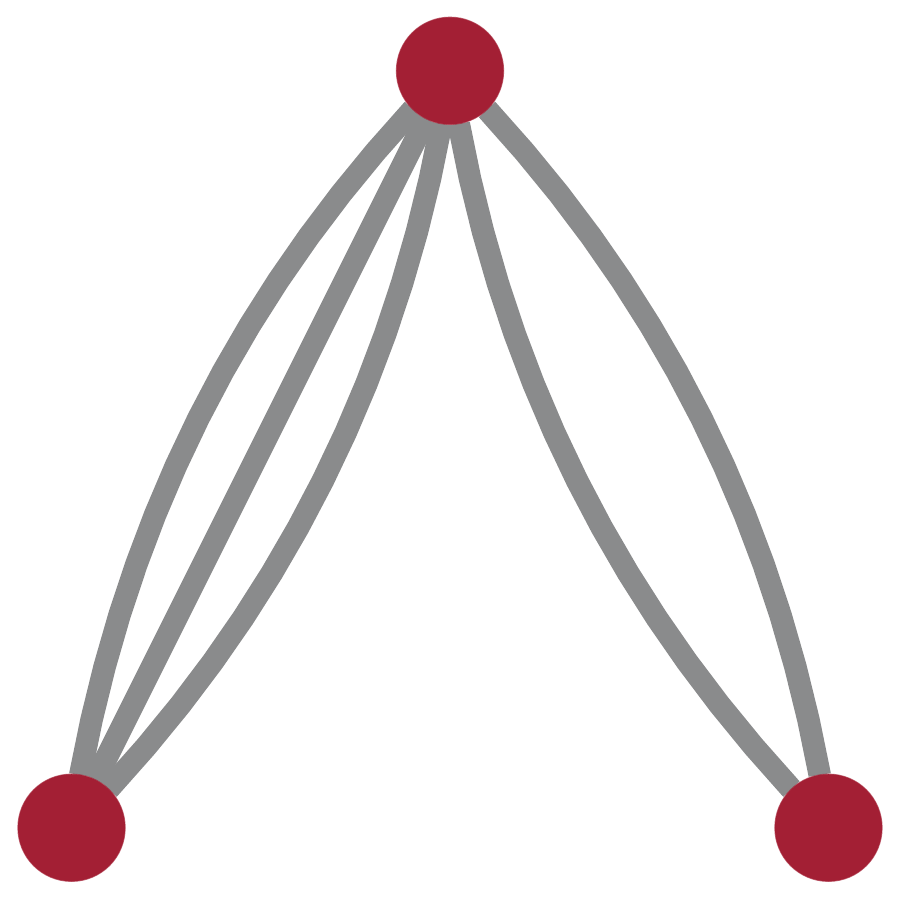}
                                     \rotatebox[origin=t]{270}{\includegraphics[scale=0.165]{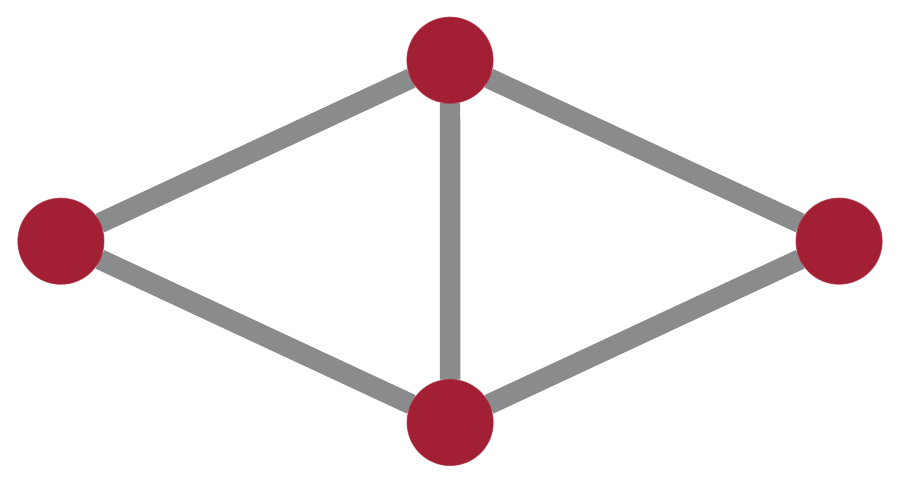}}
                                     \includegraphics[scale=0.175]{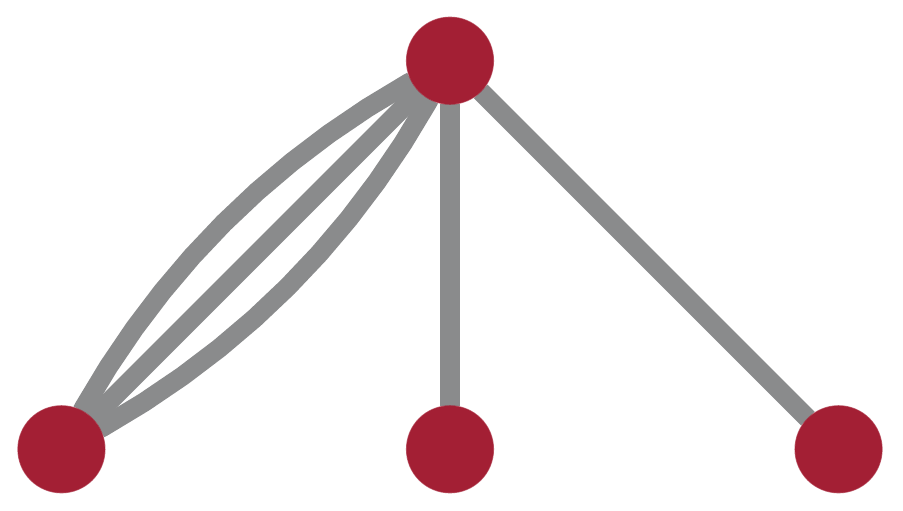}
                                     \includegraphics[scale=0.175]{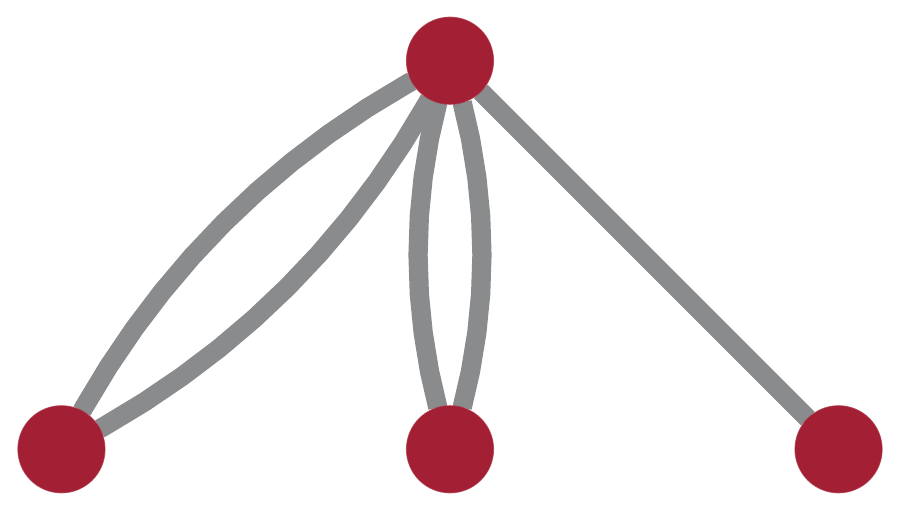}
                                     \tabularnewline &
                                     \includegraphics[scale=0.15]{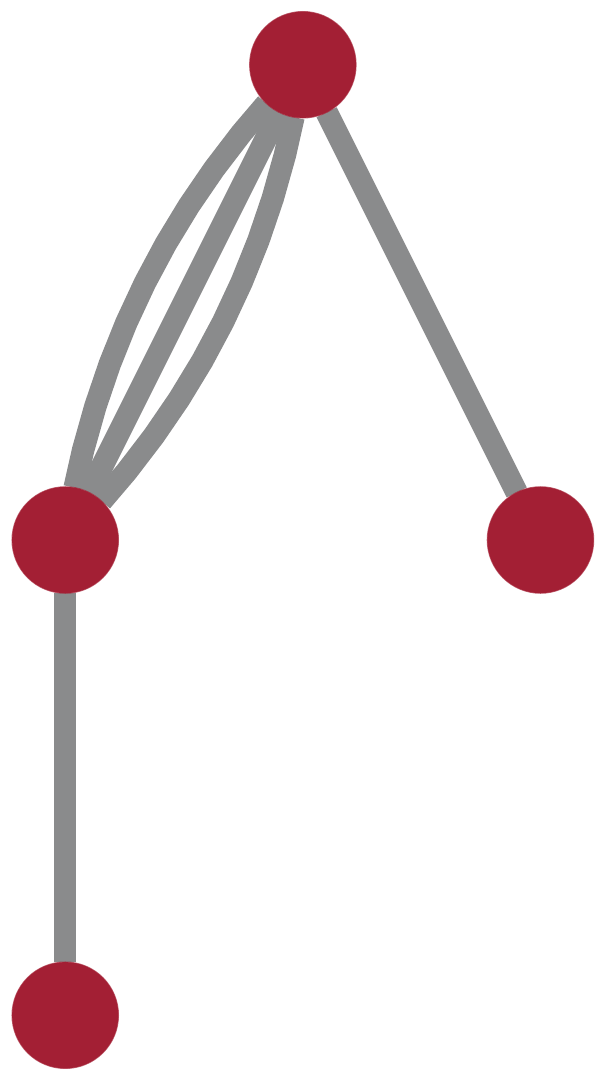}
                                     \rotatebox[origin=t]{270}{\includegraphics[scale=0.175]{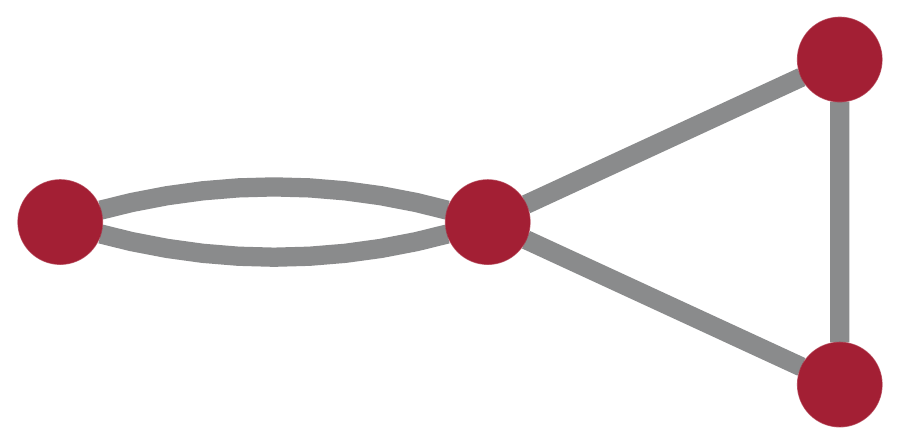}}
                                     \rotatebox[origin=t]{270}{\includegraphics[scale=0.175]{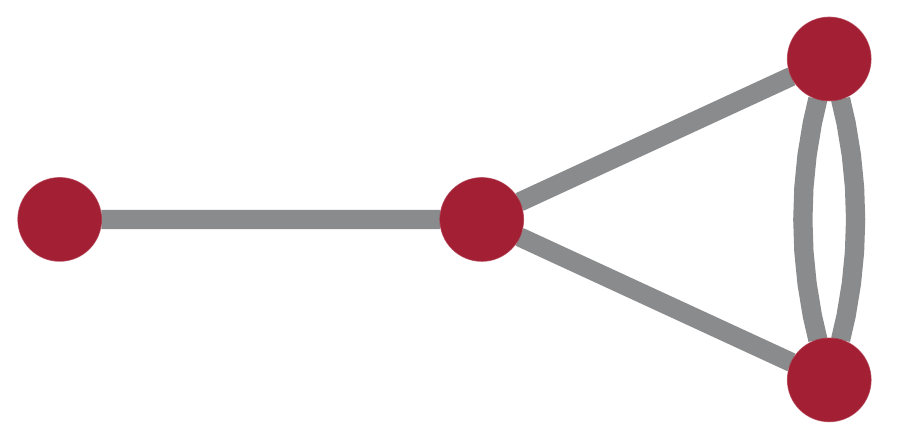}}
                                     \rotatebox[origin=t]{270}{\includegraphics[scale=0.175]{graphs/4_5_9}}
                                     \includegraphics[scale=0.15]{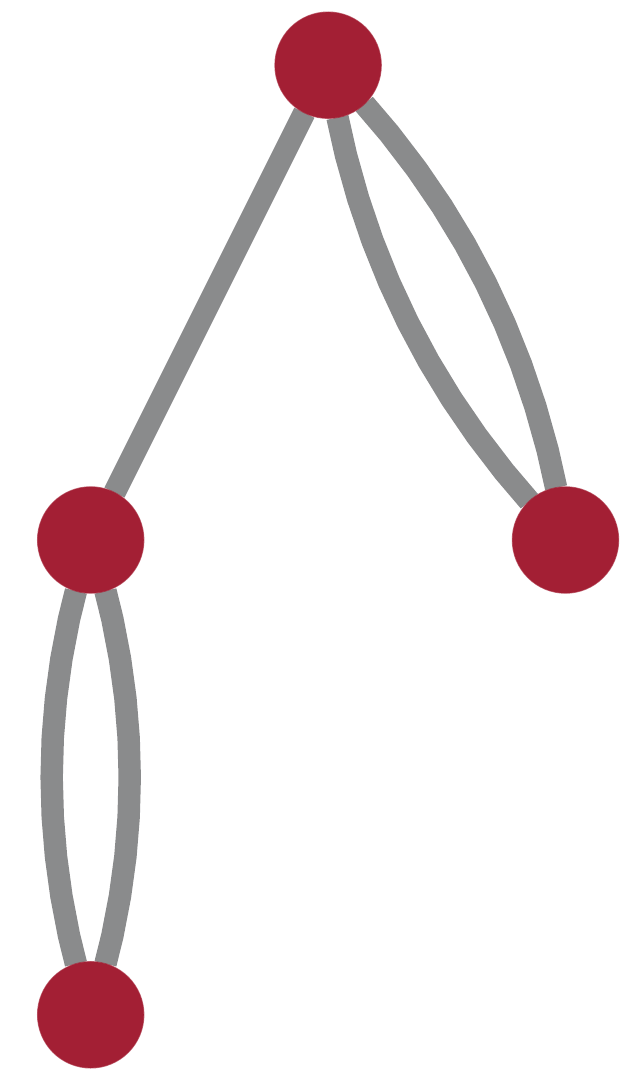}
                                     \includegraphics[scale=0.15]{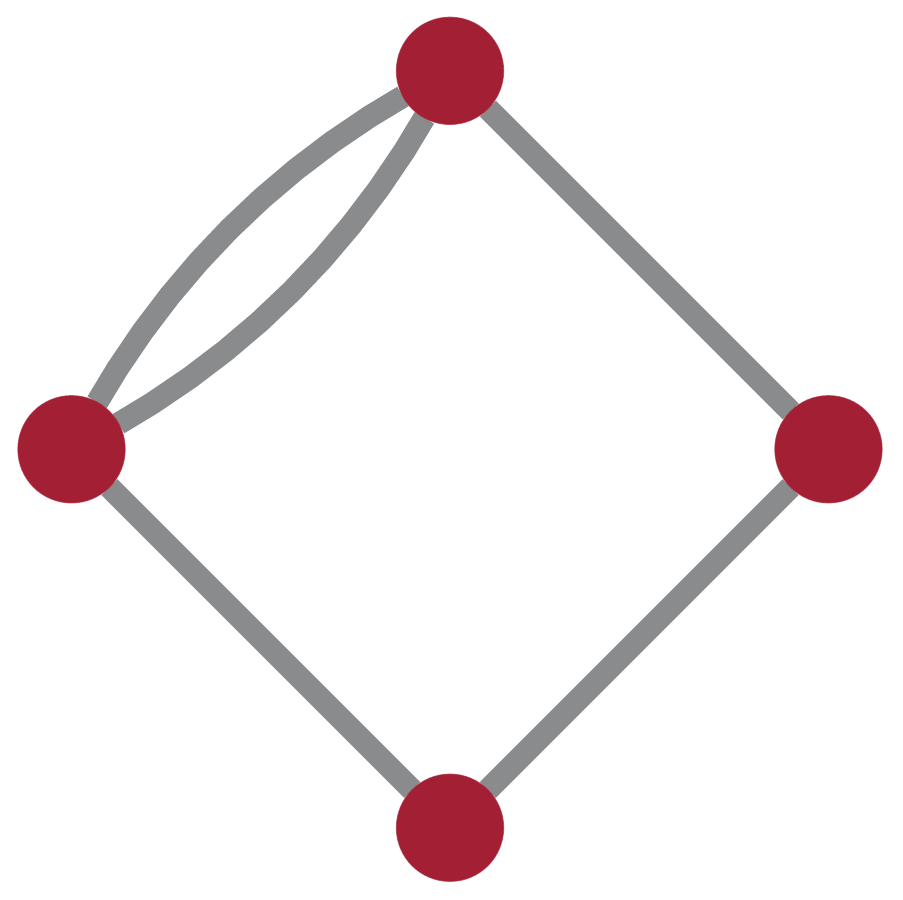}
                                     \includegraphics[scale=0.15]{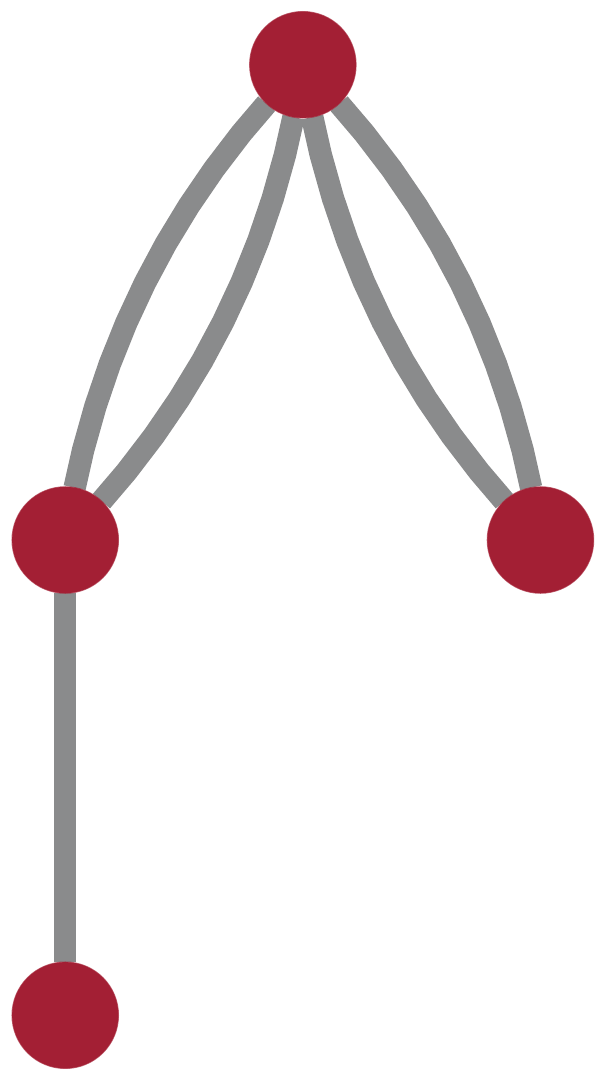}
                                     \includegraphics[scale=0.15]{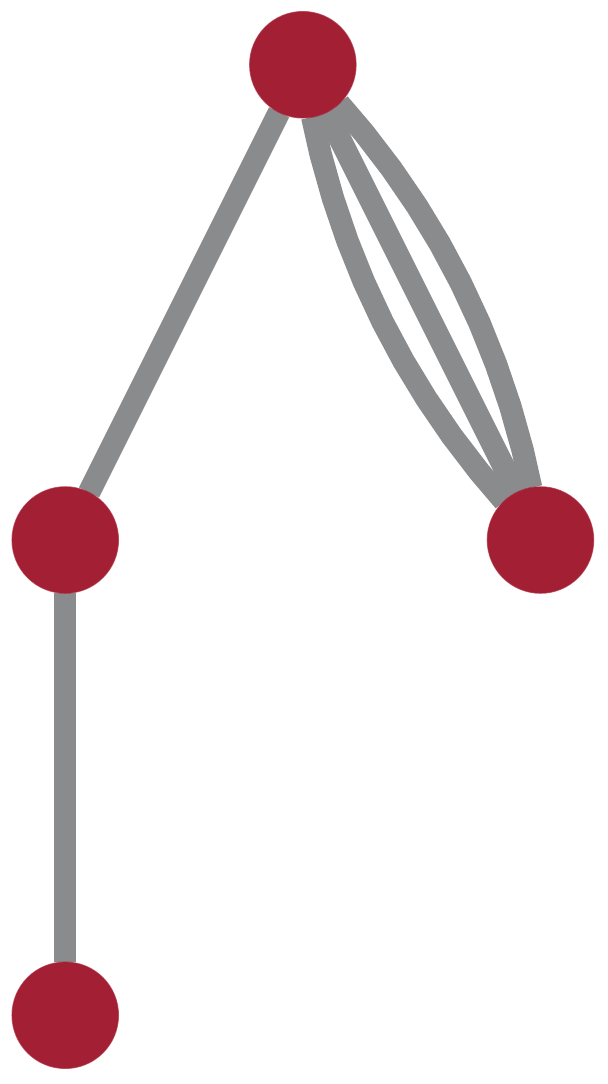}
                                     \includegraphics[scale=0.15]{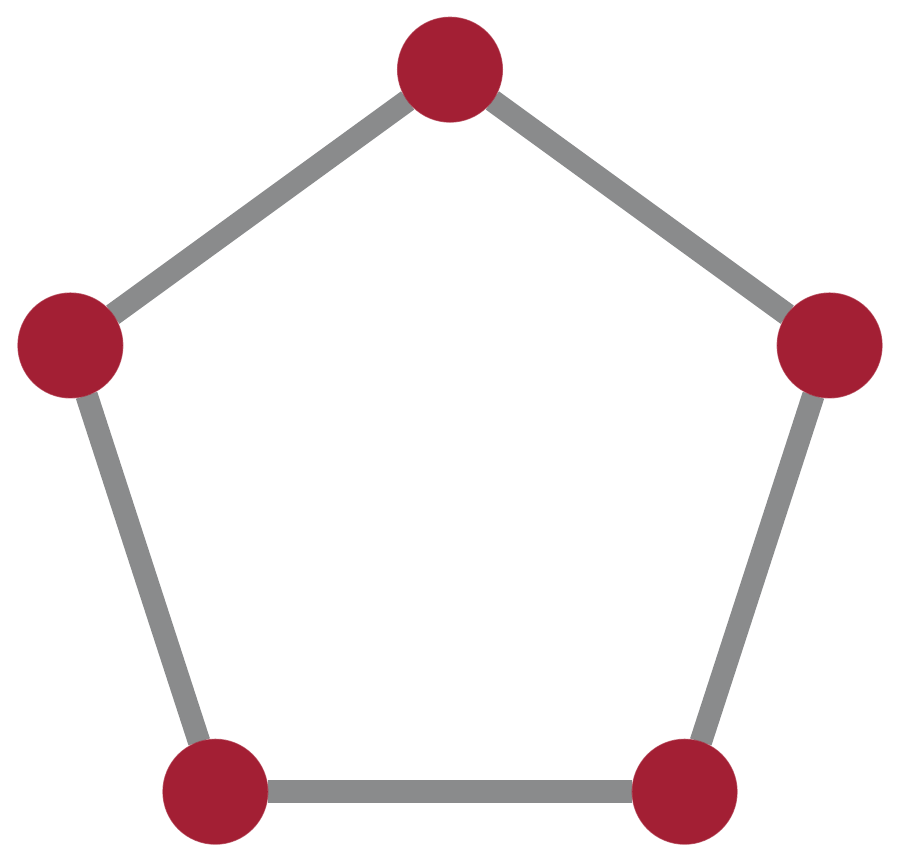}
                                     \includegraphics[scale=0.175]{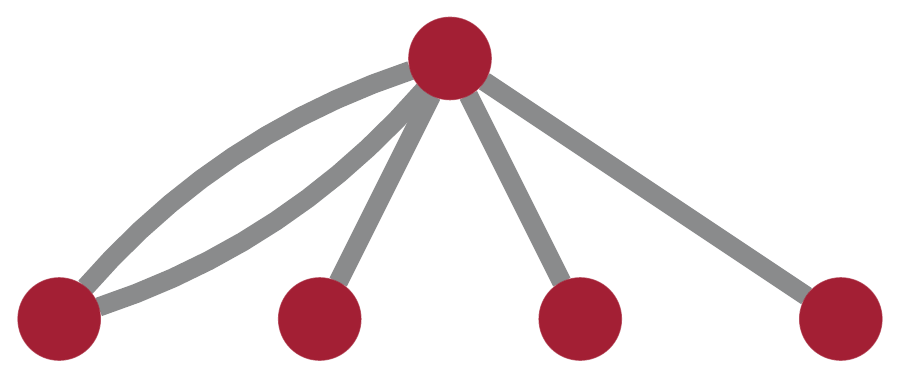}
                                     \tabularnewline &
                                     \rotatebox[origin=c]{270}{\includegraphics[scale=0.15]{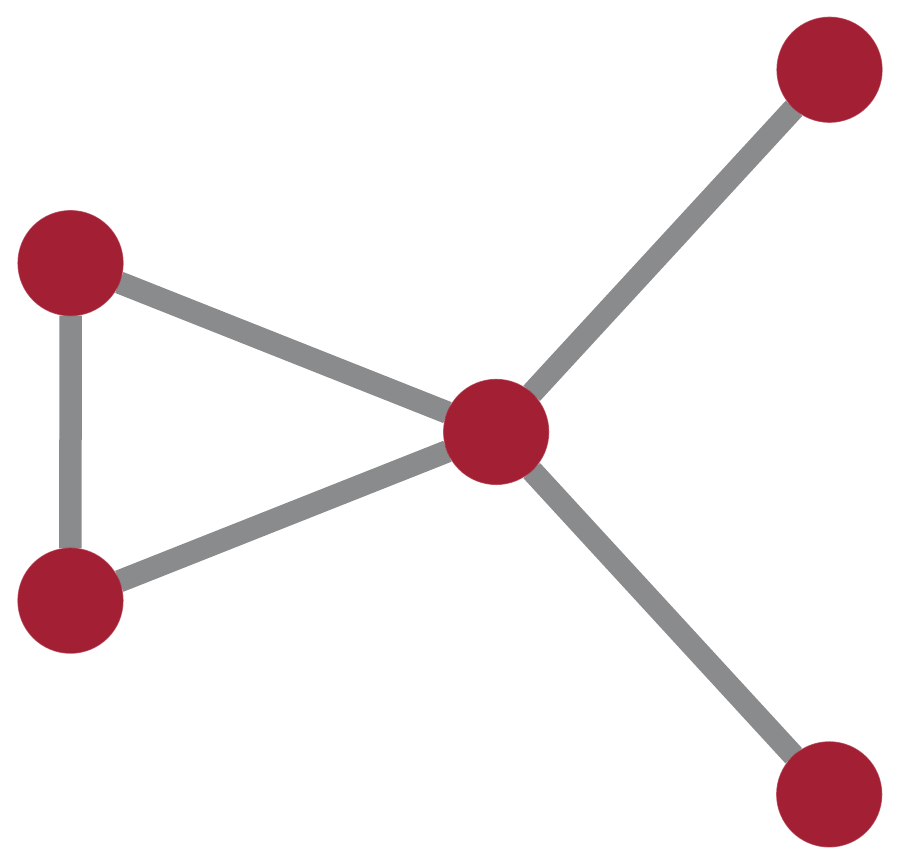}}
                                     \includegraphics[scale=0.15]{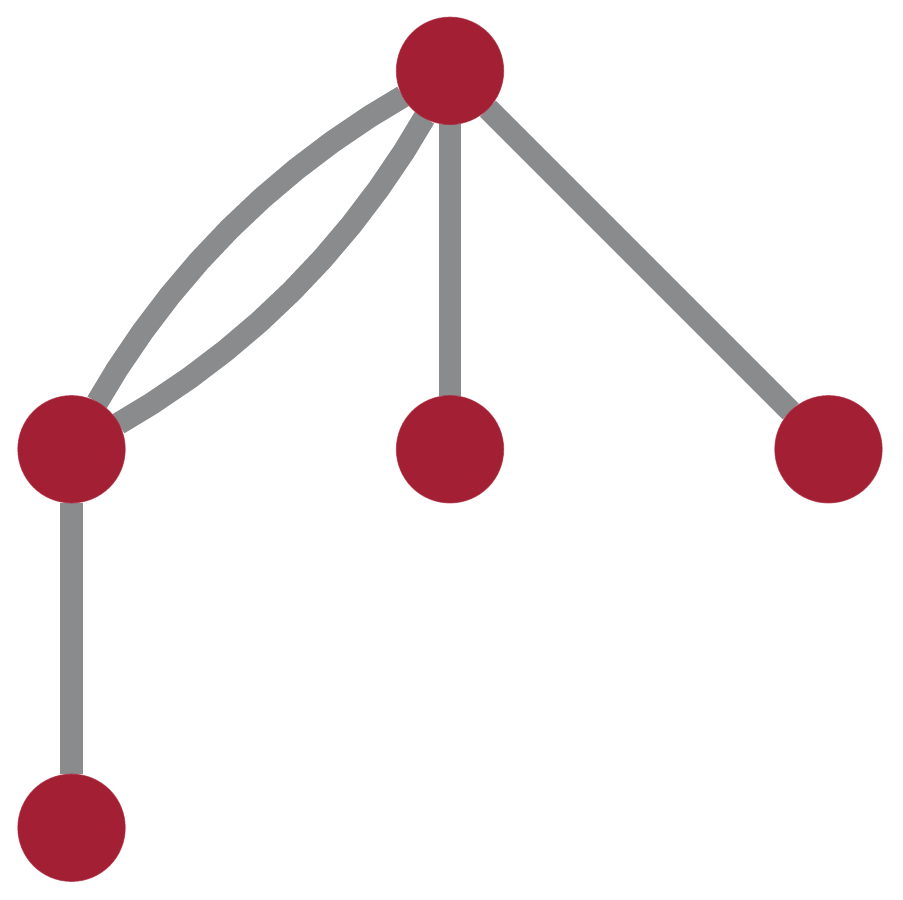}
                                     \includegraphics[scale=0.15]{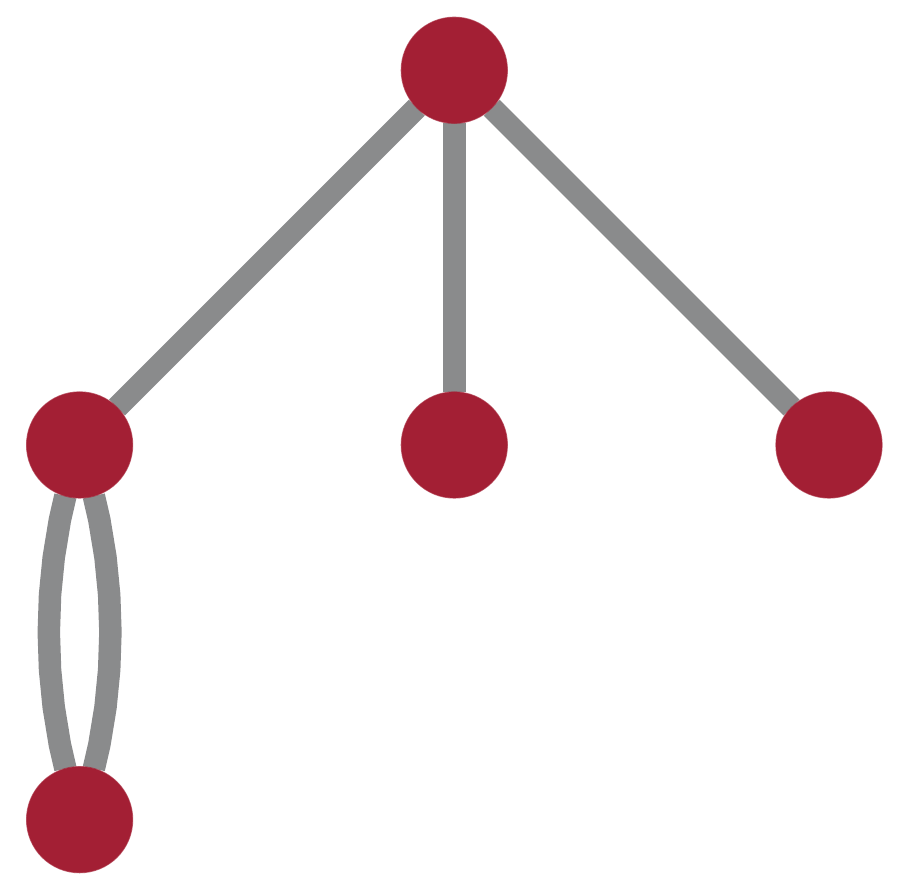}
                                     \rotatebox[origin=t]{270}{\includegraphics[scale=0.175]{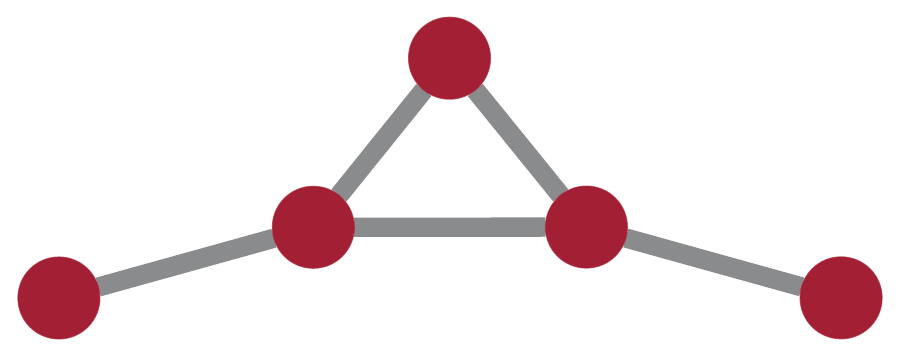}}
                                     \rotatebox[origin=t]{270}{\includegraphics[scale=0.175]{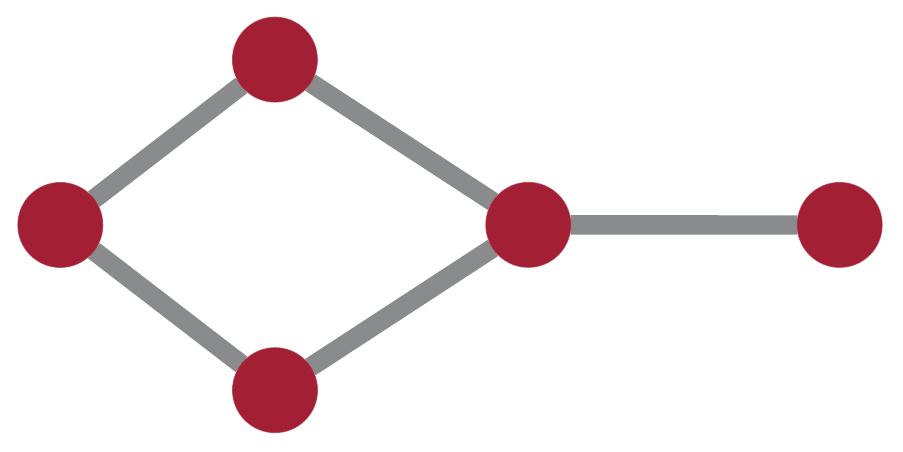}}
                                     \rotatebox[origin=t]{90}{\includegraphics[scale=0.175]{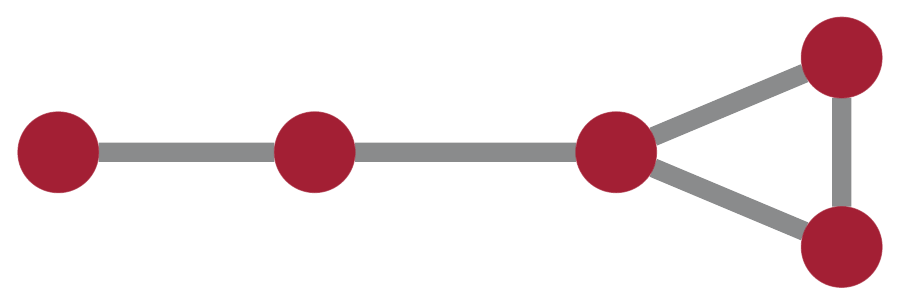}}
                                     \includegraphics[scale=0.15]{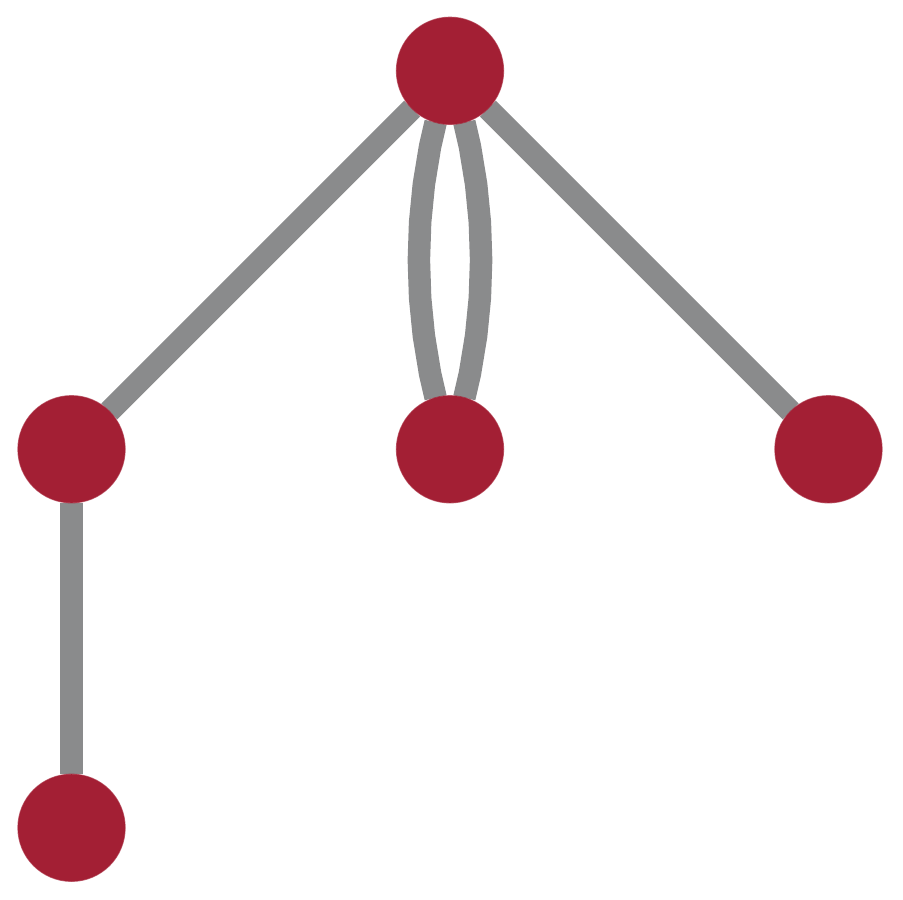}
                                     \includegraphics[scale=0.15]{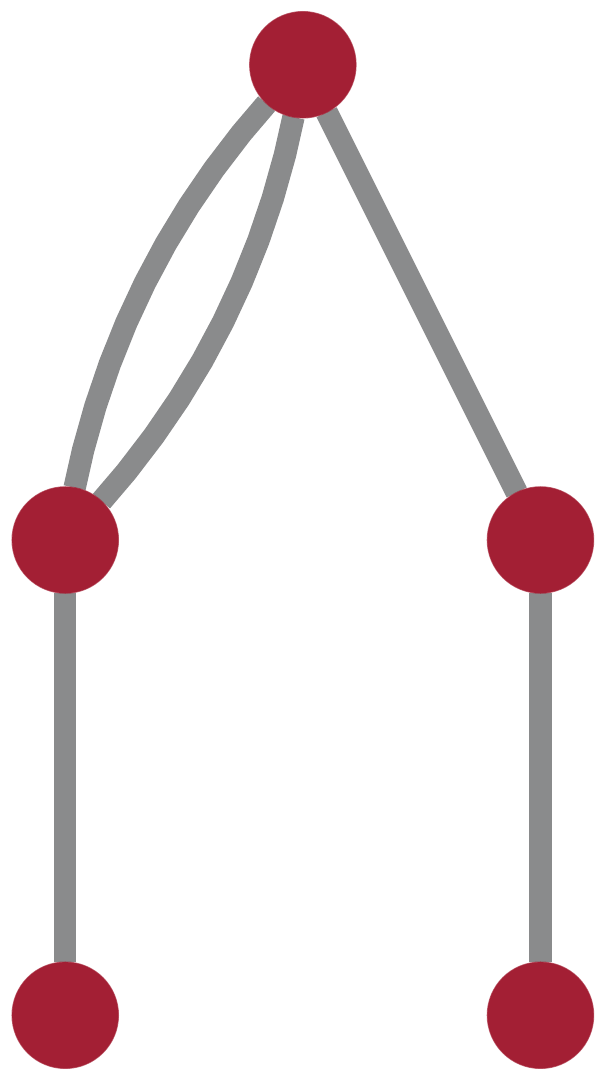}
                                     \includegraphics[scale=0.15]{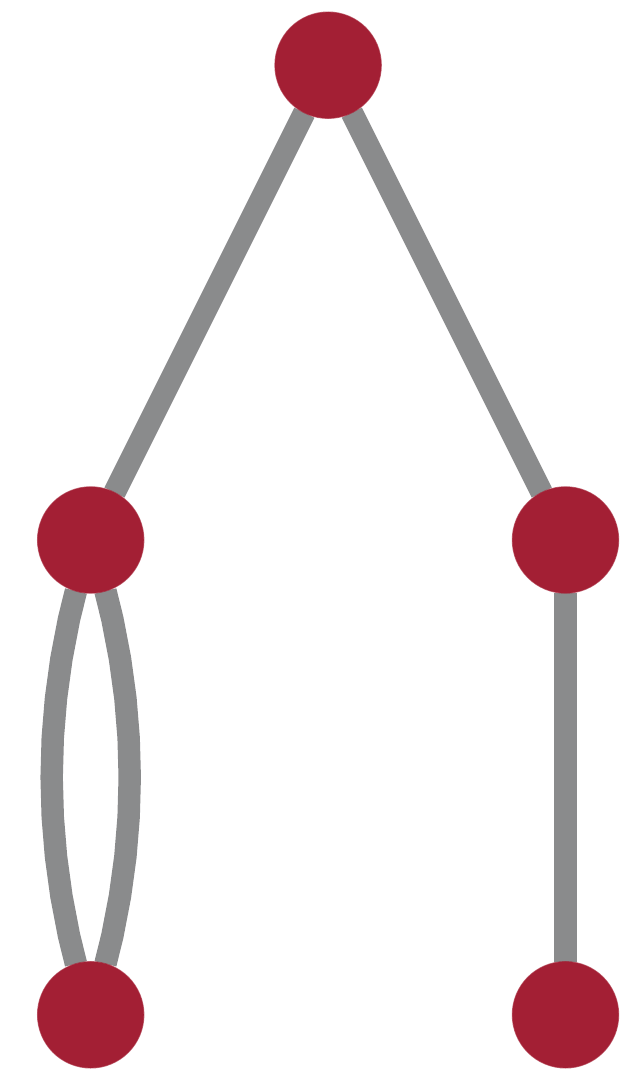}
                                     \includegraphics[scale=0.175]{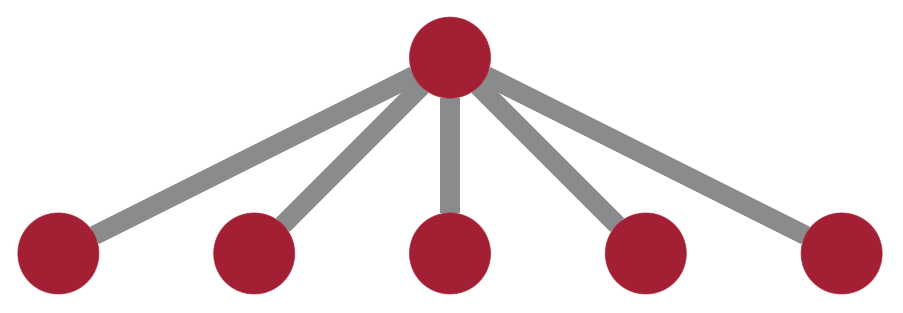}
                                     \tabularnewline & 
                                     \includegraphics[scale=0.15]{graphs/6_5_2}
                                     \includegraphics[scale=0.175]{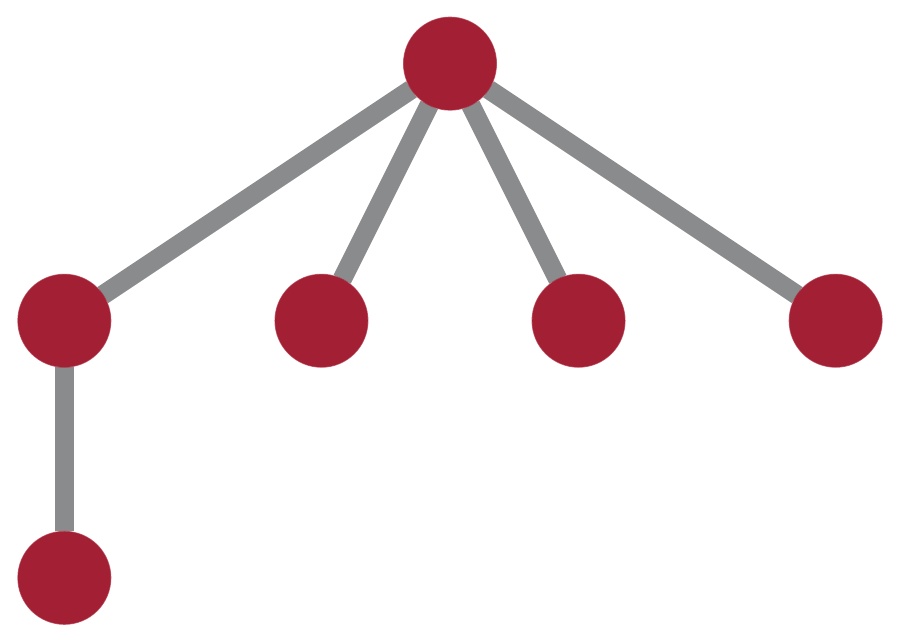}
                                     \includegraphics[scale=0.15]{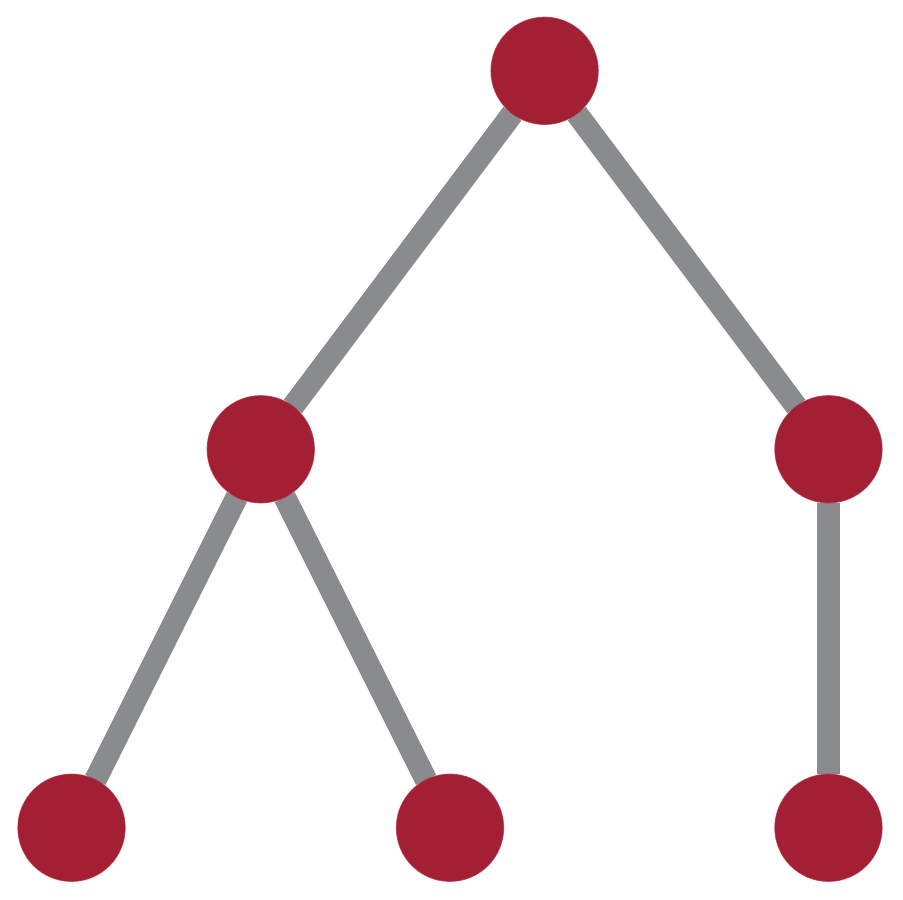}
                                     \includegraphics[scale=0.15]{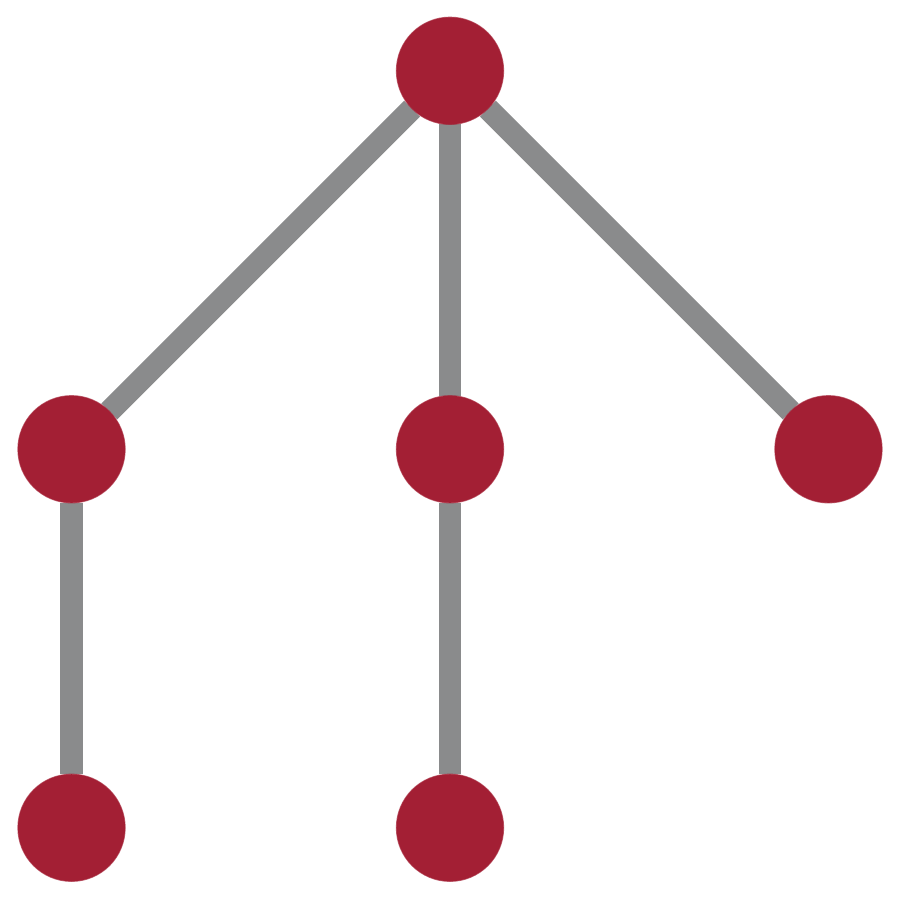}
                                     \includegraphics[scale=0.15]{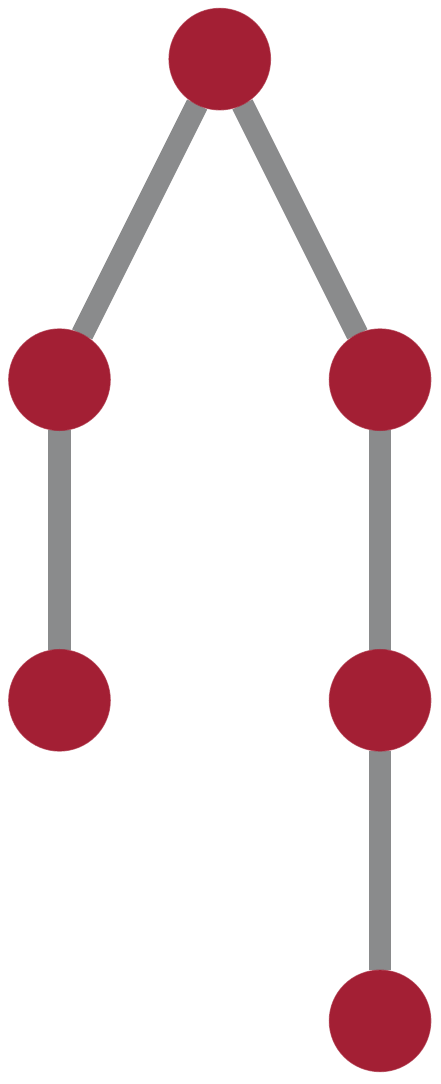}
\tabularnewline\hline
\end{tabular}
\caption{All non-isomorphic, loopless, connected multigraphs organized by the total number of edges $d$, up to $d=5$, sorted by their number of vertices $N$.  Note that for a fixed number of edges $d$, the total number of multigraphs (connected or not) is finite. These graphs correspond to the $d\le5$ prime \Bs counted in \Tab{tab:efpcounts:a}. Image files for all of the prime \B multigraphs up to $d=7$ are available \href{https://github.com/pkomiske/EnergyFlow/tree/images/graphs}{here}.}
\label{tab:graphs}
\end{table}

\afterpage{\clearpage}

\subsection{Energy and angular measures }
\label{sec:measures}

There are many possible choices for the energy fraction $z_i$ and angular measure $\theta_{ij}$ used to define the \Bs.
In the analysis of \Sec{sec:basis}, this choice arises because there are many systematic expansions of IRC-safe observables in terms of energy-like and angular-like quantities.
Typically, one wants to work with observables that respect the appropriate Lorentz subgroup for the collision type of interest.
For $e^+e^-$ colliders, the symmetries are the group of rotations about the interaction point, and for hadron colliders they are rotations about and boosts along the beam axis (sometimes with a reflection in the plane perpendicular to the beam). 
Therefore, the energy fractions $z_i$ usually use particle energies $E_i$ at an $e^+e^-$ collider and particle transverse momenta $p_{T,i}$ at a hadron collider.

For the angular weighting function $f_N$, though, there are many different angular structures one can build out of the particle directions $\hat{p}^\mu_i$.
The \Bs use the simplest and arguably most natural choice to expand the angular behavior:  \emph{pairwise} angular distances $\theta_{ij}$,  determined using spherical coordinates $(\theta,\phi)$ at an $e^+e^-$ collider and rapidity-azimuth coordinates $(y,\phi)$ at a hadron collider.
Other classes of observables, such as ECFs~\cite{Larkoski:2013eya} and ECFGs~\cite{Moult:2016cvt}, also use pairwise angles since they manifestly respect the underlying Lorentz subgroup.
For building the \Bs, is important that the $\theta_{ij}$, or any other choice of geometric object, be sufficient to reconstruct the value of the original function $f_N$ in terms of the $\hat p_i^\mu$.
For pairwise angles, this property can be shown by triangulation, under the assumption that the observable in question does not depend on the overall jet direction nor on rotations or reflections about the jet axis.
Since jets are collimated sprays of particles, the $\theta_{ij}$ are typically small and are good expansion parameters.

At various points in this paper, we explore three different energy/angular measures.  For $e^+e^-$ collisions, our default is:
\begin{equation}\label{eq:eemeasure}
\boxed{\text{$e^+e^-$ Default}} \qquad
\begin{aligned}
\quad z_i &= \frac{E_i}{E_J}, \qquad  E_J\equiv\sum_{i=1}^M E_i, \\
\theta_{ij} &= \left(\frac{2\,p_i^\mu p_{j\mu}}{E_i  E_j} \right)^{\beta/2},
\end{aligned}
\end{equation}
where $\beta>0$ is an angular weighting factor. 
For the hadron collider studies in \Secs{sec:linreg}{sec:linclass}, we use:
\begin{equation}\label{eq:hadronicmeasure}
\boxed{\text{Hadronic Default}} \qquad
\begin{aligned}
\quad z_i &= \frac{p_{T,i}}{p_{T,J}}, \qquad  p_{T,J}\equiv\sum_{i=1}^Mp_{T,i}, \\
\theta_{ij} &= \left(\Delta y_{ij}^2+\Delta\phi_{ij}^2\right)^{\beta/2},
\end{aligned}
\end{equation}
where $\Delta y_{ij}\equiv y_i-y_j$, $\Delta\phi_{ij}\equiv\phi_i-\phi_j$ are determined by the rapidity $y_i$ and azimuth $\phi_i$ of particle $i$.
This measure is rotationally-symmetric in the $(y,\phi)$ plane, which is the most commonly used case in jet substructure.
For situations where this rotational symmetry is not desirable (such as for jet pull~\cite{Gallicchio:2010sw}), we can instead use a two-dimensional measure that treats the rapidity and azimuthal directions separately:
\begin{equation}\label{eq:2Dhadronicmeasure}
\boxed{\text{Hadronic Two-Dimensional}} \qquad
\begin{aligned}
\quad z_i &= \frac{p_{T,i}}{p_{T,J}}, \qquad  p_{T,J}\equiv\sum_{i=1}^Mp_{T,i}, \\
\theta_{ij} &= \Delta y_{ij} \text{ or } \Delta\phi_{ij},
\end{aligned}
\end{equation}
where each line on the multigraph now has an additional decoration to indicate whether it corresponds to $\Delta y$ or $\Delta \phi$.

We emphasize that the choice of measure is not unique, though it is constrained by the IRC-safety arguments in \Sec{sec:basis}.
For example, IRC safety requires that the energy-like quantities appear linearly in $z_i$.
For the default measures, the angular exponent $\beta$ can take on any positive value and still be consistent with IRC safety.
Depending on the context, different choices of $\beta$ can lead to faster or slower convergence of the \B expansion, with $\beta<1$ emphasizing smaller values of $\theta_{ij}$ and $\beta>1$ emphasizing larger values of $\theta_{ij}$.
For special choices of $z_i$ and $\theta_{ij}$, some \Bs may be linearly related, a point we return to briefly in \Sec{sec:algebraic}.

\subsection{Relation to existing substructure observables}
\label{sec:jetobs}

Many familiar jet observables can be nicely interpreted in the energy flow basis.
When an observable can be written as a simple expression in terms of particle four-momenta or in terms of energies and angles, the energy flow decomposition can often be performed exactly.
Some of the most well-known observables, such as jet mass and energy correlation functions, are exactly finite linear combinations of \Bs (with appropriate choice of measure), which one might expect since they also correspond to natural $C$-correlators.
Unless otherwise specified, the analysis below uses the default hadronic measure in \Eq{eq:hadronicmeasure} with $\beta=1$ and treats all particles as massless.\footnote{A proper treatment of non-zero particle masses would require an additional expansion in the \emph{velocities} of the particles (see related discussion in \Refs{Salam:2001bd,Mateu:2012nk}). To avoid these complications, one can interpret all particles as being massless in the $E$-scheme~\cite{Salam:2001bd}, i.e.\ $p_{\text{rescaled}}^\mu = E \, (1, \hat{p})$ with $\hat{p} = \vec{p} / |\vec{p}^{}|$.}

\subsubsection{Jet mass}
\label{sec:relation_jetmass}

Jet mass is most basic jet substructure observable, and not surprisingly, it has a nice expansion in the energy flow basis.
In particular, the squared jet mass divided by the jet energy squared is an exact $N=2$ \B using the $e^+e^-$ measure in \Eq{eq:eemeasure} with $\beta = 1$:
\begin{equation}\label{eq:massexp}
e^+e^-: \qquad \frac{m_J^2}{E_J^2}=\frac12\sum_{i_1 = 1}^M \sum_{i_2 = 1}^M z_{i_1}z_{i_2} \left(\frac{2\,p_{i_1}^\mu p_{i_2\mu}}{E_{i_1} E_{i_2}} \right) =\frac12\times
\begin{gathered}
\includegraphics[scale=.2]{graphs/2_2_1}
\end{gathered}.
\end{equation}
Note that mass is exactly an \B for any $\beta = 2 / N$ measure choice.

For the hadronic measure in \Eq{eq:hadronicmeasure} with $\beta= 1$, there is an approximate equivalence with the squared jet mass divided by the jet (scalar) transverse momentum:
\begin{equation}\label{eq:massexph}
\text{Hadronic}: \qquad \frac{m_J^2}{p_{TJ}^2}=\sum_{i_1 = 1}^M\sum_{i_2 = 1}^M z_{i_1} z_{i_2} (\cosh(\Delta y_{i_1i_2}) - \cos(\Delta\phi_{i_1i_2})) = \frac12\times
\begin{gathered}
\includegraphics[scale=.2]{graphs/2_2_1}
\end{gathered}
+ \cdots.
\end{equation}
Since the jet mass is not exactly rotationally symmetric in the rapidity-azimuth plane, the subleading terms in \Eq{eq:massexph} are not fully encompassed by the simplified set of hadronic observables depending only on $\{\Delta y_{ij}^2+\Delta\phi_{ij}^2\}$, but could be fully encompassed by using an expansion in $\{\Delta y_{ij},\Delta\phi_{ij}\}$ as in \Eq{eq:2Dhadronicmeasure}.
For narrow jets, these higher-order terms in the expansion become less relevant since $\Delta y_{ij},\,\Delta\phi_{ij}\ll1$.\footnote{Alternatively, we could use a measure with $\theta_{ij} = \left(\frac{2\,p_i^\mu p_{j\mu}}{p_{T,i}  p_{T,j}} \right)^{\beta/2}$, similar in spirit to the Conical Geometric measure of \Ref{Stewart:2015waa}, to exactly recover the jet mass.}

\subsubsection{Energy correlation functions}
\label{sec:equivECF}

The ECFs are designed to be sensitive to $N$-prong jet substructure~\cite{Larkoski:2013eya}.
They can be written as a $C$-correlator, \Eq{eq:genccorr}, with a particular choice of angular weighting function:
\begin{equation}
\label{eq:ecfs}
f_N^{(\beta)}(\{\theta_{ij}\})= \prod_{i<j} \theta^\beta_{ij},
\end{equation}
where $\theta_{ij}=(\Delta y_{ij}^2+\Delta\phi_{ij}^2)^{1/2}$.
In terms of multigraphs, the ECFs correspond to complete graphs on $N$ vertices:
\begin{equation}\label{eq:ecfgraphs}
e_2^{(\beta)} = \begin{gathered}
\includegraphics[scale=.2]{graphs/2_1_1}
\end{gathered}\,,\hspace{.5in}
e_3^{(\beta)} = \begin{gathered}
\includegraphics[scale=.2]{graphs/3_3_1}
\end{gathered}\,,\hspace{.5in}
e_4^{(\beta)} = \begin{gathered}
\includegraphics[scale=.2]{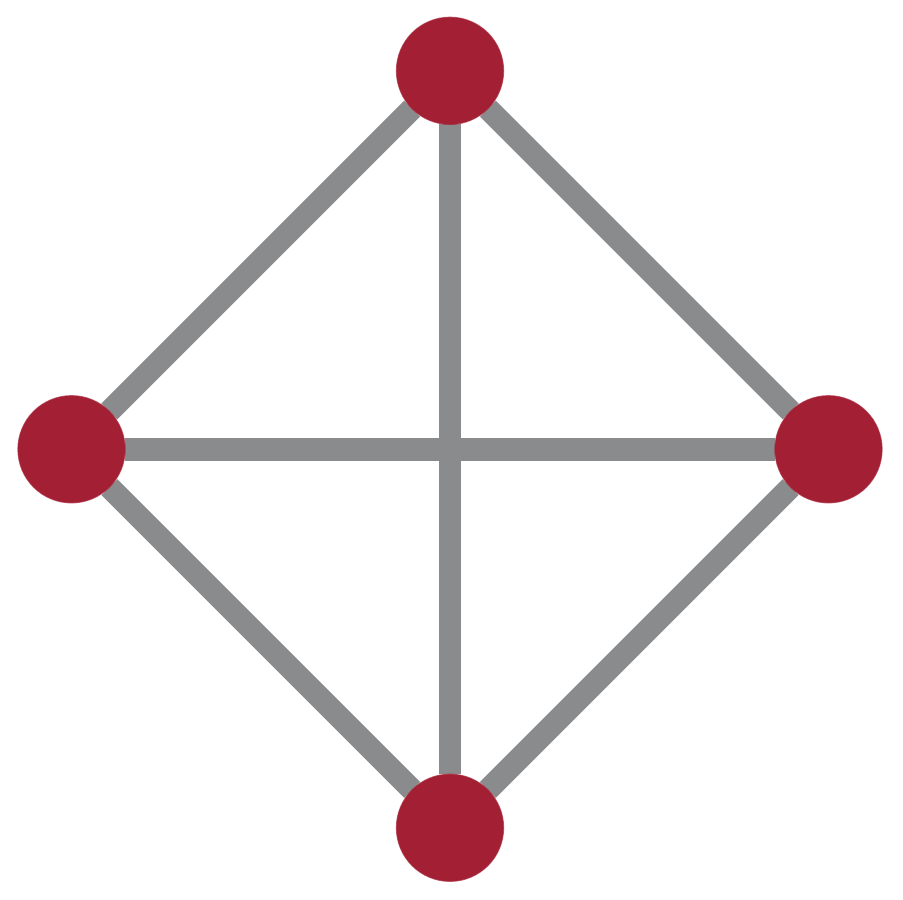}
\end{gathered},
\end{equation}
which are \Bs using the measure in \Eq{eq:hadronicmeasure} with exponent $\beta$.

The ECFs have since been expanded to a more flexible set of observables referred to as the ECFGs~\cite{Moult:2016cvt}. 
Letting $\min^{(m)}$ indicate the $m$-th smallest element in a set, the ECFGs are also $C$-correlators with angular weighting function:
\begin{equation}\label{eq:ecfgs}
\,_vf_N^{(\beta)}(\{\theta_{ij}\}) =  \prod_{m=1}^v \min_{i<j}^{(m)}\{ \theta^\beta_{ij}\}.
\end{equation}
The ECFGs do not have an exact multigraph correspondence due to the presence of the min function, but are evidently closely related to the \Bs since they share a common energy structure. 
The min function itself can be approximated by polynomials in its arguments, which induces an approximating series for the ECFGs in terms of \Bs when plugged into the common energy structure.

Both the \Bs and the ECFGs represent natural extensions of the ECFs but in different directions.
From our graph-theoretic perspective, the \Bs extend the ECFs to non-fully-connected graphs.
The ECFGs extend the scaling properties of the ECFs into observables with independent energy and angular scalings.
As discussed in \Sec{subsec:goingbeyond}, there are angular structures possible in the \Bs that are not possible in the ECFGs.
As with any jet substructure analysis, the choice of which set of observables to use depends on the physics of interest, with the \Bs designed for linear completeness and the ECFGs designed for nice power-counting properties.

\subsubsection{Angularities}
\label{sec:angularities}

Next, we consider the IRC-safe jet angularities~\cite{Larkoski:2014pca} (see also \Refs{Berger:2003iw,Almeida:2008yp,Ellis:2010rwa,Larkoski:2014uqa}) defined by:
\begin{equation}\lambda^{(\alpha)} = \sum_{i = 1}^M z_{i} \, \theta_{i}^\alpha,\end{equation}
where $\alpha>0$ is an angular exponent and $\theta_i$ denotes the distance of particle $i$ to the jet axis.
For concreteness and analytic tractability, we take the jet axis to be the $p_T$-weighted centroid in $( y,\phi)$-space, such that the jet axis is located at:
\begin{equation}
\label{eq:jetaxis}
y_J = \sum_{j=1}^M z_j  y_j, \qquad \phi_J = \sum_{j=1}^M z_j \phi_j.
\end{equation}
With this, the angularities can be expressed as:
\begin{align}
\lambda^{(\alpha)} &=\sum_{i_1=1}^Mz_{i_1}\left(( y_{i_1}- y_J)^2+(\phi_{i_1}-\phi_J)^2\right)^{\alpha/2}\nonumber
\\&=\sum_{i_1=1}^M z_{i_1} \left(\left(\sum_{i_2 = 1}^Mz_{i_2}\Delta y_{i_1i_2}\right)^2 + \left(\sum_{i_2 = 1}^Mz_{i_2}\Delta\phi_{i_1i_2}\right)^2\right)^{\alpha/2}\nonumber
\\&= \sum_{i_1=1}^M z_{i_1} \left(\sum_{i_2 = 1}^Mz_{i_2}\theta_{i_1i_2}^2 - \frac12 \sum_{i_2 = 1}^M\sum_{i_3 = 1}^M z_{i_2}z_{i_3} \theta_{i_2i_3}^2\right)^{\alpha/2}.\label{eq:angthetaijs}
\end{align}

For even $\alpha$, the parenthetical in \Eq{eq:angthetaijs} can be expanded and identified to be a linear combination of \Bs with $N = \alpha$ and $d=\alpha$ (see \Ref{GurAri:2011vx} for a related discussion).
For $\alpha = 2$, \Eq{eq:angthetaijs} implies:
\begin{equation}\label{eq:lam2}
\lambda^{(2)} = \frac12\sum_{i\in J} \sum_{j\in J} z_{i}z_{j} \theta_{ij}^2 =\frac12\times \begin{gathered}
\includegraphics[scale=.2]{graphs/2_2_1}
\end{gathered}.
\end{equation}
For $\alpha = 4$ and $\alpha = 6$,  \Eq{eq:angthetaijs} implies:
\begin{align}\label{eq:lam4}
\lambda^{(4)} &=
\begin{gathered}
\includegraphics[scale=.2]{graphs/3_4_1}
\end{gathered}
-\frac34 \times
\begin{gathered}
\includegraphics[scale=.2]{graphs/2_2_1}\hspace{0mm}
\includegraphics[scale=.2]{graphs/2_2_1}
\end{gathered}, 
\\
\label{eq:lam6}
\lambda^{(6)}&=
\begin{gathered}
\includegraphics[scale=.25]{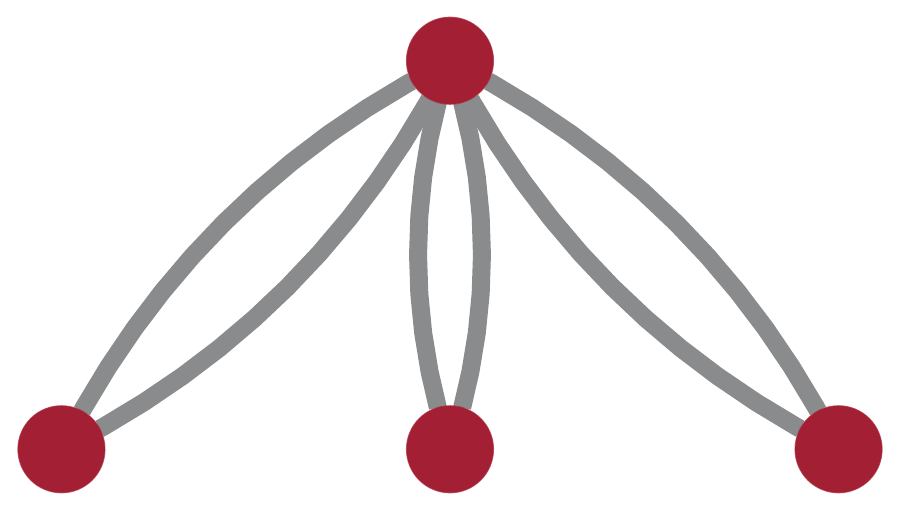}
\end{gathered}
-\frac32 \times
\begin{gathered}
\includegraphics[scale=.2]{graphs/3_4_1}
\end{gathered}
\hspace{0mm}
\begin{gathered}
\includegraphics[scale=.2]{graphs/2_2_1}
\end{gathered}\,
+\frac58\times
\begin{gathered}
\includegraphics[scale=.2]{graphs/2_2_1}\hspace{0mm}
\includegraphics[scale=.2]{graphs/2_2_1}\hspace{0mm}
\includegraphics[scale=.2]{graphs/2_2_1}
\end{gathered}\,.
\end{align}
This can be continued for arbitrarily high, even $\alpha$.
Thus, the even $\alpha$ angularities are exact, non-trivial linear combinations of \Bs, illustrating the close connections between the two classes of observables.
While angularities with odd or non-integer $\alpha$ do not have the same analytic tractability, the specific case of $\alpha=1/2$ is shown to be numerically well approximated by \Bs in \Sec{sec:regression}.

\subsubsection{Geometric moment tensors}
\label{sec:geometricmomenttensors}

Next, we consider observables based on the two-dimensional geometric moment tensor of the energy distribution in the $(y,\phi)$-plane~\cite{GurAri:2011vx,Gallicchio:2012ez}: 
\begin{align}
{\bf C}& = \sum_{i\in J} z_{i} \begin{pmatrix} \Delta y_i^2 & \Delta y_i\Delta\phi_i \\ \Delta\phi_i\Delta y_i & \Delta\phi_i^2 \end{pmatrix}= \begin{pmatrix}\frac12\sum_{i,j} z_{i}z_{j} \Delta y_{ij}^2 &\frac12\sum_{i,j} z_{i}z_{j} \Delta y_{ij}\Delta\phi_{ij} \\ \frac12\sum_{i,j} z_{i}z_{j} \Delta\phi_{ij}\Delta y_{ij} &\frac12\sum_{i,j} z_{i}z_{j} \Delta\phi_{ij}^2\end{pmatrix},
\label{eq:2dmomtens}
\end{align}
where the distances are measured with respect to the $p_T$-weighted centroid axis $(y_J,\phi_J)$ from \Eq{eq:jetaxis}.
Useful observables can be constructed from the trace and determinant of {\bf C}, such as planar flow $\text{Pf} = 4 \det{\bf C}/(\tr {\bf C})^2$~\cite{Thaler:2008ju,Almeida:2008yp}, which is a ratio of two IRC-safe observables.

We see that \Eq{eq:2dmomtens} is exactly a matrix of EFPs with $N=2$ and the two-dimensional hadronic measure from \Eq{eq:2Dhadronicmeasure}.
The trace $\tr {\bf C}$ and determinant $\det{\bf C}$ have the rotational symmetry in the $(y,\phi)$-plane of the default hadronic measure from \Eq{eq:hadronicmeasure}, allowing them to be written as linear combinations of EFPs with that measure:
\begin{align}\label{eq:momentpoly}
\tr \bf C&=\frac12\times
\begin{gathered}
\includegraphics[scale=.2]{graphs/2_2_1}
\end{gathered}\,,&\hspace{-.5in}
4\,\det{\bf C}&= 
\begin{gathered}
\includegraphics[scale=.2]{graphs/3_4_1}
\end{gathered}\,
-\frac12\times
\begin{gathered}
\includegraphics[scale=.2]{graphs/2_4_1}
\end{gathered}\,.
\end{align}

In \Ref{GurAri:2011vx}, a general class of energy flow moments was explored and categorized, with the goal of classifying observables according to their energy flow distributions.
These energy flow moments are defined with respect to a specified jet axis:
\begin{equation}\label{eq:hightens}
I_{k_1\cdots k_N} \equiv \sum_{i=1}^M z_i \, x_{k_1}^{(i)}\cdots x_{k_N}^{(i)},
\end{equation}
where $k_i \in \{1,2\}$, $x_1^{(i)} = \Delta  y_i= y_i- y_J$ and $x_2^{(i)} = \Delta\phi_i=\phi_i-\phi_J$.
Using the $p_T$-weighted centroid axis, this is the natural generalization of \Eq{eq:2dmomtens}, with the special case of $I_{k_1k_2} = ({\bf C})_{k_1k_2}$.
By performing a similar analysis to the one used to arrive at \Eq{eq:momentpoly}, one can show that any scalar constructed by contracting the indices of a product of objects in \Eq{eq:hightens} can be decomposed into an exact linear combination of \Bs.

\subsection{Going beyond existing substructure observables}
\label{subsec:goingbeyond}

Because the \Bs are $C$-correlators that span the space of IRC-safe observables, their angular structures should encompass all possible behaviors of $C$-correlators.
By contrast, the ECFs and ECFGs mentioned in \Sec{sec:equivECF} have more restricted behaviors, and it is illuminating to understand the new kinds of structures present in the \Bs.

Without loss of generality, the angular weighting function $f_N$ in \Eq{eq:genccorr} can be taken to be a symmetric function of the particle directions $\hat{p}_i^\mu$ due to the symmetrization provided by the sum structure (see \Eq{eq:symf} below).
The ECFs and ECFGs exhibit a stronger symmetry, though, since the angular functions in \Eqs{eq:ecfs}{eq:ecfgs} are invariant under the swapping any two pairwise angles $\theta_{ij}$.
This symmetry is manifested in the ECFs multigraphs in \Eq{eq:ecfgraphs} by the fact that all pairs of indices are connected by the same number of edges.

We can easily see that the pairwise swap symmetry of the ECFs is stronger than the full permutation symmetry of the \Bs: the group of permutations of the angular distances $\theta_{ij}$ has $\binom{N}{2}!$ elements, whereas the group of permutations of the indices $\{i_a\}$ has $N!$ elements.
An example of an \B that does not satisfy the stronger symmetry is the following $N=4$ graph:
\begin{equation}\label{eq:thebirdfoot}
\begin{gathered}
\includegraphics[scale=0.25]{graphs/4_3_1}
\end{gathered}
= \sum_{i_1=1}^M\sum_{i_2=1}^M\sum_{i_3=1}^M \sum_{i_4=1}^M z_{i_1}z_{i_2}z_{i_3} z_{i_4} \theta_{i_1i_2}\theta_{i_1 i_3}\theta_{i_1i_4}.
\end{equation}
The angular weighting function of the \B in \Eq{eq:thebirdfoot} is symmetric under the $4!$ permutations in the indices (vertices) $i_a \to i_{\sigma(a)}$ but not under the exchange of pairwise angles (edges) $\theta_{i_1 i_3} \to \theta_{i_2 i_3}$ which would result in a different \B, namely:
\begin{equation}
\begin{gathered}
\includegraphics[scale=0.2,angle=90]{graphs/4_3_2}
\end{gathered}
\neq
\begin{gathered}
\includegraphics[scale=0.25,angle=0]{graphs/4_3_1}
\end{gathered}.
\end{equation}

Another feature of the ECFs and ECFGs is that their angular weighting function $f_N$ vanishes whenever two of its arguments become collinear.
Indeed, one of the present authors made the erroneous claim in \Ref{Moult:2016cvt} that this vanishing behavior was required by collinear safety.\footnote{If the sums are taken over distinct $N$-tuples as in \Ref{Moult:2016cvt}, then the angular function does have to vanish on collinearity for C safety. In general, non-collinearly-vanishing angular functions are C safe if the sum is taken over all $N$-tuples of particles, including sets with repeated indices.}
Instead, the argument in \Sec{sec:collinearsafety} shows this not to be the case, and observables defined by \Eq{eq:genccorr} are IRC safe for any sufficiently smooth and non-singular $f_N$.
An example of an \B that does not necessarily vanish when two of its arguments become collinear is the following $N=3$ graph:
\begin{equation}\label{eq:thewedge}
\begin{gathered}
\includegraphics[scale=0.2]{graphs/3_2_1}
\end{gathered}
= \sum_{i_1=1}^M\sum_{i_2=1}^M\sum_{i_3=1}^M z_{i_1}z_{i_2}z_{i_3}  \theta_{i_1i_2}\theta_{i_1 i_3},
\end{equation}
which does not vanish when $\hat p_{i_2}^\mu\to\hat p_{i_3}^\mu$.  
More generally, any non-fully-connected graph will not vanish in every collinear limit, but the corresponding \B will still be collinear safe.

By relaxing the restrictions on the angular weighting function $f_N$ to those minimally required by IRC safety, the energy flow basis captures all topological structures which can possibly appear in a $C$-correlator, beyond just the ones described by ECFs and ECFGs.

\section{Constructing a linear basis of IRC-safe observables}
\label{sec:basis}

Having introduced the \Bs, we now give a detailed argument that they linearly span the space of IRC-safe observables.
Due to its more technical nature, this section can be omitted on a first reading, and the reader may skip to \Sec{sec:complexity}.
\Refs{Tkachov:1995kk,Sveshnikov:1995vi,Cherzor:1997ak,Tkachov:1999py} argue that, from the point of view of quantum field theory, all IRC-safe information about the jet structure should be contained in the $C$-correlators.
In \Sec{sec:Eexpansion}, we independently arrive at the same conclusion by a direct application of IRC safety.
We then go on in \Sec{sec:angleexpansion} to expand the angular structure of the $C$-correlators to find a correspondence between multigraphs and \Bs.

An IRC-safe observable $\mathcal S$ depends only on the unordered set of particle four-momenta $\{p^\mu_i\}_{i=1}^M$, and not any non-kinematic quantum numbers.
An observable defined on $\{p^\mu_i\}_{i=1}^M$ can alternatively be thought of as a collection of functions, one for each number of particles $M$.
IRC safety then imposes constraints on this collection and thereby induces relations between the functions.
The requirement of IR safety imposes the constraint~\cite{sterman1995handbook}:
\begin{align}\label{eq:IRsafety}
\mathcal S(\{p_1^\mu,\ldots,p_M^\mu\}) & = \lim_{\varepsilon \to 0}\mathcal S(\{p_1^\mu,\ldots, p_M^\mu, \varepsilon \, p_{M+1}^\mu\}),&&\hspace{-.5in}\forall p_{M+1}^\mu,
\end{align}
while the requirement of C safety imposes the constraint:
\begin{align}
\label{eq:Csafety}
\mathcal S(\{p_1^\mu,\ldots,p_M^\mu\}) & = \mathcal S(\{p_1^\mu,\ldots,(1 - \lambda) p_M^\mu, \lambda p_M^\mu\}),&&\hspace{-.5in}\forall \lambda\in[0,1].
\end{align}
\Eq{eq:IRsafety} says that the observable is unchanged by the addition of infinitesimally soft particles, while \Eq{eq:Csafety} guarantees that the observable is insensitive to a collinear splitting of particles.

As written, only particle $M$ is affected in \Eq{eq:Csafety}.
The indexing used to identify particles, however, is arbitrary and these properties continue to hold when the particles are reindexed.
This \emph{particle relabeling symmetry} is not an additional constraint that is imposed but rather a consequence of assigning labels to an unordered set of particles.
These three restrictions---IR safety, C safety, and particle relabeling symmetry---are necessary and sufficient conditions for obtaining the energy flow basis.

Throughout this analysis, particles are treated as massless, $p^\mu_i = E_i \,\hat p_i^\mu$, where $\hat p_i^\mu$ is purely geometric.
Note that we could replace $E_i$ with any quantity linearly dependent on energy, such as the transverse momentum $p_{T,i}$, which corresponds to making a different choice of measure in \Sec{sec:measures}.

\subsection{Expansion in energy}
\label{sec:Eexpansion}

Consider an arbitrary IRC-safe observable $\mathcal{S}$, expanded in terms of the particle energies.
If the observable has a simple analytic dependence on the energies, then the usual Taylor expansion can be used:
\begin{align}\label{eq:taylorexpand}
\mathcal S=\left.\mathcal S_M\right|_{\{E\}=0}+\sum_{i_1=1}^ME_i\left.\pd{\mathcal S_M}{E_{i_1}}\right|_{\{E\}=0}+\frac{1}{2}\sum_{i_1=1}^M\sum_{i_2=1}^ME_{i_1}E_{i_2}\left.\pd{^2\mathcal S_M}{E_{i_1}\partial E_{i_2}}\right|_{\{E\}=0}+\cdots,
\end{align}
where $M$ is the particle multiplicity and the derivatives are evaluated at vanishing energies. 
An example of this is the jet mass from \Eq{eq:massexp}:
\begin{align}\label{eq:jetmass}
m_J^2&=\sum_{i=1}^M\sum_{j=1}^M \eta_{\mu\nu}p_i^\mu p_{j}^\nu=\sum_{i=1}^M\sum_{j=1}^ME_iE_j\eta_{\mu\nu}\, \hat p_i^\mu\hat p_j^\nu,
\end{align}
where $\eta_{\mu \nu}$ is the Minkowski metric.  
This expression is already in the form of \Eq{eq:taylorexpand} with:
\begin{align}
\pd{^2m_J^2}{E_{i_1}\partial E_{i_2}}=2\eta_{\mu\nu}\hat p_{i_1}^\mu\hat p_{i_2}^\nu,
\end{align}
and all other Taylor coefficients zero.
See \Sec{sec:jetobs} for additional examples of observables with explicit formulas for which \Eq{eq:taylorexpand} can be applied.

For some observables, though, a Taylor expansion may be difficult or impossible to obtain. 
The simplest example is a non-differentiable observable.
This is the case for $m_J$ (rather than $m_J^2$); the presence of the square root spoils the existence of a Taylor expansion, but the square root can be nonetheless approximated by polynomials arbitrarily well in a bounded interval.
A more complicated case is if the observable is defined in terms of an algorithm, such as a groomed jet mass~\cite{Butterworth:2008iy,Ellis:2009su,Ellis:2009me,Krohn:2009th,Dasgupta:2013ihk,Larkoski:2014wba}, and an explicit formula in terms of particle four-momenta would not be practical to differentiate or write down.
Similarly, the observable could be a non-obvious function of the particles, i.e.\ the optimal observable to accomplish some task.  

In cases without a Taylor expansion, the Stone-Weierstrass theorem~\cite{stone1948generalized} still guarantees that the observable can be approximated over some bounded energy range by polynomials in the energies.\footnote{A version of this theorem that suffices for our purposes can be phrased as follows: for any continuous, real-valued function $f$ defined on a compact subset  $X\subset\mathbb R^n$, for all $\epsilon>0$ there exists a polynomial $p$ of finite degree at most $N_{\rm max}$ such that $|p(\mathbf x)-f(\mathbf x)|<\epsilon$ for all $\mathbf x\in X$. Conceptually, this theorem is used to approximate any continuous function on a bounded region by a polynomial.} 
We write down such an expansion by considering all possible polynomials in the energies and multiplying each one by a different geometric function.
Combining all terms of degree $N$ into $\mathcal C_N$, the expansion is:
\begin{align}\label{eq:SWexpand}
\mathcal S&\simeq\sum_{N=0}^{N_{\rm max}}\mathcal C_N,\qquad
\mathcal C_N\equiv\sum_{i_1=1}^M \cdots \sum_{i_N=1}^M C^{(M)}_{i_1 \cdots i_N}(\hat p_1^\mu,\ldots, \hat p_M^\mu)\prod_{j=1}^N E_{i_j},
\end{align}
where $C_{i_1\cdots i_n}^{(M)}(\hat p_1^\mu,\ldots,\hat p_M^\mu)$ are geometric angular functions, which depend on the indices of the energy factors $i_1\cdots i_n$ and could in general be different for different multiplicities $M$.
The Stone-Weierstrass theorem guarantees that there is a maximum degree $N_{\rm max}$ in this energy expansion for any given desired accuracy, but places no further restrictions on the $\mathcal C_N$.

To derive constraints on these angular functions $C^{(M)}_{i_1 \cdots i_N}$, we impose the three key properties of IR safety in \Sec{sec:irsafety}, particle relabeling invariance in \Sec{sec:relabelsym}, and C safety in \Sec{sec:collinearsafety}, which we summarize in \Sec{subsec:final}.
In applying these properties, we will often use the fact that when setting two expressions for the observable $\mathcal S$ equal to each other, we can read off term-by-term equality by treating the particle energies as independent quantities:
\begin{equation}\label{eq:equality}
\mathcal S=\mathcal S'\quad\implies\quad \mathcal C_N=\mathcal C'_N,\quad\forall N\le N_{\rm max}.
\end{equation}
Note that the sum structure in \Eq{eq:SWexpand} implies that, without loss of generality, the angular functions can be taken to depend only on the labels $i_1,\ldots,i_N$ as an unordered set.

\subsubsection{Infrared safety}
\label{sec:irsafety}

IR safety constrains the angular functions appearing in the expansion of \Eq{eq:SWexpand} in two ways: by restricting which particle directions contribute to a particular term in the sum and by relating angular functions of different multiplicities.

First, consider a particular angular function, $C^{(M)}_{i_1\cdots i_N}$ in \Eq{eq:SWexpand}, and some particle $j\not\in\{ i_1,\ldots, i_N\}$. Consider particle $j$ in the soft limit: if $C^{(M)}_{i_1\cdots i_N}$ depends on $\hat p_j^\mu$ in any way, then IR safety is violated because $E_j$ does not appear in the product of energies but the value of the observable changes as the direction of $j$ is changed.
Hence, IR safety imposes the requirement that
\begin{align}\label{eq:IRresult1}
C^{(M)}_{i_1\cdots i_N}(\hat p_1^\mu,\ldots,\hat p_M^\mu)=C^{(M)}_{i_1\cdots i_N}(\hat p_{i_1}^\mu,\ldots,\hat p_{i_N}^\mu),
\end{align}
namely the indices of the arguments must match those of the angular function.
Note that we must always write $C^{(M)}_{i_1\cdots i_N}$ with $N$ arguments, even if some are equal due to indices coinciding.

Next, consider two polynomial approximations of the same observable: one as a function of $M$ particles and the other as a function of $M+1$ particles.
In the soft limit of particle $M+1$, $E_{M+1}\to 0$, the IR safety of $\mathcal S$, written formally in \Eq{eq:IRsafety}, guarantees that the function of $M+1$ particles approaches the function of $M$ particles. In terms of the corresponding polynomial approximations, we have that:
\begin{align}\label{eq:IRM+1}
\sum_{i_1=1}^{M+1} \hspace{-.05in}\cdots \hspace{-.05in}\sum_{i_N=1}^{M+1}C^{(M+1)}_{i_1 \cdots i_N}(\hat p_{i_1}^\mu,\ldots, \hat p_{i_N}^\mu) \prod_{j=1}^N E_{i_j}=\sum_{i_1=1}^{M} \hspace{-.05in}\cdots \hspace{-.05in}\sum_{i_N=1}^{M} C^{(M)}_{i_1 \cdots i_N} (\hat p_{i_1}^\mu,\ldots, \hat p_{i_N}^\mu)\prod_{j=1}^N E_{i_j}+\mathcal O(E_{M+1}).
\end{align}

We see from \Eq{eq:IRM+1} that the same angular coefficients from the polynomial approximation of the function of $M+1$ particles can be validly chosen for the approximation of the function of $M$ particles, with the following equality of angular functions:
\begin{align}\label{eq:IRresult2}
C^{(M+1)}_{i_1 \cdots i_N} (\hat p_{i_1}^\mu,\ldots,\hat p_{i_N}^\mu)=C^{(M)}_{i_1\cdots i_N} (\hat p_{i_1}^\mu, \ldots, \hat p_{i_N}^\mu)\equiv C_{i_1\cdots i_N}(\hat p_{i_1}^\mu,\ldots,\hat p_{i_N}^\mu),
\end{align}
which says that the multiplicity label on the angular functions can be dropped.

As a result of enforcing IR safety, the dependence of the angular functions on multiplicity has been eliminated, as well as the dependence of a given angular function on any particles with indices not appearing in its subscripts.

\subsubsection{Particle relabeling symmetry }
\label{sec:relabelsym}

Now, using particle relabeling symmetry, for all $\sigma\in S_M$, where $S_M$ is the group of permutations of $M$ objects, we have that $\mathcal C_N$ is unchanged by the replacement $E_{i_j} \to E_{\sigma(i_j)}$ and $\hat p^\mu_{i_j} \to \hat p^\mu_{\sigma(i_j)}$.
With the angular functions as constrained by IR safety, the particle relabeling invariance of $\mathcal C_N$ can be written as:
\begin{align}\label{eq:updatedCn}
\mathcal C_N&=\sum_{i_1=1}^M\cdots\sum_{i_N=1}^MC_{i_1\cdots i_N}(\hat p_{i_1}^\mu,\ldots,\hat p_{i_N}^\mu)\prod_{j=1}^NE_{i_j}\\
&=\sum_{i_1=1}^M\cdots\sum_{i_N=1}^MC_{i_1\cdots i_N}(\hat p_{\sigma(i_1)}^\mu,\ldots, \hat p_{\sigma(i_N)}^\mu)\prod_{j=1}^NE_{\sigma(i_j)}\nonumber
\\&=\sum_{i_1=1}^M\cdots\sum_{i_N=1}^MC_{\sigma^{-1}(i_1)\cdots \sigma^{-1}(i_N)}(\hat p_{i_1}^\mu ,\ldots,\hat p_{i_N}^\mu)\prod_{j=1}^NE_{i_j},\label{eq:relabelsym1}
\end{align}
where the sums were reindexed according to $\sigma^{-1}$. In particular, from \Eq{eq:relabelsym1}, we have for any $\sigma \in S_M$ that:
\begin{align}\label{eq:relabelsym2}
C_{i_1\cdots i_N}(\hat p_{i_1}^\mu,\ldots,\hat p_{i_N}^\mu)=C_{\sigma(i_1)\cdots \sigma(i_N)}(\hat p_{i_1}^\mu,\ldots, \hat p_{i_N}^\mu).
\end{align}
\Eq{eq:relabelsym2} allows us to permute the indices of $C_{i_1\cdots i_N}$ within $S_M$, equating previously unrelated angular functions.

As written, $C_{i_1\cdots i_N}$ is not necessarily symmetric in its arguments.
Without loss of generality, though, we can symmetrize $C_{i_1\cdots i_N}$ without changing the value of $\mathcal C_N$ as follows:
\begin{align}\label{eq:symf}
\mathcal C_N&=\sum_{i_1=1}^M\cdots\sum_{i_N=1}^M\underbrace{\frac{1}{N!}\sum_{\sigma\in S_N}C_{i_1\cdots i_N}(\hat p_{\sigma(i_1)}^\mu,\ldots,\hat p_{\sigma(i_N)}^\mu)}_{C'_{i_1\cdots i_N}(\hat p_{i_1}^\mu,\ldots,\hat p_{i_N}^\mu)}\prod_{j=1}^NE_{i_j},
\end{align}
where $C'_{i_1\cdots i_N}$ is now symmetric in its arguments.
We assume in the next step of the derivation that the angular weighting functions are symmetric in their arguments.

\subsubsection{Collinear safety}
\label{sec:collinearsafety}

The key requirement for restricting the form of $C_{i_1\cdots i_N}$ is C safety.
If the angular weighting function(s) were required to vanish whenever two of the inputs were collinear, then the observable would be manifestly C safe~(see e.g.\ \cite{Moult:2016cvt}); this is a sufficient condition for C safety but not a necessary one.
More generally, one can have non-zero angular functions of $N$ arguments even when subsets of the arguments are collinear.

Using the IR safety argument of \Eq{eq:IRresult2} and the particle relabeling symmetry of \Eq{eq:relabelsym2}, we can relate any angular function $C_{i_1\cdots i_N}$ to one of the following:
\begin{align}\label{eq:Cexample}
&C_{123\cdots N}(\hat p_{i_1}^\mu,\hat p_{i_2}^\mu,\hat p_{i_3}^\mu,\ldots,\hat p_{i_N}^\mu),\nonumber\\
&C_{1123\cdots (N-1)}(\hat p_{i_1}^\mu,\hat p_{i_1}^\mu,\hat p_{i_2}^\mu,\hat p_{i_3}^\mu,\ldots,\hat p_{i_{N-1}}^\mu),\nonumber\\
&C_{112234\cdots (N-2)}(\hat p_{i_1}^\mu,\hat p_{i_1}^\mu,\hat p_{i_2}^\mu,\hat p_{i_2}^\mu,\hat p_{i_3}^\mu,\hat p_{i_4}^\mu,\ldots,\hat p_{i_{N-2}}^\mu),\nonumber\\
&C_{1112234\cdots(N-3)}(\hat p_{i_1}^\mu,\hat p_{i_1}^\mu,\hat p_{i_1}^\mu,\hat p_{i_2}^\mu,\hat p_{i_2}^\mu,\hat p_{i_3}^\mu,\hat p_{i_4}^\mu,\ldots,\hat p_{i_{N-3}}^\mu),\nonumber\\
&\vdots\nonumber\\
&C_{11\cdots 1}(\hat p_{i_1}^\mu,\hat p_{i_1}^\mu,\ldots,\hat p_{i_1}^\mu),
\end{align}
where there is one of these ``standard'' angular functions for each integer partition of $N$.
In particular, the length of the integer partition is how many unique indices appear in the subscript and the values of the partition indicate how many times each index is repeated.

The role of C safety is to impose relationships between these standard angular functions, eventually showing that the only required function is $C_{123\cdots N}$.
Intuitively, this means that as any set of particles become collinear, the angular dependence is that of collinear limit of $N$ arbitrary directions.
The proof that this follows from C safety, however, is the most technically involved step of this derivation.

The requirement of C safety in \Eq{eq:Csafety} implies that $\mathcal S$ is unchanged whether one considers $\{E_i,\hat p_i^\mu\}_{i=1}^M$ or the same particles with a collinear splitting of the first particle, $\{\tilde E_i, \hat p_i^\mu\}_{i=0}^M$, where:
\begin{equation}
\label{eq:csplit}
\tilde E_0=(1- \lambda) E_1,\qquad \tilde E_1 = \lambda E_1,\qquad \hat p_0^\mu=\hat p_1^\mu,\qquad \tilde E_i = E_i,
\end{equation}
for all $\lambda\in[0,1]$ and $i>1$.  Rewriting \Eq{eq:updatedCn}, we can explicitly separate out the terms of the sums involving $k$ collinearly split indices $\{0,1\}$:
\begin{align}
\mathcal C_N&=\sum_{i_1=0}^M\cdots\sum_{i_N=0}^MC_{i_1\cdots i_N}(\hat p_{i_1}^\mu,\ldots,\hat p_{i_N}^\mu)\prod_{j=1}^N\tilde E_{i_j}\label{eq:step1}\\
&=\sum_{k=0}^N\binom{N}{k}\sum_{i_1=0}^1\cdots\sum_{i_k=0}^1\sum_{i_{k+1}=2}^M\cdots\sum_{i_N=2}^MC_{i_1\cdots i_N}(\hat p_{i_1}^\mu,\ldots,\hat p_{i_N}^\mu)\prod_{j=1}^N\tilde E_{i_j}, \label{eq:step2}
\end{align}
where in going to this last expression, we have used the symmetry of $C_{i_1\cdots i_N}$ in its arguments and accounted for the degeneracy of such terms using the binomial factor $\binom{N}{k}$.
We then insert the collinear splitting kinematics of \Eq{eq:csplit} into \Eq{eq:step2},
\begin{align}
\mathcal C_N&=\sum_{k=0}^N\binom{N}{k}\sum_{i_1=0}^1\cdots\sum_{i_k=0}^1\lambda^{\sum_{a=1}^ki_a}(1-\lambda)^{k-\sum_{a=1}^ki_a}E_1^k\label{eq:step3}\\
&\hspace{.5in}\times\sum_{i_{k+1}=2}^M\cdots\sum_{i_N=2}^MC_{i_1\cdots i_N}(\hat p_{i_1}^\mu,\ldots,\hat p_{i_N}^\mu)\prod_{j=k+1}^N E_{i_j} \nonumber\\
&=\sum_{k=0}^N\binom{N}{k}\sum_{\ell=0}^k\binom{k}{\ell}\lambda^\ell(1-\lambda)^{k-\ell}E_1^k\label{eq:collinear}\\
&\hspace{.5in}\times\sum_{i_{k+1}=2}^M\cdots\sum_{i_N=2}^M C_{\underbrace{{}_{0\cdots0}}_{\ell}\underbrace{{}_{1\cdots1}}_{k-\ell}i_{k+1}\cdots i_n}(\hat p_{1}^\mu,\ldots,\hat p_{1}^\mu,\hat p_{i_{k+1}}^\mu,\ldots,\hat p_{i_N}^\mu)\prod_{j=k+1}^N E_{i_j},\nonumber
\end{align}
where in going to this last expression, we have used the particle relabeling symmetry of \Eq{eq:relabelsym2} to sort the $\{0,1\}$ subscript indices of the angular functions.

The constraint of C safety says that \Eq{eq:collinear} is equal to \Eq{eq:updatedCn} on the non-collinearly split event.
To make this constraint more useful, we use the binomial theorem to write 1 in a suggestive way:
\begin{align}\label{eq:binomial1}
1=(\lambda+1-\lambda)^k=\sum_{\ell=0}^k\binom{k}{\ell}\lambda^\ell(1-\lambda)^{k-\ell},
\end{align}
and insert this expression into \Eq{eq:updatedCn}, separating out factors where $k$ of the indices are equal to 1:
\begin{align}\label{eq:Fnmanip}
\mathcal C_N&=\sum_{k=0}^N\binom{N}{k}\sum_{\ell=0}^k\binom{k}{\ell}\lambda^\ell(1-\lambda)^{k-\ell}E_1^k\\
&\nonumber\hspace{.5in}\times \sum_{i_{k+1}=2}^M\cdots\sum_{i_N=2}^MC_{\underbrace{{}_{1\cdots1}}_{k}i_{k+1}\cdots i_N}(\hat p_{1}^\mu,\ldots,\hat p_{1}^\mu,\hat p_{i_{k+1}}^\mu,\ldots,\hat p_{i_N}^\mu)\prod_{j=k+1}^NE_{i_j}.
\end{align}
Subtracting~\Eq{eq:Fnmanip} from~\Eq{eq:collinear} and treating the energies as independent quantities, the following constraint can be read off:
\begin{align}\label{eq:Cconstraint}
\sum_{\ell=0}^k\binom{k}{\ell}\lambda^\ell(1-\lambda)^{k-\ell}\left(C_{\underbrace{{}_{0\cdots0}}_{\ell}\underbrace{{}_{1\cdots1}}_{k-\ell}i_{k+1}\cdots i_N}-C_{\underbrace{{}_{1\cdots 1}}_{k}i_{k+1}\cdots i_N}\right)=0,
\end{align}
where the identical arguments of the angular functions are suppressed for compactness.  

We would like to obtain that the quantity in parentheses in \Eq{eq:Cconstraint} vanishes since the equation holds for all $\lambda$. 
To see this, suppose that the quantity in parentheses does not vanish, and let $\hat\ell$ be the smallest such $\ell$ where this happens.
Consider the regime $0<\lambda\ll1$: by the definition of $\hat\ell$, there are no $\mathcal O(\lambda^\ell)$ terms for $\ell<\hat \ell$ and thus the left-hand side of \Eq{eq:Cconstraint} is $\mathcal O(\lambda^{\hat\ell})\neq 0$, contradicting \Eq{eq:Cconstraint}. 
We thus obtain:
\begin{align}\label{eq:Cresult}
C_{\underbrace{{}_{0\cdots0}}_{\ell}\underbrace{{}_{1\cdots1}}_{k-\ell}i_{k+1}\cdots i_N}(\hat p_{1}^\mu,\ldots,\hat p_{1}^\mu,\hat p_{i_{k+1}}^\mu,\ldots,\hat p_{i_n}^\mu)=C_{\underbrace{{}_{1\cdots 1}}_{k}i_{k+1}\cdots i_N}(\hat p_{1}^\mu,\ldots,\hat p_{1}^\mu,\hat p_{i_{k+1}}^\mu,\ldots,\hat p_{i_N}^\mu),
\end{align}
for $0\le\ell\le k$.  Note that in this expression, the first $k$ arguments of the functions are identical.

The constraint in~\Eq{eq:Cresult} is very powerful, especially when combined with the relabeling symmetry of~\Eq{eq:relabelsym2}.
While we obtained \Eq{eq:Cresult} using the collinear limit, the particle direction $\hat{p}_{0}$ appears nowhere in this expression, so the $0$ subscript is simply an index on the angular function.
Therefore, when any $k$ arguments of one of the angular functions become collinear, any $\ell\le k$ of the corresponding subscript labels may be swapped out for values not appearing anywhere else in the indices.  
A concrete example of this is
\begin{align}\label{eq:relabelex}
C_{1123\cdots N-1}(\hat p_{i_1}^\mu,\hat p_{i_1}^\mu,\hat p_{i_2}^\mu,\ldots,\hat p_{i_{N-1}}^\mu)=C_{1234\cdots N}(\hat p_{i_1}^\mu,\hat p_{i_1}^\mu,\hat p_{i_2}^\mu,\ldots,\hat p_{i_{N-1}}^\mu),
\end{align}
where the $N$ index here plays the role of the $0$ index in \Eq{eq:Cresult}.
This then implies that all of the angular functions in~\Eq{eq:Cexample} can related to a single function:
\begin{align}\label{eq:fNresult}
C_{i_1\cdots i_N}(\hat p_{i_1}^\mu, \ldots, \hat p_{i_N}^\mu) = C_{123\cdots N}(\hat p_{i_1}^\mu,\ldots,\hat p_{i_N}^\mu) \equiv f_N(\hat p_{i_1}^\mu,\ldots,\hat p_{i_N}^\mu),
\end{align}
yielding the intuitive result that the angular dependence when some number of particles become collinear should follow from the collinear limit of $N$ arbitrary directions.

\subsubsection{A new derivation of $C$-correlators}
\label{subsec:final}

Finally, substituting \Eq{eq:fNresult} into \Eq{eq:updatedCn} implies that 
\begin{align}\label{eq:Eexpansionresult}
\mathcal S\simeq\sum_{N=0}^{N_{\rm max}}\mathcal C_N^{f_N},\qquad \mathcal C_N^{f_N}=\sum_{i_1=1}^M\cdots\sum_{i_N=1}^M E_{i_1}\cdots E_{i_N} f_N(\hat p_{i_1}^\mu,\ldots,\hat p_{i_N}^\mu),
\end{align}
where we recognize $\mathcal C_N^{f_N}$ as the $C$-correlators of \Eq{eq:genccorr}.
This expression says that an arbitrary IRC-safe observable can be approximated arbitrarily well by a linear combination of $C$-correlators.
In this way, we have given a new derivation that $C$-correlators linearly span the space of IRC-safe observables by directly imposing the constraints of IRC safety and particle relabeling symmetry on an arbitrary observable.

The argument presented here suffices to show the IRC-safety of the $C$-correlators with any continuous angular weighting function, even if it is not symmetric.
Though we used the symmetrization in \Eq{eq:symf} to aid the C-safety derivation in \Sec{sec:collinearsafety}, it is now perfectly valid to relax this constraint on $f_N$.
In particular, we can simply consider \Eq{eq:symf} applied in reverse and select a single term in the symmetrization sum to represent $f_N$.
Thus we are not constrained merely to symmetric $f_N$, which will be helpful in obtaining the \Bs.

\subsection{Expansion in geometry}
\label{sec:angleexpansion}

Having now established that the $C$-correlators linearly span the space of IRC-safe observables, we now expand the angular weighting function $f_N$ in \Eq{eq:Eexpansionresult} in terms of a discrete linear angular basis.\footnote{Our approach here turns out to be similar to the construction of kinematic polynomial rings for operator bases in \Ref{Henning:2017fpj}.}
By virtue of the sum structure of the $C$-correlators, this angular basis directly translates into a basis of IRC-safe observables, i.e.\ the energy flow basis.

Following the discussion in \Sec{sec:measures}, we take the angular function $f_N$ to depend only on the pairwise angular distances $\theta_{ij}$.
Note that the results of \Sec{sec:Eexpansion} continue to hold with pairwise angular distances in place of particle directions, as long as $\theta_{ij}$ is a dimensionless function of $\hat{p}_i^\mu$ and $\hat{p}_j^\mu$ with no residual dependence on energy.
Of course, this choice would not be valid for expanding IRC-safe observables that do not respect the symmetries implied by $\theta_{ij}$, such as trying to use the default hadronic measure in \Eq{eq:hadronicmeasure} for observables that depend on the overall jet rapidity.
In such cases, one can perform an expansion directly in the $\hat p_{i}^\mu$, though we will not pursue that here.

Expanding the angular function $f_N$ in terms of polynomials up to order $d_{\rm max}$ in the pairwise angular distances yields:
\begin{align}\label{eq:angexp}
f_N(\hat p_{i_1}^\mu,\ldots,\hat p_{i_N}^\mu)&\simeq\sum_{d=0}^{d_{\rm max}} \sum_{\mathcal M\in \Theta_d} b_{\mathcal M} \,\mathcal M,
\end{align}
where $\Theta_d$ is the set of monomials in $\{\theta_{ij}\,|\,i<j\in \{i_1,\ldots,i_N\}\}$ of degree $d$, $\mathcal M$ is one of these monomials, and the $b_{\mathcal M}$ are numerical coefficients.
While this is a perfectly valid expansion, it represents a vast overcounting of the number of potential angular structures.

Our goal is to substitute \Eq{eq:angexp} into the definition of a $C$-correlator in \Eq{eq:Eexpansionresult} and identify the \emph{unique} analytic structures that emerge.
Note that two monomials $\mathcal M_1,\mathcal M_2\in \Theta_d$ that are related by a permutation $\sigma\in S_N$ with action $\theta_{i_ai_b}\to\theta_{i_{\sigma(a)} i_{\sigma(b)}}$ give rise to identical $C$-correlators, $\mathcal C^{\mathcal M_1}=\mathcal C^{\mathcal M_2}$, as a result of the relabeling symmetry in \Sec{sec:relabelsym}.
Thus, we can greatly simplify the angular expansion by summing only over equivalence classes of monomials not related by permutations, which we write as $\Theta_d/S_N$.
Writing this out in terms of $\mathcal{E} \in\Theta_d/S_N$:
\begin{align}
\mathcal C_N^{f_N} & \simeq \sum_{i_1 = 1}^M \cdots \sum_{i_N = 1}^M E_{i_1}\cdots E_{i_N} \sum_{d=0}^{d_{\rm max}} \sum_{\mathcal E\in\Theta_d/S_N}\sum_{\mathcal M\in\mathcal E} \, b_{\mathcal M} \, \mathcal M\label{eq:sub2}\\
&=\sum_{d=0}^{d_{\rm max}}\sum_{\mathcal E\in\Theta_d/S_N}b_{\mathcal E}\sum_{i_1 = 1}^M \cdots \sum_{i_N = 1}^M E_{i_1}\cdots E_{i_N}\mathcal M_{\mathcal E}\label{eq:sub3},
\end{align}
where, by the relabeling symmetry, $\mathcal M_{\mathcal E}$ can be any representative monomial in the equivalence class $\mathcal{E}$, and the coefficient $b_{\mathcal E} = |\mathcal E | \, b_{\mathcal M}$ absorbs the size $|\mathcal E |$ of the equivalence class.

As described in \Sec{sec:efbasis}, the set of monomials $\Theta_d$ is in bijection with the set of multigraphs with $d$ edges and $N$ vertices, and the set of equivalence classes $\Theta_d/S_N$ is in bijection with the set of non-isomorphic multigraphs with $d$ edges and $N$ vertices. 
In particular, each edge $(k,\ell)$ in a multigraph $G$ corresponds to a factor of $\theta_{i_ki_\ell}$ in the monomial $\mathcal M_{\mathcal E}$:
\begin{equation}\label{eq:mon2graph}
\mathcal M_{\mathcal E}=\prod_{(k,\ell)\in G}\theta_{i_ki_\ell},
\end{equation}
where $G$ corresponds to the equivalence class $\mathcal E$. 
By substituting \Eq{eq:mon2graph} into \Eq{eq:sub3} and relabeling the coefficient $b_{\mathcal E}$ to $b_G$, we can identify the resulting analytic structures that linearly span the space of $C$-correlators as the (unnormalized) \Bs:
\begin{equation}\label{eq:efpspanCN}
\mathcal C_N^{f_N}\simeq \sum_{d=0}^{d_{\rm max}}\sum_{G\in\mathcal G_{N,d}}b_G\,\B_G,\qquad \B_G\equiv\sum_{i_1=1}^M\cdots\sum_{i_N=1}^ME_{i_1}\cdots E_{i_N}\prod_{(k,\ell)\in G}\theta_{i_ki_\ell},
\end{equation}
where $\mathcal G_{N,d}$ is the set of non-isomorphic multigraphs with $d$ edges on $N$ vertices.

In \Sec{subsec:final}, it was shown that the set of IRC-safe observables is linearly spanned by the set of $C$-correlators, summarized in \Eq{eq:Eexpansionresult}. 
In this section, we have shown in \Eq{eq:efpspanCN} that the $C$-correlators themselves are linearly spanned by the \Bs, whose angular structures are efficiently encoded by multigraphs. 
By linearity, the \Bs therefore form a complete linear basis for all IRC-safe observables, completing our argument.

\section{Computational complexity of the energy flow basis}
\label{sec:complexity}

Since we would like to apply the energy flow basis in the context of jet substructure, the efficient computation of \Bs is of great practical interest. 
Naively, calculating an \B whose graph has a large number of vertices requires a prohibitively large amount of computation time, especially as the number of particles in the jet grows large.
In practice, though, we can dramatically speed up the implementation of the \Bs by making use of the correspondence with multigraphs.
Beta code to calculate the \Bs using these methods is available through our {\tt \href{https://pkomiske.github.io/EnergyFlow}{EnergyFlow}} module.

\subsection{Algebraic structure}
\label{sec:algebraic}

The set of \Bs has a rich algebraic structure which will allow in some cases for faster computation. 
Firstly, they form a monoid (a group without inverses) under multiplication. 
In analogy with the natural numbers, the composite \Bs, those with disconnected multigraphs, can be expressed as a product of the prime \Bs corresponding to the connected components of a disconnected graph:
\begin{equation}\label{eq:efpprod}
\B_G=\prod_{g\in C(G)}\B_g,
\end{equation}
where $C(G)$ is the set of connected components of the multigraph $G$. 

As a concrete example of \Eq{eq:efpprod}, consider:
\begin{align}
\begin{gathered}
\includegraphics[scale=.2]{graphs/3_3_2}
\includegraphics[scale=.2]{graphs/2_4_1}
\end{gathered}
&= \left(\sum_{i_1=1}^M\sum_{i_1=1}^M\sum_{i_3=1}^Mz_{i_1}z_{i_2}z_{i_3}\theta_{i_1i_2}^2\theta_{i_2i_3}\right)\left(\sum_{i_4=1}^M\sum_{i_5=1}^Mz_{i_4}z_{i_5}\theta_{i_4i_5}^4\right).
\end{align}
Thus, we only need to perform summations for the computation of prime \Bs, with the composite ones given by \Eq{eq:efpprod}. 
Note that if one were combining \Bs with a nonlinear method, such as a neural network, the composite \Bs would not be needed as separate inputs since the model could in principle learn to compute them on its own. 
The composite \Bs are, however, required to have a linear basis and should be included when linear methods are employed, such as those in \Secs{sec:linreg}{sec:linclass}.

The relationship between prime and composite \Bs is just the simplest example of the algebraic structure of the energy flow basis. 
The \Bs depend on $M$ energies and $M\choose2$ pairwise angles, but there are only $3M-4$ degrees of freedom for the phase space of $M$ massless particles, leading generically to additional (linear) relations among the \Bs. 
Hence, the \Bs are an \emph{over}complete linear basis. 
We leave further analysis and exploration of these relations to future work, and simply remark here that linear methods continue to work even if there are redundancies in the basis elements.

\subsection{Dispelling the $\mathcal O(M^N)$ myth for $N$-particle correlators}
\label{sec:dispel}

\begin{table}[t]
\centering
\subfloat[]{\label{tab:numchi:a} 
\begin{tabular}{|c|cc||*{10}{r}|}
\hline
\multicolumn{3}{|c}{$d$} & \multicolumn{1}{c}{\bf1}&\multicolumn{1}{c}{\bf2}&\multicolumn{1}{c}{\bf3}&\multicolumn{1}{r}{\bf4}&\multicolumn{1}{c}{\bf5}&\multicolumn{1}{c}{\,\,\bf6}&\multicolumn{1}{c}{\bf7}&\multicolumn{1}{c}{\,\,\,\bf8}&\multicolumn{1}{c}{\bf9}&\multicolumn{1}{c|}{\bf10} \\ \hhline{:===:t:*{10}{=}:}
\multirow{5}{*}{Prime} & \multirow{4}{*}{$\chi$} 
  & 2 & 1 & 2 & 4 & 9 & 21 & 55 & 146 & 415 & 1\,212 & 3\,653   \\
&& 3 &    &    & 1 & 3 & 12 & 47 & 185 & 757 & 3\,181 & 13\,691 \\
&& 4 &    &    &    &    &      & 1   & 2     & 11   & 49       & 231       \\ 
&& 5 &    &    &    &    &      &      &        &        &            & 1           \\ 
\hhline{|~|--||*{10}{-}|}
&\multicolumn{2}{c||}{Total} & 1 & 2 & 5 & 12 & 33 & 103 & 333 & 1\,183 & 4\,442 & 17\,576 \\ 
\hhline{:===::*{10}{=}:}
\multirow{5}{*}{All} & \multirow{4}{*}{$\chi$}
  & 2 & 1 & 3 & 7 & 19 & 48 & 135 & 371 & 1\,077 & 3\,161 & 9\,539  \\
&& 3 &    &    & 1 & 4  & 18 & 76   & 312 & 1\,296 & 5\,447 & 23\,268 \\
&& 4 &    &    &    &     &      & 1     & 3     & 16       & 74       & 352       \\
&& 5 &    &    &    &     &      &        &        &            &            & 1           \\ 
\hhline{|~|--||*{10}{-}|}
&\multicolumn{2}{c||}{Total} & 1 & 3 & 8 & 23 & 66 & 212 & 686 & 2\,389 & 8\,682 & 33\,160 \\ 
\hhline{---||*{10}{-}}
\end{tabular}}
\\
\subfloat[]{\label{tab:numchi:b} 
\begin{tabular}{|cc||*{15}{r}|}
\hhline{--::*{15}{-}}
&& \multicolumn{15}{c|}{$N$} \\ 
$d$ & $\chi$
\unskip\textcolor{white!5}{\makebox[0pt]{\smash{\rule[27pt]{20pt}{3pt}}}}  
& \bf2 & \bf3 & \bf4 & \bf5 & \bf6 & \bf7 & \bf8 & \bf9 & \bf10 & \bf11 & \bf12 & \bf13 & \bf14 & \bf15 & \bf16 \\ 
\hhline{|--|:*{15}{=}:}
\multirow{1}{*}{\bf1} & 2 & 1 & & & & & & & & & & & & & & \\ \hhline{|-|-||*{15}{-}|}
\multirow{1}{*}{\bf2} & 2 & 1 & 1 & 1 & & & & & & & & & & & & \\ \hhline{|-|-||*{15}{-}|}
\multirow{2}{*}{\bf3} & 2 & 1 & 1 & 3 & 1 & 1 & & & & & & & & & & \\ 
                            & 3 & & 1 & & & & & & & & & & & & & \\ \hhline{|-|-||*{15}{-}|}
\multirow{2}{*}{\bf4} & 2 & 1 & 2 & 5 & 5 & 4 & 1 & 1 & & & & & & & & \\ 
                            & 3 & & 1 & 2 & 1 & & & & & & & & & & & \\ \hhline{|-|-||*{15}{-}|}
\multirow{2}{*}{\bf5} & 2 & 1 & 2 & 8 & 10 & 14 & 7 & 4 & 1 & 1 & & & & & & \\ 
                            & 3 & & 2 & 5 & 7 & 3 & 1 & & & & & & & & & \\ \hhline{|-|-||*{15}{-}|}
\multirow{3}{*}{\bf6} & 2 & 1 & 3 & 12 & 21 & 33 & 30 & 21 & 8 & 4 & 1 & 1 & & & & \\ 
                            & 3 & & 3 & 12 & 23 & 23 & 11 & 3 & 1 & & & & & & & \\
                            & 4 & & & 1 & & & & & & & & & & & & \\ \hhline{|-|-||*{15}{-}|}
\multirow{3}{*}{\bf7} & 2 & 1 & 3 & 16 & 35 & 71 & 82 & 81 & 45 & 23 & 8 & 4 & 1 & 1 & & \\ 
                            & 3 & & 4 & 23 & 65 & 92 & 76 & 36 & 12 & 3 & 1 & & & & & \\
                            & 4 & & & 1 & 1 & 1 & & & & & & & & & & \\ \hhline{|-|-||*{15}{-}|}
\multirow{3}{*}{\bf8} & 2 & 1 & 4 & 21 & 58 & 134 & 205 & 245 & 197 & 122 & 52 & 24 & 8 & 4 & 1 & 1 \\ 
                            & 3 & & 5 & 41 & 153 & 311 & 355 & 257 & 118 & 40 & 12 & 3 & 1 & & & \\
                            & 4 & & & 3 & 5 & 5 & 2 & 1 & & & & & & & & \\ \hhline{|-|-||*{15}{-}|}
\end{tabular}}
\caption{(a) The number of prime/all \Bs binned by degree $d$ and complexity $\chi$ up to $d=10$. The complexity is that of our \href{https://pkomiske.github.io/EnergyFlow}{\tt EnergyFlow} implementation, running in time $\mathcal O(M^\chi)$. The partial sums of the ``Total'' rows are the entries of \Tab{tab:efpcounts:a}. (b) The number of \Bs binned by degree $d$, complexity $\chi$, and $N$ up to $d=8$. Note that the majority of \Bs shown here have $N>4$, which would be computationally intractable without algorithmic speedups such as VE.}
\label{tab:numchi} 
\end{table}

\afterpage{\clearpage}

It is useful to analyze the complexity of computing an \B.\footnote{The title of this section is inspired by \Ref{Cacciari:2005hq}.}
A naive implementation of \Eq{eq:introefp} runs in $\mathcal O(M^N)$ due to the $N$ nested sums over $M$ particles.
There is a computational simplification, however, that can be used to tremendously speed up calculations of certain \Bs by making use of the graph structure of $G$.
As an example, consider the following EFP:
\begin{equation}\label{eq:wedgesimple}
\begin{gathered}
\includegraphics[scale=.3]{graphs/4_3_1}
\end{gathered}
=\sum_{i_1=1}^M \sum_{i_2=1}^M \sum_{i_3=1}^M\sum_{i_4=1}^M z_{i_1}z_{i_2}z_{i_3}z_{i_4}\theta_{i_1i_2}\theta_{i_1i_3}\theta_{i_1i_4} = \sum_{i_1=1}^M z_{i_1} \left(\sum_{i_2=1}^M z_{i_2}\theta_{i_1i_2}\right)^3,
\end{equation}
which can be computed in $\mathcal O(M^2)$ rather than $\mathcal O(M^4)$ by first computing the $M$ objects in parentheses in \Eq{eq:wedgesimple} and then performing the overall sum.

In general, since the summand is a product of factors, the distributive property allows one to put parentheses around combinations of sum operators and factors. 
A clever choice of such parentheses, known as an \emph{elimination ordering}, can often be used to perform the $N$ sums of \Eq{eq:introefp} in a way which greatly reduces the number of operations needed to obtain the value of the \B for a given set of particles. 
This technique is known as the Variable Elimination (VE) algorithm~\cite{zhang1996exploiting} (see also \Ref{murphy2012machine} for a review). 

When run optimally, the VE algorithm reduces the complexity of computing $\B_G$ to $\mathcal{O}(M^{\text{tw}(G)+1})$ where $\text{tw}(G)$ is the \emph{treewidth} of the graph $G$, neglecting multiple edges in the case of multigraphs.
The treewidth is a measure which captures how tangled a graph is, with trees (graphs with no cycles) being the least tangled (with treewidth 1) and complete graphs the most tangled (with treewidth $N-1$). 
Additionally, we have that for graphs with a single cycle the treewidth is 2 and for complete graphs minus one edge the treewidth is $N-2$. 
Thus the \Bs corresponding to tree multigraphs can be computed with VE in $\mathcal O(M^2)$ whereas complete graphs do require the naive $\mathcal O(M^N)$ to compute with VE. 
Since the ECFs correspond to complete graphs (see \Eq{eq:ecfgraphs}), they do not benefit from VE.  Similarly, VE cannot speed up the computation of ECFGs, since the ECFGs do not have a factorable summand.

Finding the optimal elimination ordering and computing the treewidth for a graph $G$ are both NP-hard. 
In practice, heuristics are used to decide on a pretty-good elimination ordering (which for the small graphs we consider here is often optimal) and to approximate the treewidth. 
In principle, these orderings need only be computed once for a fixed set of graphs of interest. 
Similarly, many algebraic structures reappear when computing a set of \Bs for the same set of particles, making dynamic programming a viable technique for further improving the computational complexity of the method.

\begin{figure}[t]
\centering
\includegraphics[scale=.76]{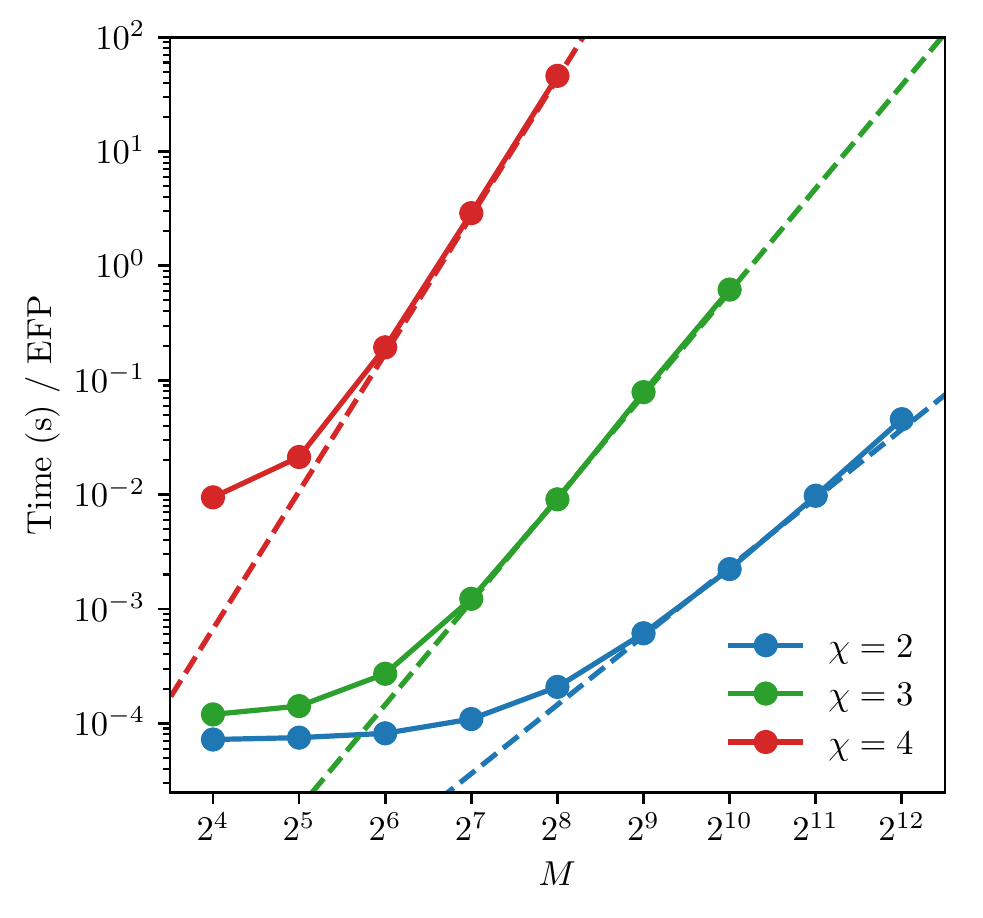}
\caption{Compute time (in seconds) per \B for different VE complexities $\chi$ as a function of the number of inputs $M$.  The quoted value is based on all \Bs with $d\le7$, and each data point is the average of 10 computations.  The dashed lines show the expected $\mathcal{O}(M^\chi)$ scaling behavior.  As $\chi$ increases, the relative amount of overhead decreases and the asymptotic behavior is achieved more rapidly than for smaller $\chi$. Computations were run with Python 3.5.2 and {\tt NumPy} 1.13.3 on a 2.3 GHz Intel Xeon E5-2673 v4 (Broadwell) processor on Microsoft Azure using our \href{https://pkomiske.github.io/EnergyFlow}{\tt EnergyFlow} module.}
\label{fig:times}
\end{figure}

\Tab{tab:numchi:a} shows the number of \Bs listed by degree $d$ and VE complexity $\chi$ (with respect to the heuristics used in our implementation), and \Tab{tab:numchi:b} further breaks up the \B counts by $N$. 
\Fig{fig:times} shows the time to compute the average $d\le7$ \B as a function of multiplicity $M$ for different VE complexity $\chi$. 
Finally, we note that though VE often provides a significant speedup over the naive algorithm, there may be even faster ways of computing the \Bs.\footnote{At the risk of burying the lede in a footnote, we have found that with certain choices of the angular measure, it is possible to compute all \Bs in $\mathcal O(M)$.  We leave a further exploration of these interesting special cases to future work.}

\section{Linear regression with jet observables}
\label{sec:linreg}

Regression, classification, and generation are three dominant machine learning paradigms. 
Machine learning applications in collider physics have been largely focused on classification (e.g.\ jet tagging)~\cite{Komiske:2016rsd,Almeida:2015jua,Baldi:2016fql,Kasieczka:2017nvn,Pearkes:2017hku,Butter:2017cot,Aguilar-Saavedra:2017rzt,Guest:2016iqz,Louppe:2017ipp} with recent developments in regression~\cite{Komiske:2017ubm} and generation~\cite{deOliveira:2017pjk, Paganini:2017hrr}. 
For a more complete review of modern machine learning techniques in jet substructure, see \Ref{Larkoski:2017jix}. 
The lack of established regression problems in jet physics is due in part to the difficulty of theoretically probing multivariate combinations as well as the challenges associated with extracting physics information from trained regressions models.

In this section, we show that the linearity of the energy flow basis mitigates many of these problems, providing a natural regression framework using simple linear models, probing the learned observable combinations, and gaining insight into the physics of the target observables. 
Since regression requires training samples, we observe how the regression performance compares on jets with three characteristic phase-space configurations: one-prong QCD jets, two-prong boosted $W$ jets, and three-prong boosted top jets. 
We use linear regression to demonstrate convergence of the energy flow basis on IRC-safe observables, while illustrating their less-performant behavior for non-IRC-safe observables.

\subsection{Linear models with the energy flow basis}
\label{sec:linmod}

Linear models assume a linear relationship between the input and target variables, making them the natural choice for (machine) learning with the energy flow basis for both regression and classification. 
A linear model $\mathscr M$ with \Bs as the inputs is defined by a finite set $\mathcal G$ of multigraphs and numerical coefficients ${\bf w}=\{w_G\}_{G\in\mathcal G}$:
\begin{equation}\label{eq:linmod}
\mathscr M = \sum_{G\in\mathcal G}w_G \,  \B_G.
\end{equation}
The fundamental relationship between \Bs, linear models, and IRC-safe observables is highlighted by comparing \Eq{eq:linmod} to \Eq{eq:introefpsspan}, where the linear model $\mathscr M$ in \Eq{eq:linmod} takes the place of the IRC-safe observable $\mathcal S$ in \Eq{eq:introefpsspan}. 
Because the \Bs are a complete linear basis, $\mathscr M$ is capable of approximating any $\mathcal S$ for a sufficiently large set of \Bs.

The linear structure of \Eq{eq:linmod} allows for an avenue to ``open the box'' and interpret the learned coefficients as defining a unique multiparticle correlator for each $N$. 
To see this, partition the set $\mathcal G$ into subsets $\mathcal G_N$ of graphs with $N$ vertices.
The sum in \Eq{eq:linmod} can be broken into two sums, one over $N$ and the other over all graphs in $\mathcal G_N$. 
The linear energy structure of the \Bs in \Eq{eq:introefp} allows for the second sum to be pushed inside the product of energies onto the angular weighting function:
\begin{equation}\label{eq:linmodefp}
\mathscr M=\sum_{N=0}^{N_{\rm max}}\sum_{i_1=1}^M\cdots\sum_{i_N=1}^Mz_{i_1}\cdots z_{i_N}\left(\sum_{G\in\mathcal G_N}w_G\prod_{(k,\ell)\in G}\theta_{i_ki_\ell}\right),
\end{equation}
where $N_{\rm max}$ is the maximum number of vertices of any graph in $\mathcal G$. 
The quantity in parentheses in \Eq{eq:linmodefp} may be though of as a \emph{single} angular weighting function. 
The linear model written in this way reveals itself to be a sum of $C$-correlators (similar to \Eq{eq:Eexpansionresult}), one for each $N$, where the linear coefficients within each $\mathcal G_N$ parameterize the angular weighting function $f_N$ of that $C$-correlator. 
This arrangement of the learned parameters of the linear model into $N_{\rm max}$ $C$-correlators contrasts sharply with the lack of a physical organization of parameters in nonlinear methods such as neural networks or boosted decision trees.

\subsection{Event generation and \B computation}
\label{sec:eventgen}

For the studies in this section and in \Sec{sec:linclass}, we generate events using \textsc{Pythia} 8.226~\cite{Sjostrand:2006za,Sjostrand:2007gs,Sjostrand:2014zea} with the default tunings and shower parameters at $\sqrt{s}=\SI{13}{TeV}$.
Hadronization and multiple parton interactions (i.e.\ underlying event) are included, and a $\SI{400}{GeV}$ parton-level $p_T$ cut is applied.
For quark/gluon distribution, quark (signal) jets are generated through $pp \to q Z(\to \nu\bar\nu)$, and gluon (background) jets through $pp \to g Z(\to \nu\bar\nu)$, where only light-quarks ($uds$) appear in the quark sample.
For $W$ and top tagging, signal jets are generated through $pp\to W^+W^-(\to\text{hadrons})$ and $pp\to t\bar t(\to\text{hadrons})$, respectively.
For both $W$ and top events, the background consists of QCD dijets.

Final state, non-neutrino particles were made massless, keeping $y$, $\phi$, and $p_T$ fixed,\footnote{Using massless inputs is not a requirement for using the \Bs, but for these initial \B studies, we wanted to avoid the caveats associated with massive inputs for the validity of \Sec{sec:basis}.} and then were clustered with \textsc{FastJet 3.3.0}~\cite{Cacciari:2011ma} using the anti-$k_T$ algorithm~\cite{Cacciari:2008gp} with a jet radius of $R=0.4$ for quark/gluon samples and $R = 0.8$ for $W$ and top samples (and the relevant dijet background). 
The hardest jet with rapidity $|y|<1.7$ and $500\text{ GeV}\le p_T\le550$ GeV was kept. 
For each type of sample, 200k jets were generated. 
For the regression models, 75\% were used for training and 25\% for testing.

For these events, all \Bs up to degree $d\le7$ were computed in Python using our \href{https://pkomiske.github.io/EnergyFlow}{\tt EnergyFlow} module making use of {\tt NumPy}'s einsum function. 
See \Tabs{tab:efpcounts}{tab:numchi} for counts of \Bs tabulated by various properties such as $N$, $d$, and $\chi$. 
Note that all but 4 of the 1000 $d\le 7$ \Bs can be computed in $\mathcal O(M^2)$ or $\mathcal O(M^3)$ in the VE paradigm, making the set of \Bs with $d\le 7$ efficient to compute.

\subsection{Spanning substructure observables with linear regression}
\label{sec:regression}

We now consider the specific case of training linear models to approximate substructure observables with linear combinations of \Bs. 
For an arbitrary observable $O$, we use least-squares regression to find a suitable set of coefficients ${\bf w}^*$:
\begin{equation}\label{eq:linregdef}
{\bf w}^* = \underset{\bf w}{\arg\min} \left\{\sum_{J\in\text{jets}}\left(O(J) - \sum_{G\in\mathcal G} w_G \,\B_G(J)\right)^2\right\},
\end{equation}
where $O(J)$ is the value of the observable and $\B_G(J)$ the value of the \B given by multigraph $G$ on jet $J$. 
There are possible modifications to \Eq{eq:linregdef} which introduce penalties proportional to $\|{\bf w}\|_1$ or $\|{\bf w}\|_2^2$ where $\|\cdot\|_1$ is the 1-norm and $\|\cdot\|_2$ is the 2-norm. 
The first of these choices, referred to as lasso regression~\cite{tibshirani1996regression}, may be particularly interesting because of the variable selection behavior of this model, which would aid in selecting the most important \Bs to approximate a particular observable. 
We leave such investigation to future work. See \Ref{bishop2006pattern} for a review of linear models for regression.

We use the {\tt LinearRegression} class of the {\tt scikit-learn} python module~\cite{scikit-learn} to implement \Eq{eq:linregdef} with no regularization on the samples described in \Sec{sec:eventgen}. 
In general, the smallest possible regularization which prevents overfitting (if any) should be used.
Because of the linear nature of linear regression and the analytic tractability of \Eq{eq:linregdef}, the ${\bf w}^*$ corresponding to the global minimum of the squared loss function can be found efficiently using convex optimization techniques. 
Such techniques include closed-form solutions or convergent iterative methods.

\begin{table}[t]
\centering
\begin{tabular}{|r||l|l|}
\hhline{~|-|-|}
\multicolumn{1}{c|}{}& \textbf{Observable} & \textbf{Properties}  \\ \hhline{-::==:}
$m_J$/$p_{T,J}$ 
\unskip\textcolor{white!5}{\makebox[0pt]{\smash{\rule[12pt]{14pt}{3pt}}}}
& Scaled jet mass &  No Taylor expansion about zero energy limit \\ 
$\lambda^{(\alpha=1/2)}$ & Les Houches Angularity & No analytic relationship beyond even integer $\alpha$ \\ 
$\tau_2^{(\beta = 1)}$ & $2$-subjettiness & Algorithmically defined IRC-safe observable \\
\hhline{|-||-|-|}
$\tau_{21}^{(\beta=1)}$ & $N$-subjettiness ratio & Sudakov safe, safe for two-prong kinematics\\
$\tau_{32}^{(\beta = 1)}$ & $N$-subjettiness ratio & Sudakov safe, safe for three-prong kinematics\\
\hhline{|-||-|-|}
$M$ & Particle multiplicity & IRC unsafe\\
\hhline{|-||-|-|}
\end{tabular}
\caption{The six substructure observables used as targets for linear regression, listed with relevant properties. The first three are IRC safe, the next two are Sudakov safe in general (and IRC safe in the noted regions of phase space), and particle multiplicity is IRC unsafe. The Les Houches Angularity~\cite{Badger:2016bpw,Gras:2017jty} is calculated with respect to the $p_T$-weighted centroid axis in \Eq{eq:jetaxis}, and the $N$-subjettiness observables~\cite{Thaler:2010tr,Thaler:2011gf} are calculated using $k_T$ axes.}
\label{tab:observables}
\end{table}

\begin{figure}[t]
\centering
\subfloat[]{\includegraphics[scale=.52]{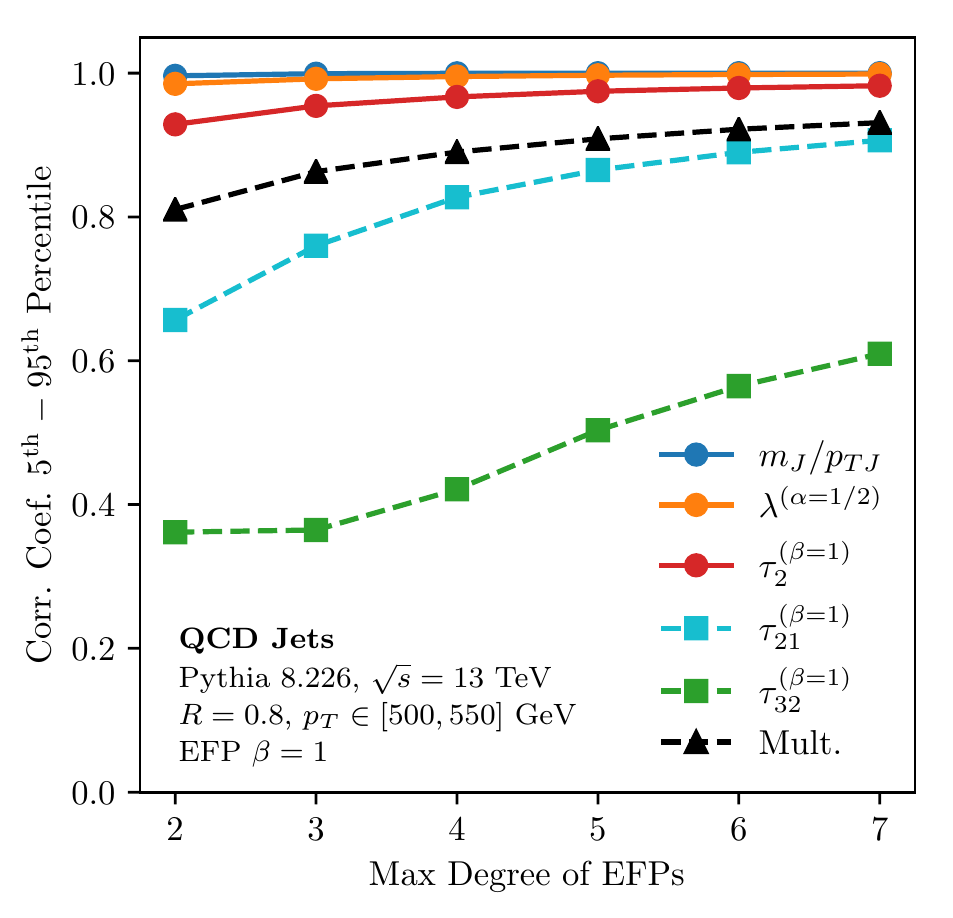}}
\subfloat[]{\includegraphics[scale=.52]{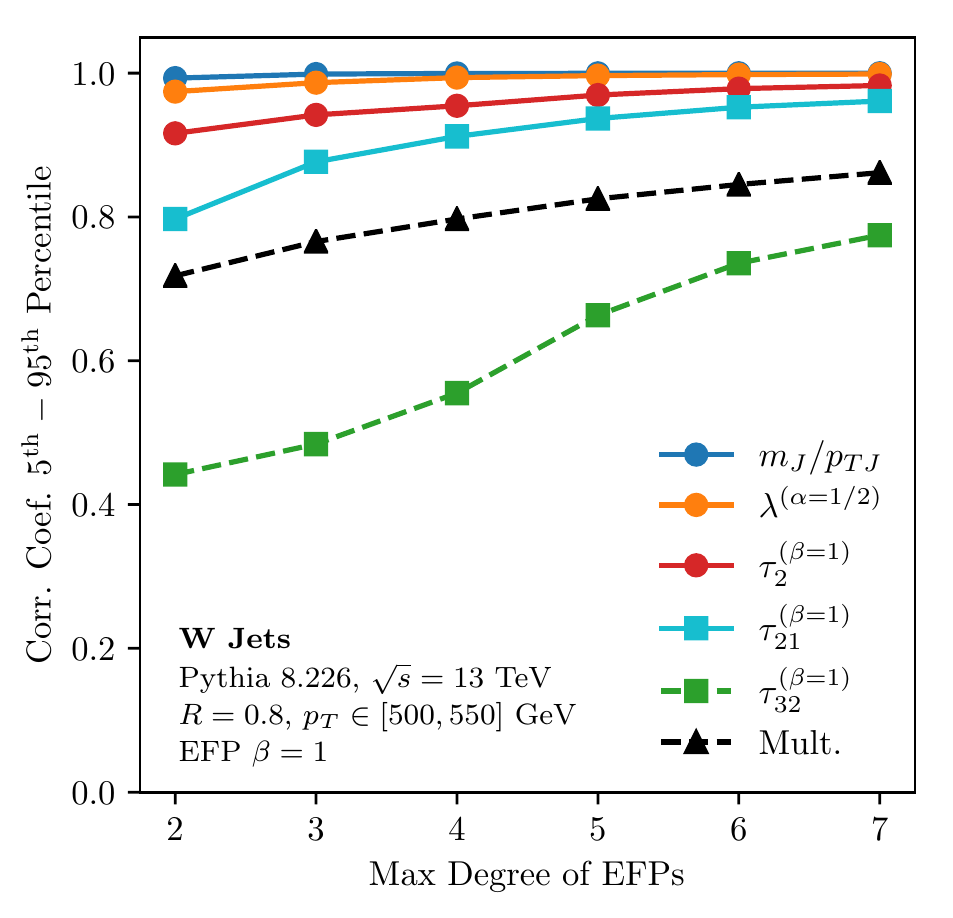}}
\subfloat[]{\includegraphics[scale=.52]{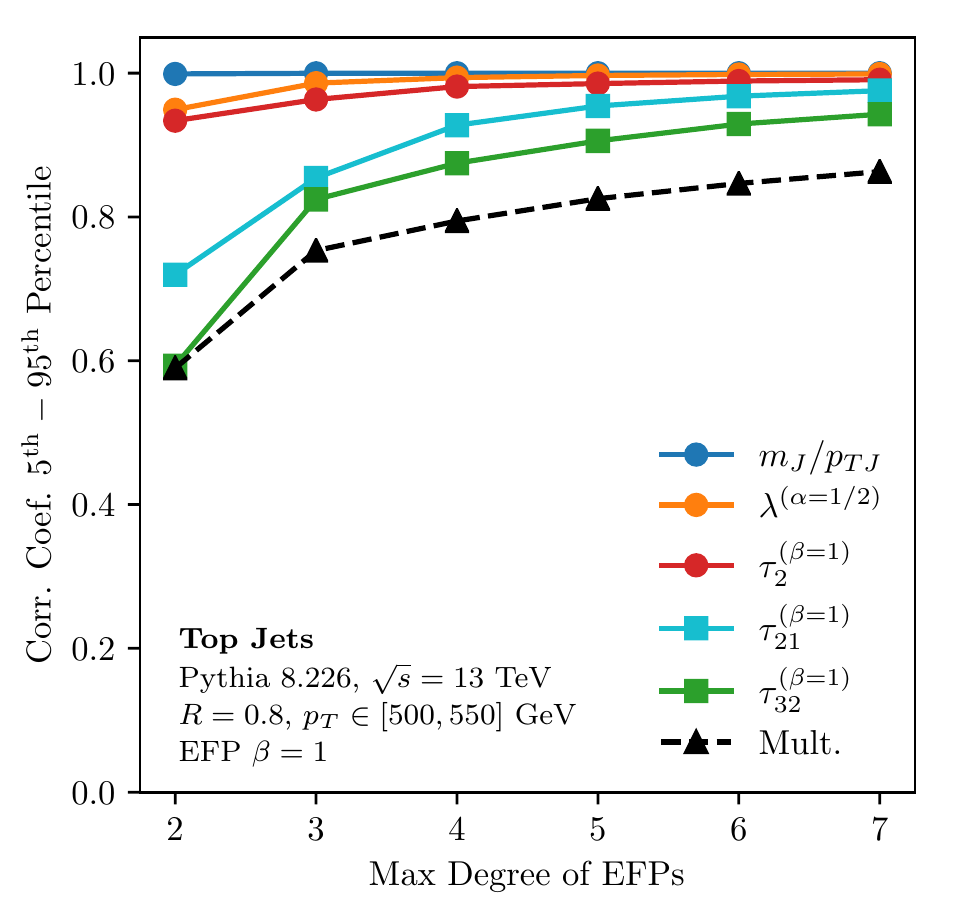}}
\caption{Correlation coefficients between true and predicted values for the jet observables in \Tab{tab:observables}, plotted as a function of maximum \B degree. Shown are the (a) QCD dijet, (b) $W$ jet, and (c) top jet samples, and as explained in the text, we restrict to predictions in the 5$^\text{th}$--95$^\text{th}$ percentiles. Observables in IRC-safe regions of phase space are shown with solid lines and those in IRC-unsafe regions (including Sudakov-safe regions) are shown with dashed lines. The IRC-safe observables are all learned with correlation coefficient above 0.98 in all three cases by $d=7$. Multiplicity (black triangles) sets the scale for the regression performance on IRC-unsafe observables. Note that $\tau_{21}$ has performance similar to the IRC-safe observables only when jets are characteristically two-pronged or higher ($W$ and top jets), and similarly for $\tau_{32}$ when the jets are characteristically three-pronged (top jets).}
\label{fig:robustcorr}
\end{figure}

\begin{figure}[t]
\centering
\subfloat[]{\includegraphics[scale=.485]{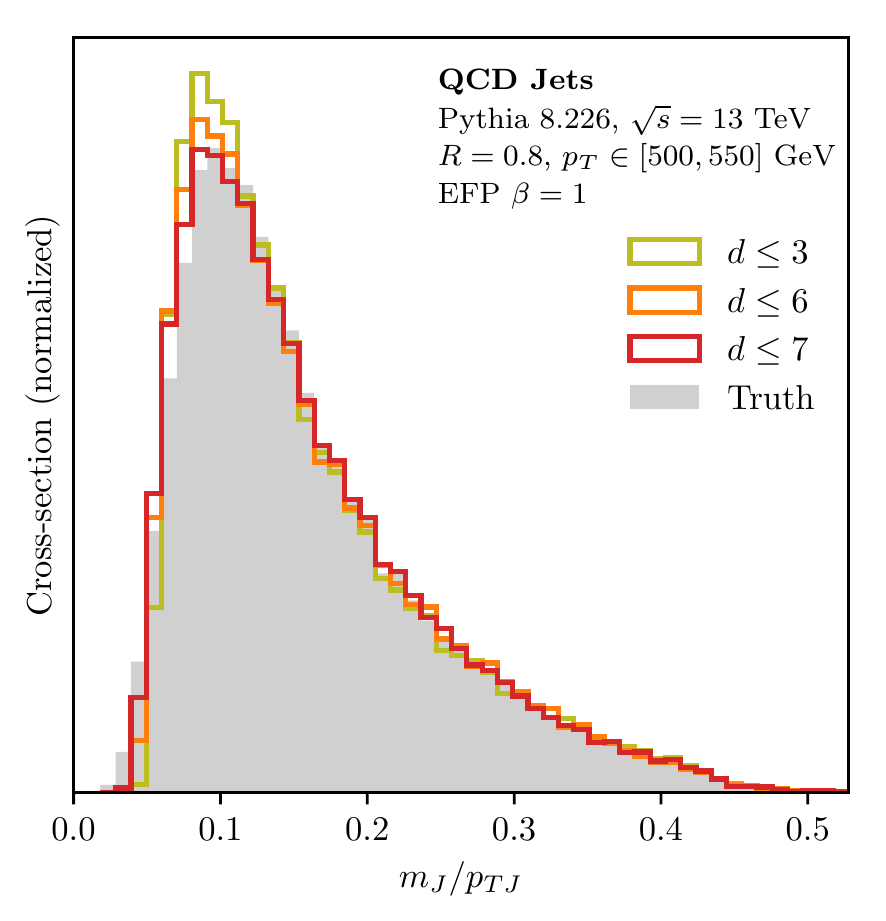}}
\subfloat[]{\includegraphics[scale=.485]{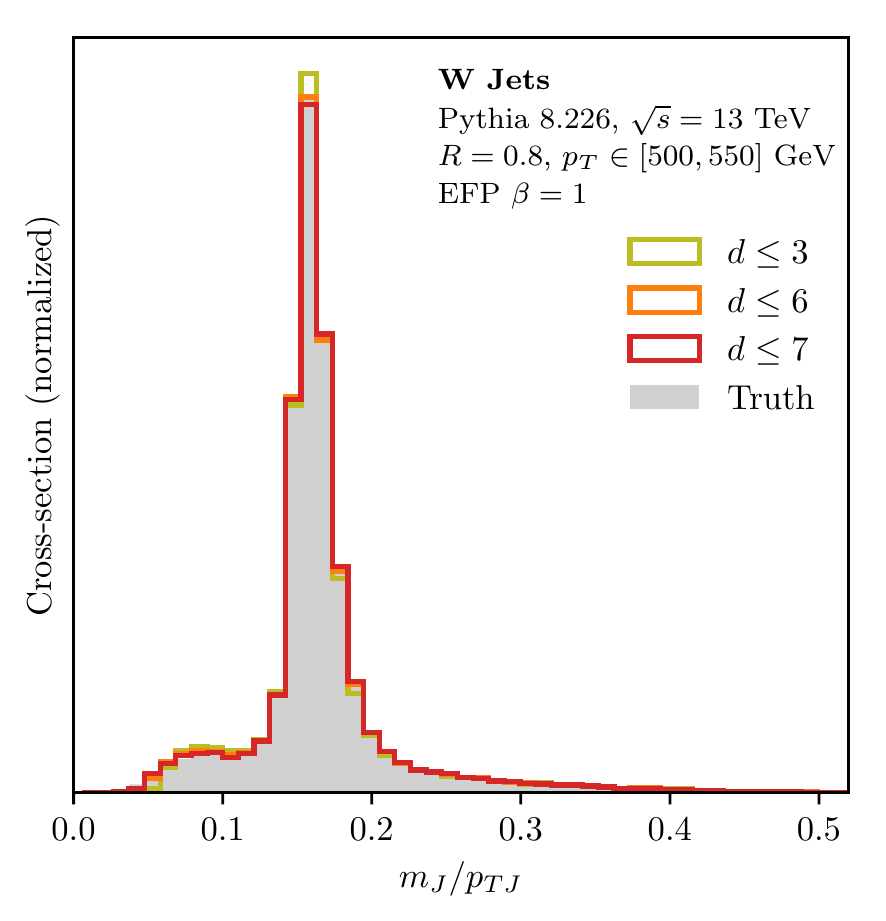}}
\subfloat[]{\includegraphics[scale=.485]{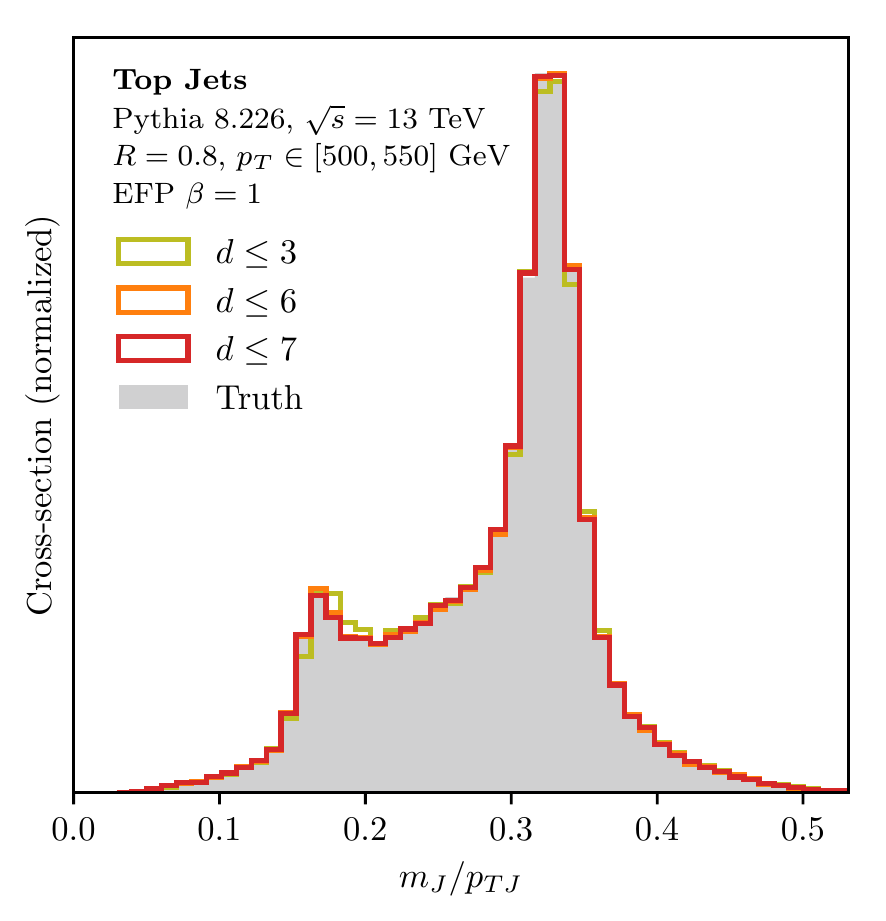}}
\\
\subfloat[]{\includegraphics[scale=.485]{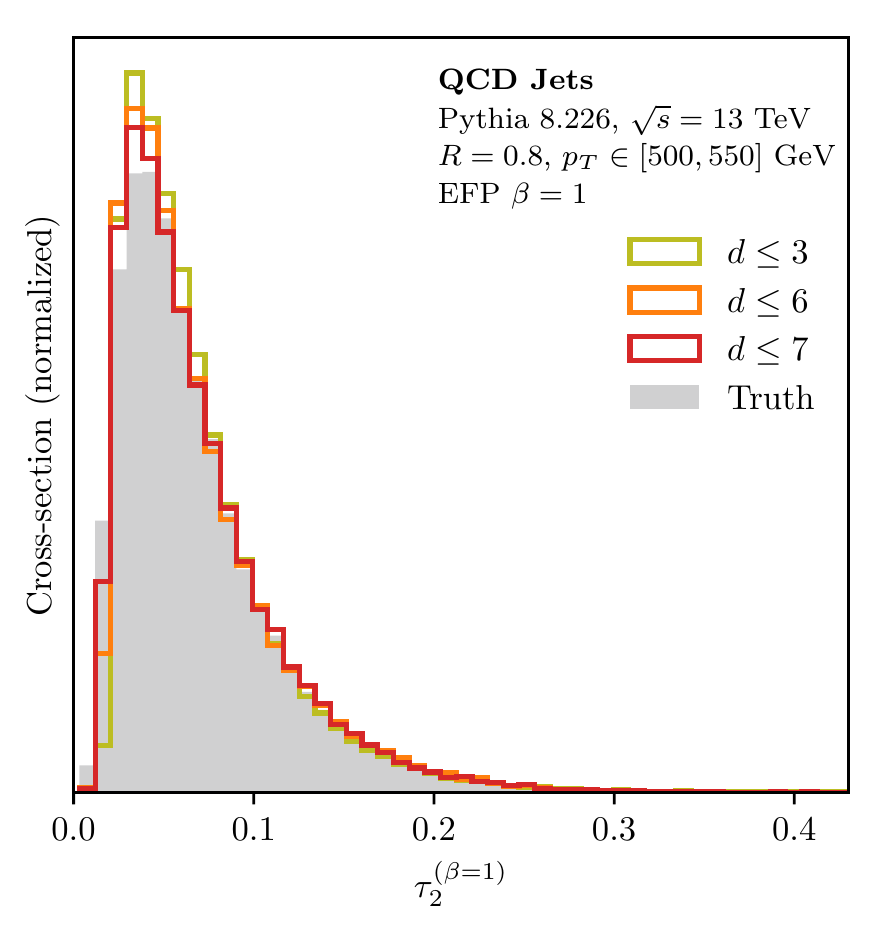}}
\subfloat[]{\includegraphics[scale=.485]{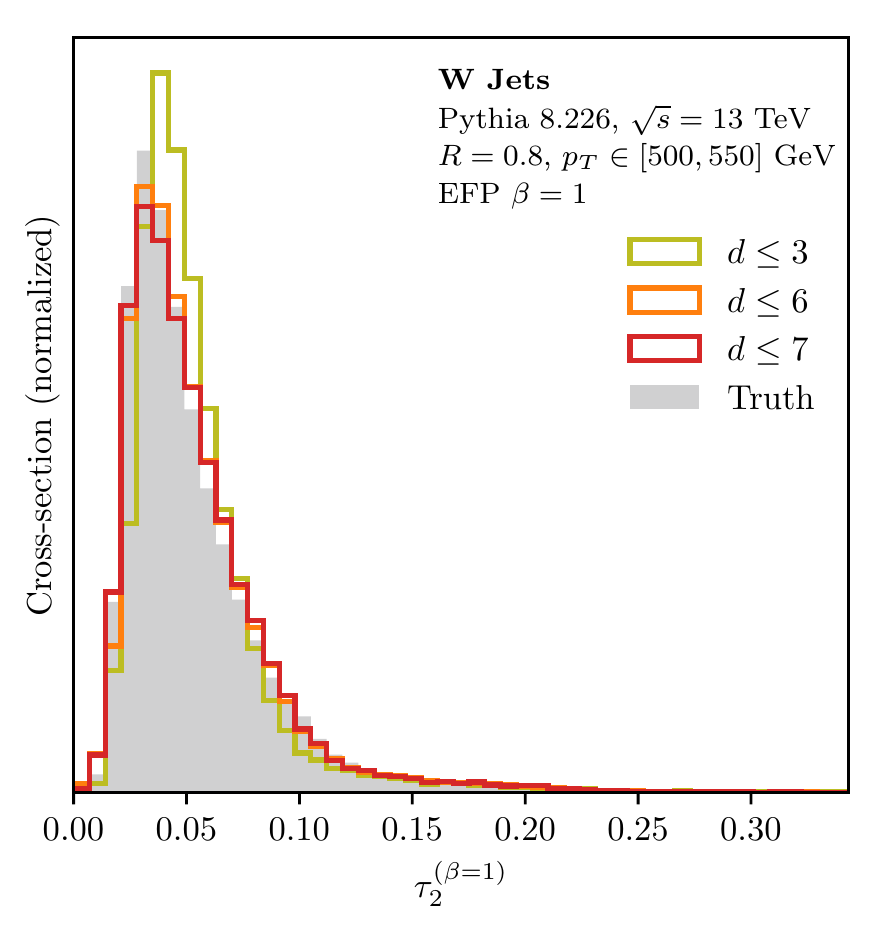}}
\subfloat[]{\includegraphics[scale=.485]{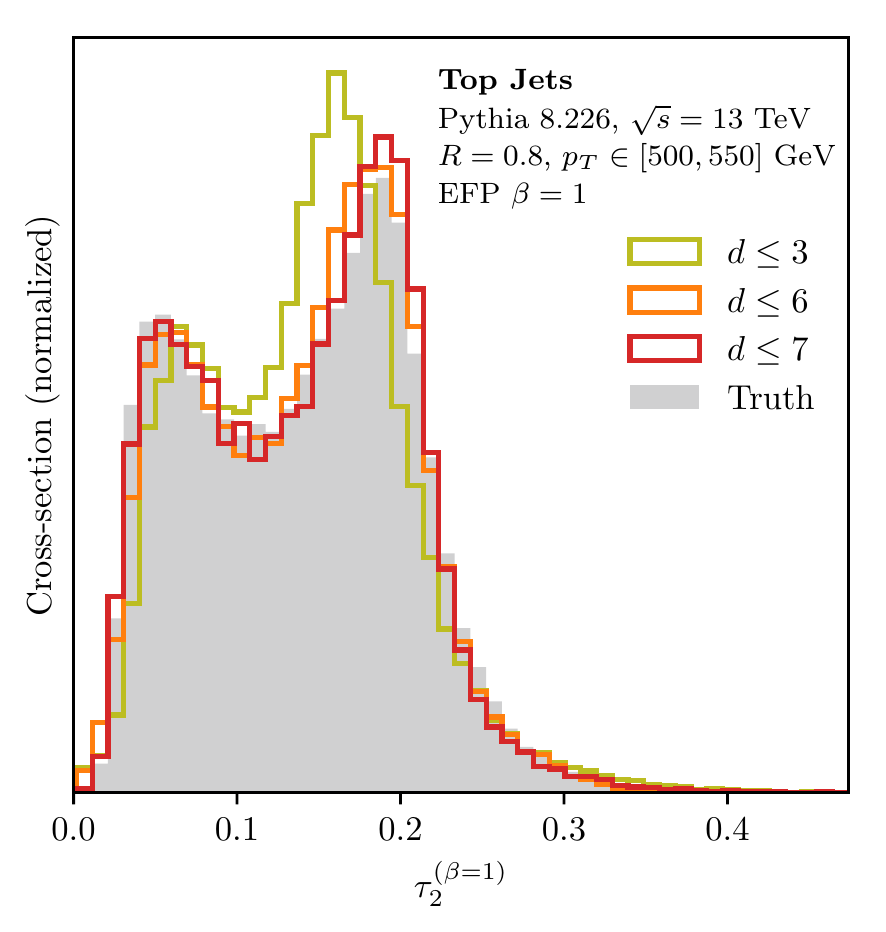}}
\\
\subfloat[]{\includegraphics[scale=.485]{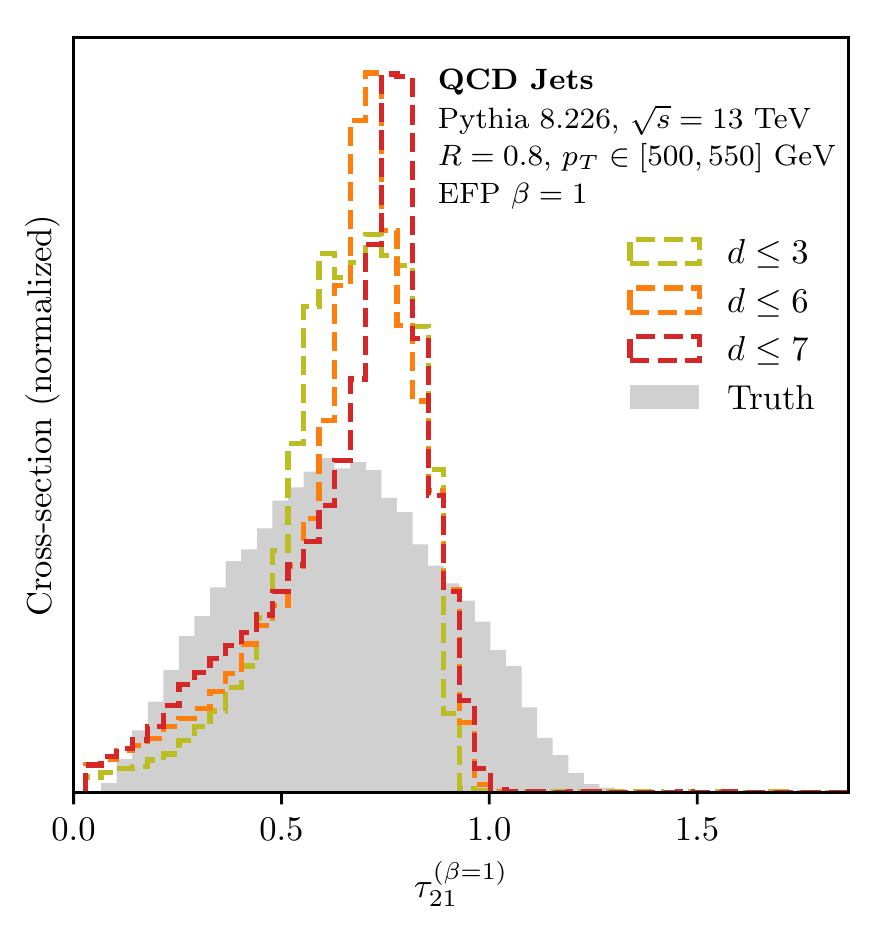}}
\subfloat[]{\includegraphics[scale=.485]{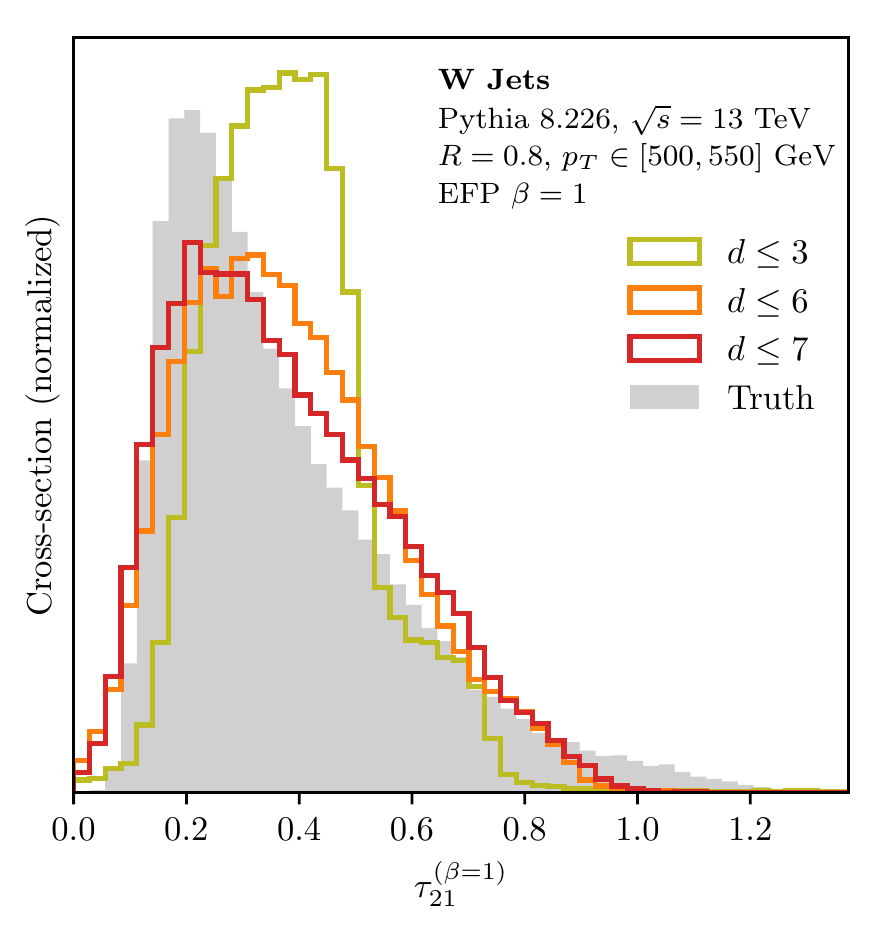}}
\subfloat[]{\includegraphics[scale=.485]{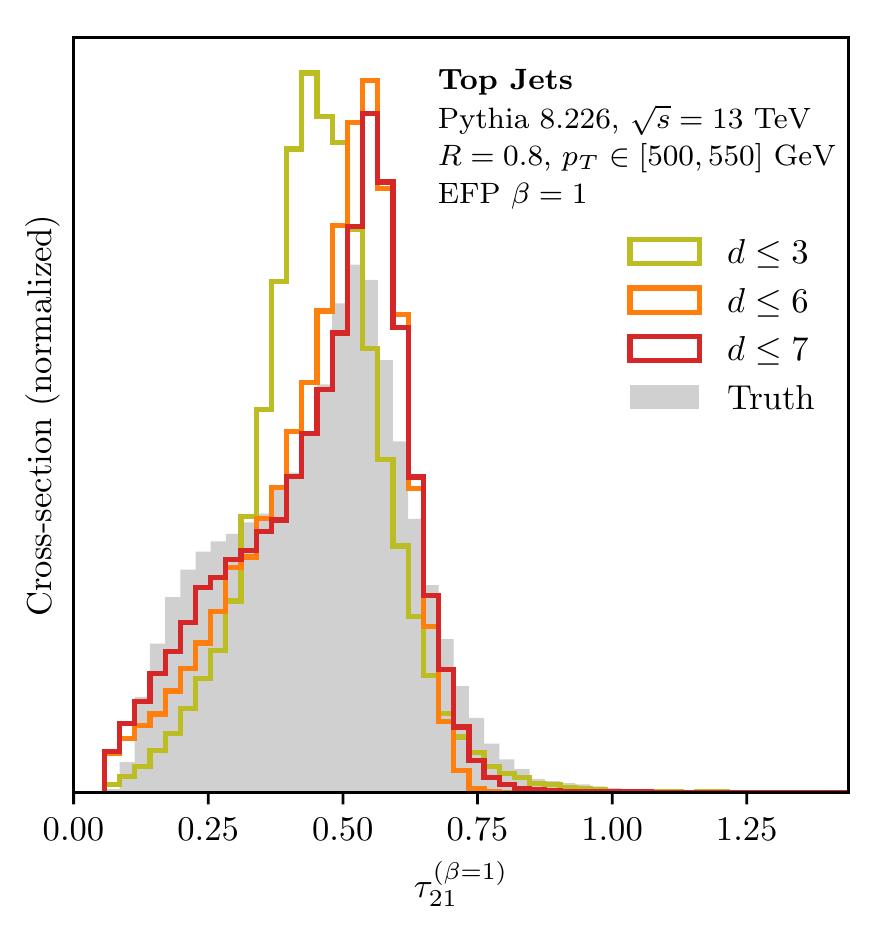}}
\caption{The distributions of true and predicted scaled jet mass (top), $\tau_2^{(\beta=1)}$ (middle), and $\tau_{21}^{(\beta=1)}$ (bottom) using linear regression with \Bs up to different maximum degrees $d$ on QCD jets (left), $W$ jets (center), and top jets (right). Note the excellent agreement for the IRC-safe observables in the first two rows. Observables in IRC-safe regions of phase space are shown with solid lines and those in IRC-unsafe regions are shown with dashed lines. The Sudakov-safe $\tau_{21}^{(\beta=1)}$ predicted distributions match the true distributions for jets typically with two or more prongs ($W$ and top jets) better than for typically one-pronged (QCD) jets.}
\label{fig:reghists}
\end{figure}

As targets for the regression, we consider the six jet observables in \Tab{tab:observables} to highlight some interesting test cases. 
As our measure of the success of the regression, we use a variant of the correlation coefficient between the true and predicted observables that is less sensitive to outliers than the unadulterated correlation coefficient.
When evaluating the trained linear model on the test set, only test samples with predicted values within the 5$^\text{th}$ and 95$^\text{th}$ percentiles of the predictions are included.
In the contexts considered in this paper, narrowing this percentile range lowers the correlation coefficient and widening the range out toward all of the test set increases the correlation coefficient.
The qualitative nature of the results are insensitive to the specific choice of percentile cutoffs. 
We perform this regression using \Bs of degree up to $d$ for $d$ from 2 to 7 on all three jet samples, with the results shown in \Fig{fig:robustcorr}.
Histograms of the true and predicted distributions for a subset of these observables are shown in \Fig{fig:reghists} for the three types of jets considered here. 

Since the learned coefficients depend on the training set, in principle different linear combinations may be learned to approximate the substructure observables in different jet contexts.
This stands in contrast to the analysis in \Sec{sec:jetobs}, where many jet substructure observables were identified as exact linear combinations of \Bs, independent of the choice of inputs.
The IRC-safe observables---mass, Les Houches angularity, and $2$-subjettiness---are all learned with a correlation coefficient above 0.98 in all three cases by $d=7$.

The IRC-unsafe multiplicity sets the scale of performance for observables that are not IRC safe.
For the $N$-subjettiness ratios, the regression performance depends on whether the observable is IRC safe or only Sudakov safe~\cite{Larkoski:2013paa,Larkoski:2015lea}.
The ratio $\tau_{21}$ is only IRC safe for regions of phase space with two prongs or more (i.e.\ the $W$ and top samples), and $\tau_{32}$ is only IRC safe for three prongs or more (i.e.\ just the top sample).
In cases where the $N$-subjettiness ratio is IRC safe, the regression performs similarly to the other IRC-safe observables, whereas for the cases where the $N$-subjettiness ratio is only Sudakov safe, the regression performance is poor (even worse than for multiplicity).
It is satisfying to see the expected behavior between the safety of the observable and the quality of the regression with \Bs.

\begin{figure}[t]
\centering
\subfloat[]{\includegraphics[scale=.535]{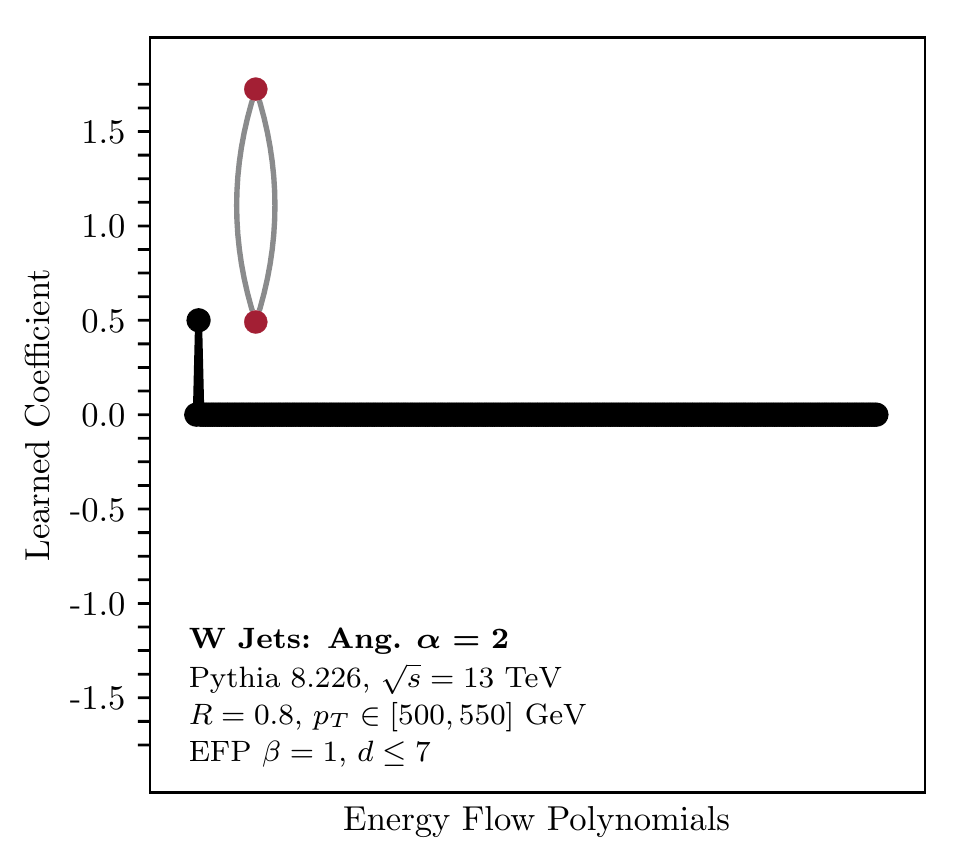}}
\subfloat[]{\includegraphics[scale=.535]{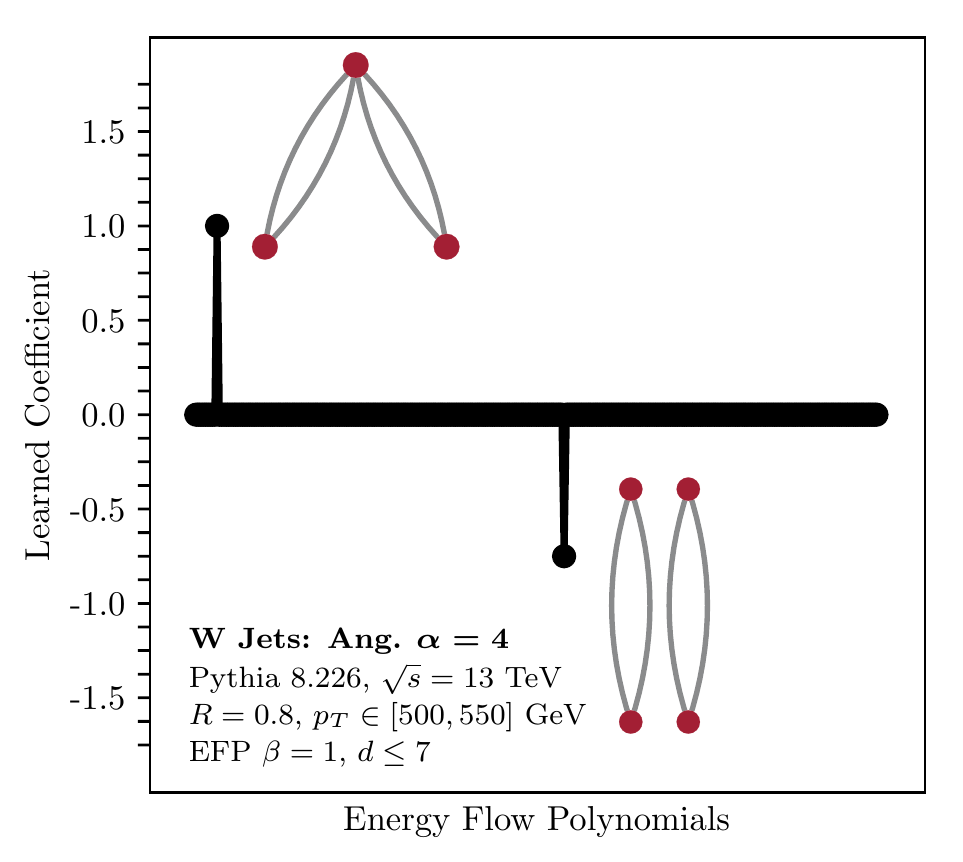}}
\subfloat[]{\includegraphics[scale=.535]{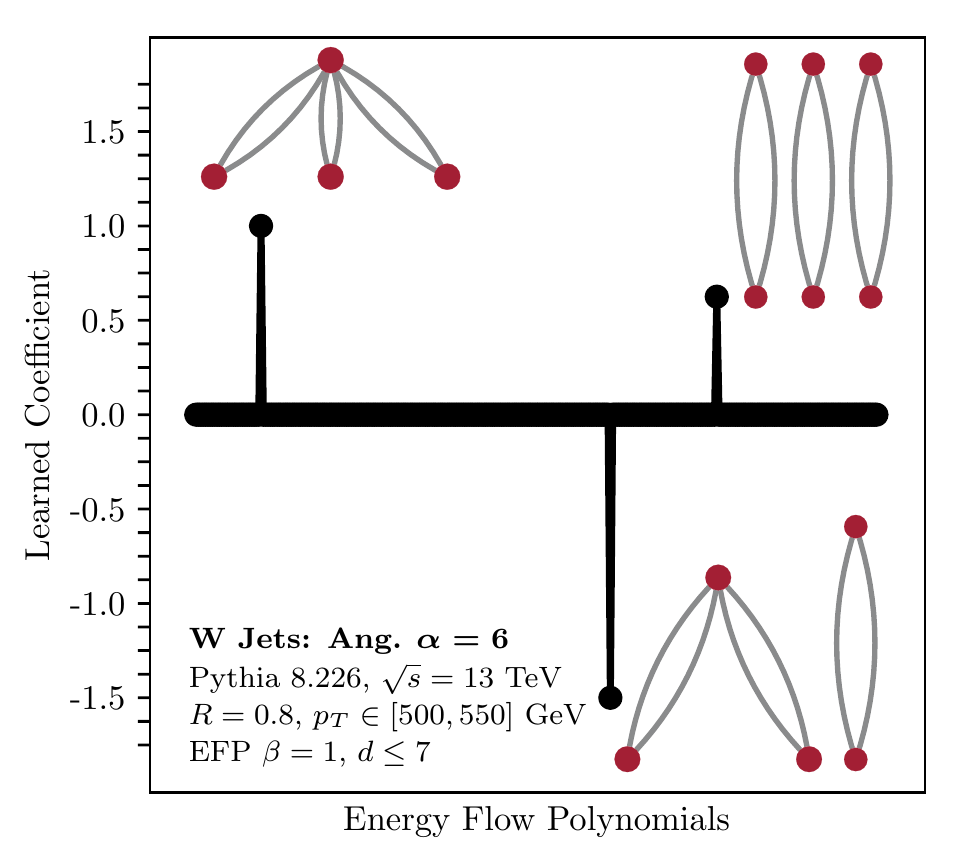}}
\caption{The linear combinations of \Bs learned by linear regression for even-$\alpha$ angularities with the $W$ jet samples.   Shown are (a) $\alpha = 2$, (b) $\alpha = 4$, and (c) $\alpha = 6$.  All but the highlighted \B coefficients are learned to be near zero. The \Bs corresponding to those non-zero coefficients are illustrated directly on the figure. The learned linear coefficients are exactly those predicted analytically in \Eqss{eq:lam2}{eq:lam4}{eq:lam6}. The same behavior is found with the QCD and top jet samples.}
\label{fig:linspec}
\end{figure}

As a final cross check of the regression, we can use the linear model in \Eq{eq:linmod} to confirm some of the analytic results of \Sec{sec:jetobs}.
Specifically, we perform a linear regression with the target observable being the even-$\alpha$ angularities with respect to the $p_T$-weighted centroid axis.
These were shown to be non-trivial linear combinations of \Bs in \Sec{sec:angularities}.
Regressing onto $\lambda^{(2)},\,\lambda^{(4)},$ and $\lambda^{(6)}$, the linear model learned the observables with effectively 100\% accuracy and the learned linear combination was exactly that predicted by \Eqss{eq:lam2}{eq:lam4}{eq:lam6}, up to a precision of $10^{-6}$.
\Fig{fig:linspec} shows the learned linear combinations of \Bs for the $W$ jet sample.

\section{Linear jet tagging}
\label{sec:linclass}

We now apply the energy flow basis to three representative jet tagging problems---light-quark/gluon classification, $W$ tagging, and top tagging---providing a broad set of contexts in which to study the \Bs.
Since the energy flow basis is linear, we can (in principle) access the optimal IRC-safe observable for jet tagging by training a linear classifier for this problem.
As mentioned in \Sec{sec:regression}, one benefit of linear models, in addition to their inherent simplicity, is that they are typically convex problems which can be solved exactly or with gradient descent to a global minimum.
See \Ref{bishop2006pattern} for a review of linear models for classification.

A (binary) linear classifier learns a vector ${\bf w}^*$ that defines a hyperplane orthogonal to the vector. 
A bias term, which can be related to the distance of this hyperplane from the origin, sets the location of the decision boundary, which is the hyperplane translated away from the origin. 
The decision function for a particular point in the input space is the normal distance to the decision boundary. 
In contrast with regression, where the target variable is usually continuous, classification predictions are classes, typically 0 or 1 for a binary classifier. 

Different methods of determining the vector ${\bf w}^*$---such as logistic regression, support vector machines, or linear discriminant analysis---may learn different linear classifiers since the methods optimize different loss functions. 
For our linear classifier, we use Fisher's linear discriminant~\cite{fisher1936use} provided by the {\tt LinearDiscriminantAnalysis} class of the {\tt scikit-learn} python module~\cite{scikit-learn}.  
The choice of logistic regression was also explored, and jet tagging performance was found to be insensitive to which type of linear classifier was used.

The details of the event generation and \B computation are the same as in \Sec{sec:eventgen}. 
To avoid a proliferation of plots, we present only the case of $W$ tagging in the text and refer to \App{app:moretagging} for the corresponding results for quark/gluon discrimination and top tagging.
Qualitatively similar results are obtained on all three tagging problems, with the conclusion that linear classification with \Bs yields comparable classification performance to other powerful machine learning techniques.
This is good evidence that the \Bs provides a suitable linear expansion of generic IRC-safe information relevant for practical jet substructure applications.

\subsection{Alternative jet representations}

In order to benchmark the \Bs, we compare them to two alternative jet tagging paradigms:
\begin{itemize}
\item The \textbf{jet images} approach~\cite{Cogan:2014oua} treats calorimeter deposits as pixels and the jet as an image, often using convolutional neural networks to determine a classifier. 
Jet images have been applied successfully to the same tagging problems considered here:  quark/gluon discrimination~\cite{Komiske:2016rsd}, $W$ tagging~\cite{deOliveira:2015xxd}, and top tagging~\cite{Kasieczka:2017nvn,Baldi:2016fql}.
\item The \textbf{$\boldsymbol{N}$-subjettiness basis} was introduced for $W$ tagging in \Ref{Datta:2017rhs} and later applied to tagging non-QCD jets~\cite{Aguilar-Saavedra:2017rzt}.
We use the same choice of $N$-subjettiness basis elements as \Ref{Datta:2017rhs}, namely:
\begin{equation}\label{eq:nsubsspan}
\{\tau_1^{(1/2)}, \tau_1^{(1)}, \tau_1^{(2)}, \tau_2^{(1/2)}, \tau_2^{(1)}, \tau_2^{(2)}, \cdots, \tau_{N-2}^{(1/2)}, \tau_{N-2}^{(1)}, \tau_{N-2}^{(2)}, \tau_{N-1}^{(1)}, \tau_{N-1}^{(2)}\},
\end{equation}
with $3N - 4$ elements needed to probe $N$-body phase space.  These are then used as inputs to a DNN.\end{itemize}
Both of these learning paradigms are expected to perform well, and we will see below that this is the case.
As a strawman, we also consider linear classification with the $N$-subjettiness basis elements in \Eq{eq:nsubsspan}, which is not expected to yield good performance.
For completeness, we also perform DNN classification with the energy flow basis.

We now summarize the technical details of these alternative jet tagging approaches.
For jet images, we create $33\times33$ jet images spanning $2R\times 2R$ in the rapidity-azimuth plane.
Motivated by \Ref{Komiske:2016rsd}, both single-channel ``grayscale'' jet images of the $p_T$ per pixel and two-channel ``color'' jet images consisting of the $p_T$ channel and particle multiplicity per pixel were used.
The $p_T$-channel of the jet image was normalized such that the sum of the pixels was one. 
Standardization was used to ensure that each pixel had zero mean and unit standard deviation by subtracting the training set mean and dividing by the training set standard deviation of each pixel in each channel.
A jet image CNN architecture similar to that used in \Ref{Komiske:2016rsd} was employed: three 36-filter convolutional layers with filter sizes of $8\times 8$, $4\times 4$, and $4\times 4$, respectively, followed by a 128-unit dense layer and a 2-unit softmaxed output. 
A rectified linear unit (ReLU) activation~\cite{nair2010rectified} was applied to the output of each internal layer.
Maxpooling of size $2\times2$ was performed after each convolutional layer with a stride length of 2.
The dropout rate was taken to be 0.1 for all layers.
He-uniform initialization~\cite{heuniform} was used to initialize the model weights. 

For the DNN (both for the $N$-subjettiness basis and for the \Bs), we use an architecture consisting of three dense layers of 100 units each connected to a 2-unit softmax output layer, with ReLU activation functions applied to the output of each internal layer.
For the training of all networks, 300k samples were used for training, 50k for validation, and 50k for testing.
Networks were trained using the Adam algorithm~\cite{adam} using categorical cross-entropy as a loss function with a learning rate of $10^{-3}$ and a batch size of 100 over a maximum of 50 epochs. 
Early stopping was employed, monitoring the validation loss, with a patience parameter of 5.
The python deep learning library {\tt Keras}~\cite{keras} with the {\tt Theano} backend~\cite{bergstra2010theano} was used to instantiate and train all neural networks.
Training of the CNNs was performed on Microsoft Azure using NVIDIA Tesla K80 GPUs and the NVIDIA CUDA framework.
Neural network performance was  checked to be mildly insensitive to these parameter choices, but these parameter choices were not tuned for optimality.
As a general rule, the neural networks used here are employed to give a sense of scale for the performance attainable with jet images and the $N$-subjettiness basis using out-of-the-box techniques; improvements in classification accuracy may be possible for these methods with additional hyperparameter tuning.

\subsection{$W$ tagging results and comparisons}
\label{sec:perf}

We present results for the $W$ tagging study here, with the other two classification problems discussed in \App{app:moretagging}.
The performance of a binary classifier is encapsulated by the background mistag rate $\varepsilon_b$ at a given signal efficiency $\varepsilon_s$.
For all of the figures below, we plot inverse receiver operator characteristic (ROC) curves, $1/\varepsilon_b$ as a function of $\varepsilon_s$, on a semi-log scale; a higher ROC curve indicates a better classifier.
The corresponding standard ROC ($\varepsilon_b$ vs.\ $\varepsilon_s$) and significance improvement ($\varepsilon_s/\sqrt{\varepsilon_b}$ vs.\ $\varepsilon_s$) curves are available in the source files of the {\tt arXiv} preprint as additional pages in the figure.

\begin{figure}[t]
\centering
\includegraphics[scale=.76]{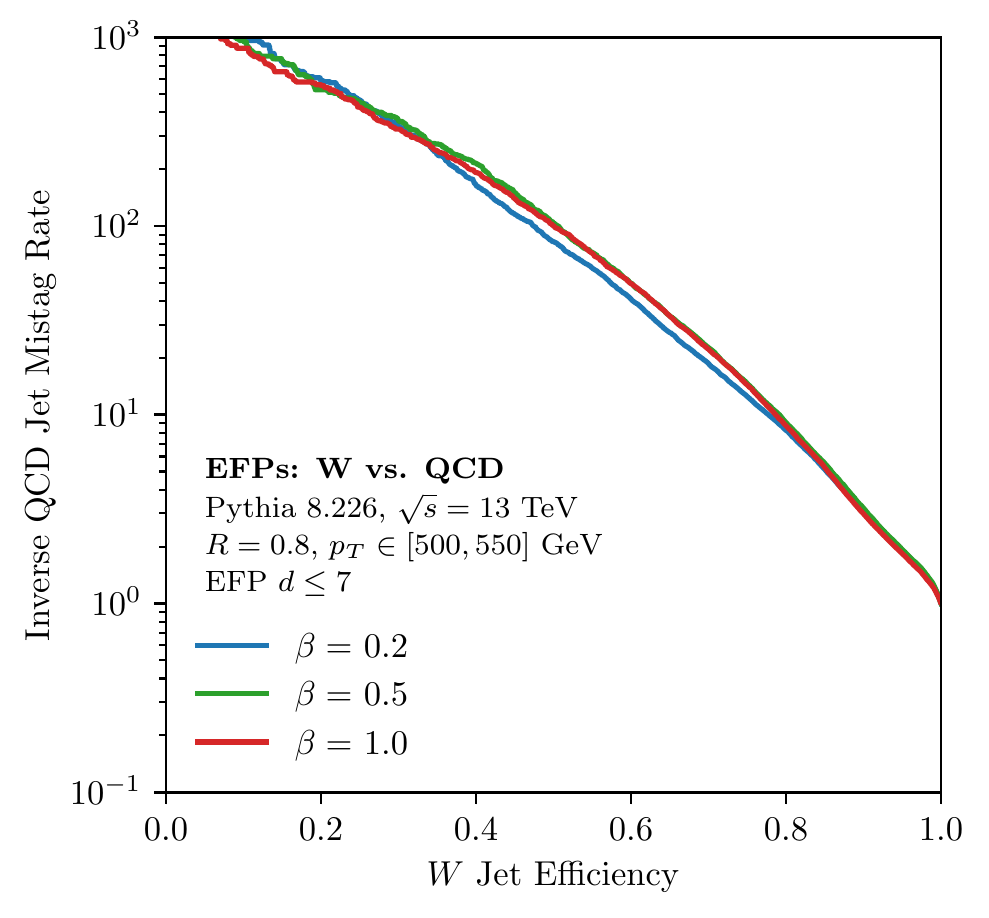}
\caption{Inverse ROC curves for linear $W$ tagging with the energy flow basis using different choices of angular exponent $\beta$ in \Eq{eq:hadronicmeasure}. Though the improvement is mild, $\beta=0.5$ shows the best overall performance. See \Fig{fig:appbetasweep} for the corresponding quark/gluon discrimination and top tagging results, where $\beta=0.5$ is also the best choice by a slight margin.}
\label{fig:Wbetasweep} 
\end{figure}

We begin by studying the performance for different choices of angular exponent $\beta$ in the default hadronic measure from \Eq{eq:hadronicmeasure}.
\Fig{fig:Wbetasweep} shows ROC curves for the choices of $\beta=0.2$, $\beta=0.5$, and $\beta=1$, using all \Bs with $d \le 7$.
The differences in performance are mild, but $\beta=0.5$ slightly improves the ROC curves for $W$ tagging, so we use $\beta = 0.5$ for the remainder of our studies.
The choice of $\beta = 0.5$ was also found to be optimal for the cases of quark/gluon and top tagging discussed in \App{app:moretagging}.

\begin{figure}[t]
\centering
\subfloat[]{\label{fig:Wefpsweep}\includegraphics[scale=.76]{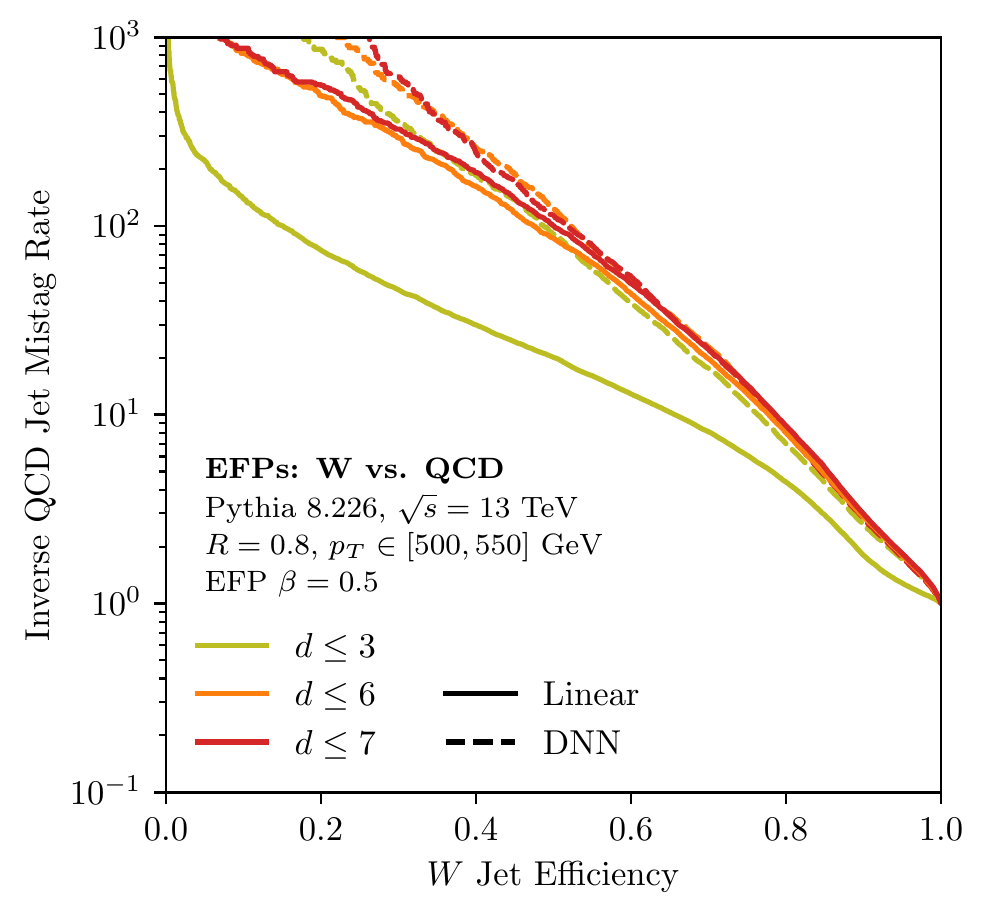}}
\subfloat[]{\label{fig:Wnsubsweep}\includegraphics[scale=.76]{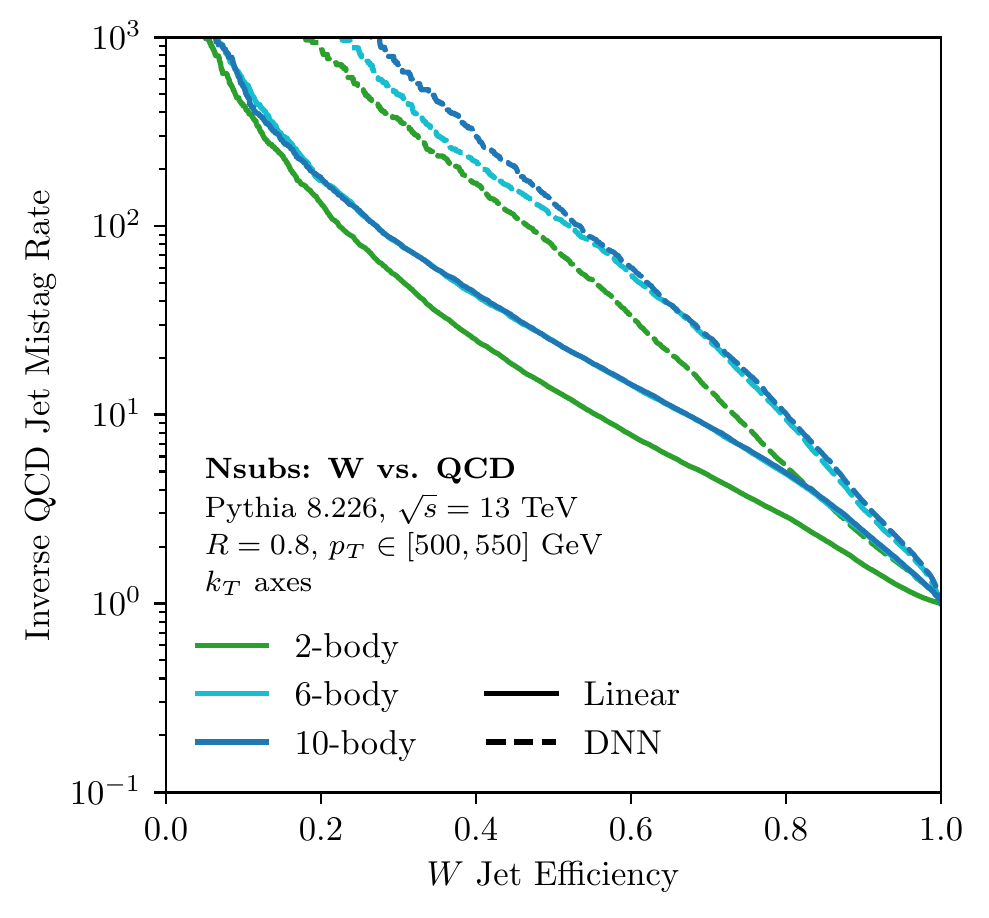}}
\caption{Inverse ROC curves for $W$ tagging with (a) the energy flow basis including degrees up to $d=7$ and (b) the $N$-subjettiness basis up to 10-body phase space information.  In both cases, we show the observables combined linearly (solid) and with a DNN (dashed).  The linear combinations of \Bs can be seen to approach the nonlinear combinations, particularly for higher signal efficiencies, while the linear combinations of the $N$-subjettiness basis can be seen to saturate well below the nonlinear combinations as the number of observables is increased. See \Fig{fig:appefpnsubsweep} for the corresponding quark/gluon discrimination and top tagging results.}
\label{fig:Wefpnsubsweep}
\end{figure}

Next, in \Fig{fig:Wefpsweep}, we test the linear spanning nature of the \Bs by comparing the ROC curves of the linear and nonlinear models trained on \Bs up to different $d$.
With linear regression, there is a large jump in performance in going from $d\le3$ (13 \Bs) to $d\le6$ (314 \Bs), and a slight increase in performance from $d\le6$ to $d\le7$ (1000 \Bs), indicating good convergence to the optimal IRC-safe observable for $W$ jet discrimination.
To avoid cluttering the plot, $d\le4$ and $d\le5$ are not shown in \Fig{fig:Wefpsweep}, but their ROC curves fall between those of $d\le3$ and $d\le6$, highlighting that the higher $d$ \Bs carry essential information for linear classification. 
By contrast, using nonlinear classification with a DNN, the \Bs performance with $d \le 3$ is already very good, since functions of the low $d$ \Bs can be combined in a nonlinear fashion to construct information contained in higher $d$ composite \Bs.
The linear and nonlinear performance is similar with the $d\le7$ \Bs for operating points of $\varepsilon_s\gtrsim0.5$, though the nonlinear DNN outperforms the linear classifier in the low signal efficiency region.
It should be noted that the linear classifier is not trained specifically for the low signal efficiency region and it may be possible that choosing a different hyperplane could boost performance there.
We leave to future work a more detailed investigation of optimizing the choice of linear classifier.

The performance of the $N$-subjettiness basis with both linear and nonlinear classifiers is shown in \Fig{fig:Wnsubsweep}.
For both linear classification and the DNN, performance appears to saturate with the 6-body (14 $\tau_N$s) phase space, with not much gained in going to 10-body (26 $\tau_N$s) phase space, except for a small increase in the low signal efficiency region for the DNN; we confirmed up to 30-body (86 $\tau_N$s) phase space that no change in ROC curves was observed compared to 10-body phase space.
As expected, there is relativity poor performance with linear classification even as the dimension of phase space is increased.
Classification with a DNN, though, shows an immense increase in performance over linear classification, as expected since the $N$-subjettiness basis is expected to nonlinearly capture all of the relevant IRC-safe kinematic information~\cite{Datta:2017rhs}.
This illustrates that nonlinear combinations of the $N$-subjettiness observables are crucial for extracting the full physics information.

The corresponding quark/gluon and top tagging plots in \Fig{fig:appefpnsubsweep} effectively tell the same story as \Fig{fig:Wefpnsubsweep}, robustly demonstrating the linear spanning nature of the \Bs used for classification across a wide variety of kinematic configurations.
As a side note, in \App{app:moretagging} there are sometimes cases where a linear combination of \Bs yields \emph{improved} performance compared to a DNN on the same inputs, particularly at medium to high signal efficiencies.
Since even a one-node DNN should theoretically be able to learn the linear combination of \Bs learned by the linear classifier, regimes where the linear classifier outperforms the DNN demonstrate the inherent difficulty of training complex multivariate models.

\begin{figure}[t]
\centering
\includegraphics[scale=.76]{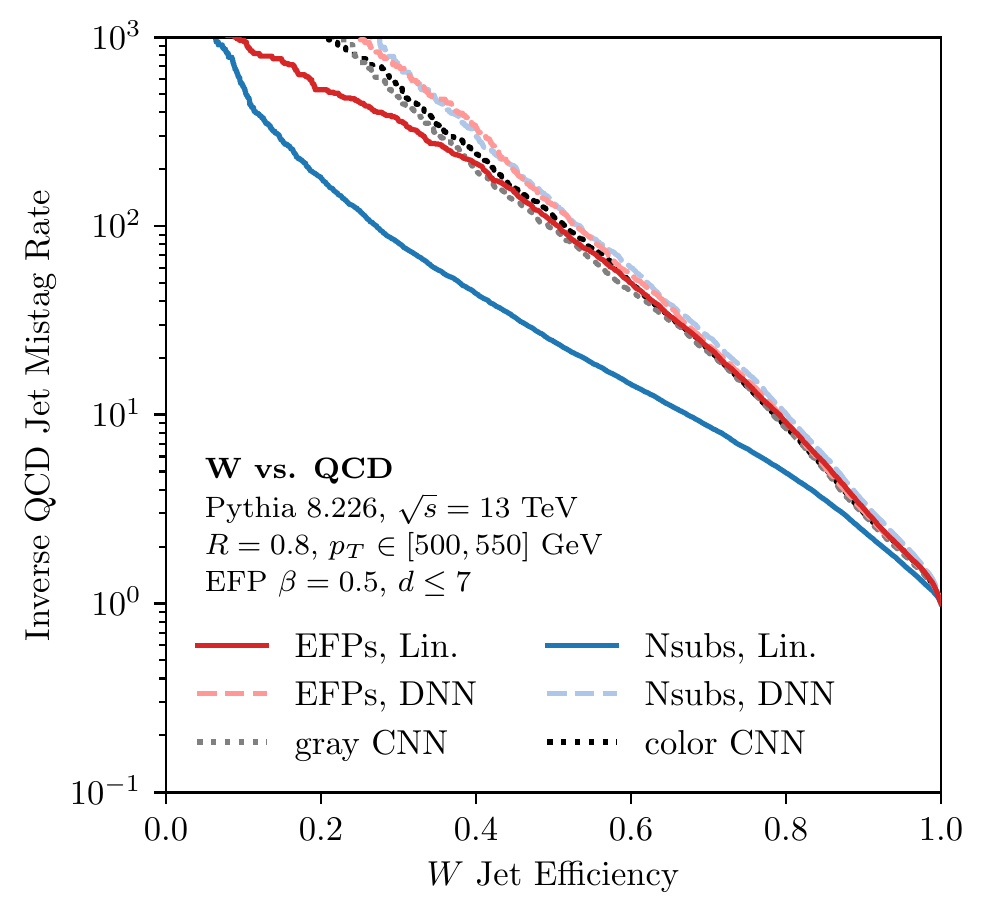}
\caption{Inverse ROC curves for $W$ tagging comparing six different methods:  linear and DNN classification with the energy flow basis up to $d\le 7$, linear and DNN classification with the $N$-subjettiness basis up to 10-body phase space, and grayscale and color jet images with CNNs. The most evident gap is between the linearly-combined $N$-subjettiness basis and the remaining curves, which achieve similar classification performance for medium and high signal efficiencies. See \Fig{fig:apptagcomp} for the corresponding quark/gluon discrimination and top tagging results.}
\label{fig:Wtagcomp}
\end{figure}

In \Fig{fig:Wtagcomp} we directly compare the \B classification power against the $N$-subjettiness basis and the 1-channel (``grayscale'') and 2-channel (``color'') CNNs. 
For operating points with $\varepsilon_s\gtrsim0.5$, all methods except the linear $N$-subjettiness classifier show essentially the same performance.
The worse performance of the linear \B classifier at low signal efficiencies is expected, since the Fisher linear discriminant is not optimized for that regime.
Overall, it is remarkable that similar classification performance can be achieved with these three very different learning paradigms, especially considering that the DNNs and grayscale CNN implicitly, and the color CNN explicitly, have access to non-IRC-safe information (including Sudakov-safe combinations of the IRC-safe inputs).
This agreement gives evidence that the tagging techniques have approached a global bound on the maximum possible discrimination power achievable, at least in the context of parton shower simulations.

Once again, the analogous quark/gluon and top tagging plots, shown in \Fig{fig:apptagcomp}, show very similar behavior to the $W$ tagging case in \Fig{fig:Wtagcomp}.
Linear classification with the \Bs performs similarly to the DNNs and CNNs, tending to slightly outperform at high signal efficiencies and underperform at low signal efficiencies.
Ultimately, the choice of tagging method comes down to a trade off between the simplicity of the inputs and the simplicity of the training method, with the \Bs presently requiring more inputs than the $N$-subjettiness basis but with the benefit of using a linear model.
In the future, we plan to study ways of reducing the size of the \B basis by exploiting linear redundancies among the \Bs and using powerful linear methods to automatically select the most important observables for a given task.

\subsection{Opening the energy flow box}
\label{sec:openbox}

\begin{figure}
\centering
\subfloat[]{\label{fig:Wefpnsweep:a}\includegraphics[scale=.76]{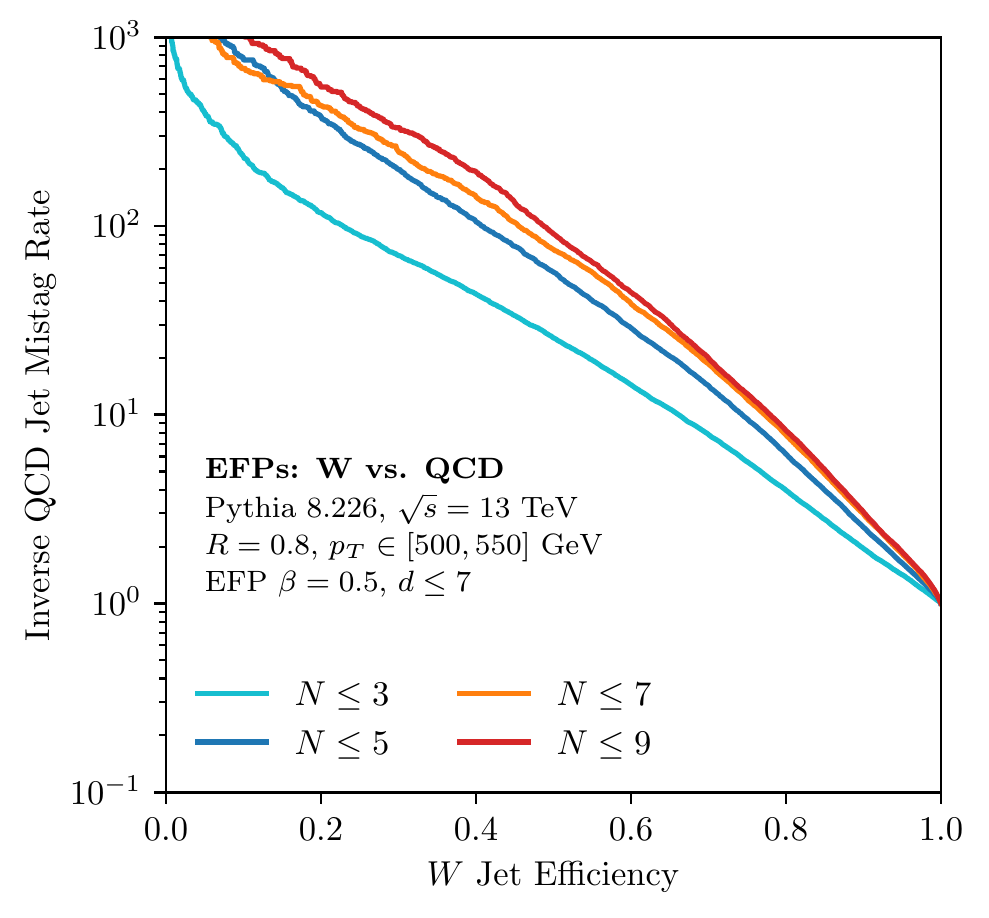}}
\subfloat[]{\label{fig:Wefpnsweep:b}\includegraphics[scale=.76]{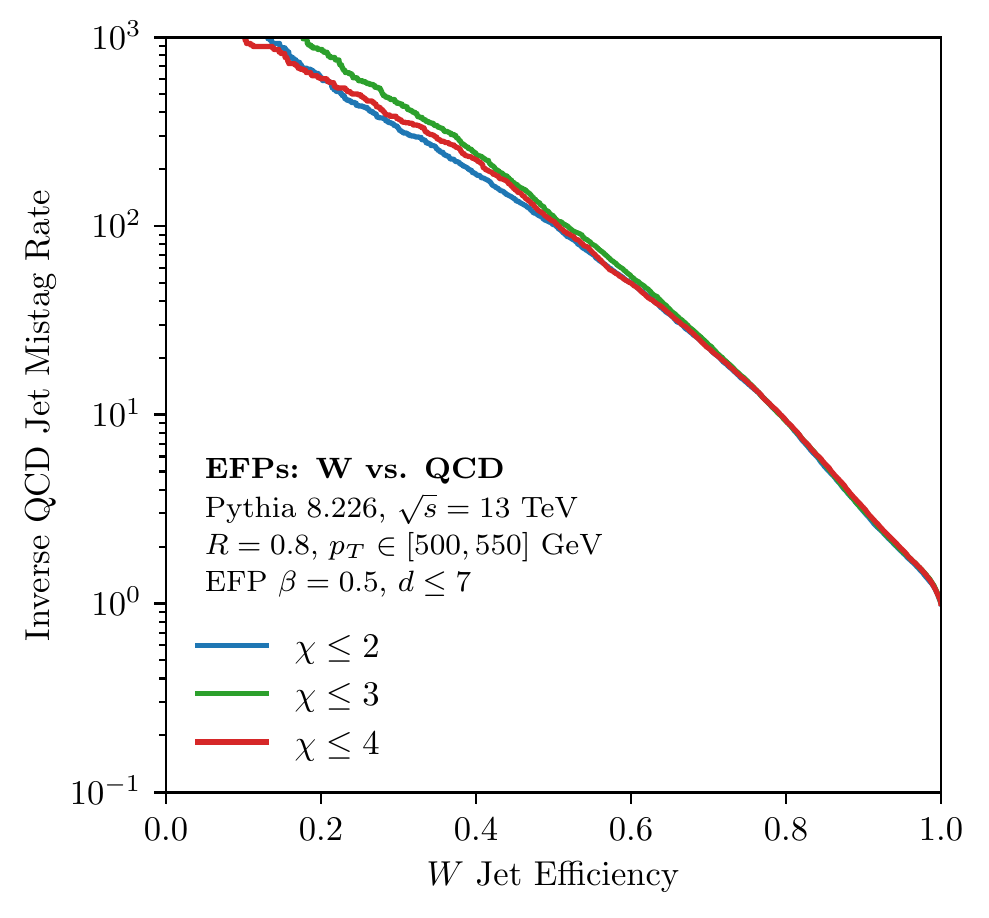}}
\caption{Inverse ROC curves for linear $W$ tagging with the energy flow basis with $d\le 7$, sweeping over (a) which $N$-point correlators and (b) observables of which VE computational complexity $\mathcal O(M^\chi)$ are included in the linear fit.  It is clear that important information is contained in the higher $N$-particle correlators, which can be included because the algorithm in \Sec{sec:complexity} evades the naive $\mathcal O(M^N)$ scaling.  Interestingly, the discrimination power appears to be almost saturated by the graphs computable in $\mathcal O(M^2)$.  See \Fig{fig:appnchisweep} for the corresponding quark/gluon discrimination and top tagging results.}
\label{fig:efpnsweep}
\end{figure}

As argued in \Eq{eq:linmodefp}, one of the main advantages of linear methods with the energy flow basis is that one can attempt to ``open the box'' and directly explore what features have been learned.
We leave to future work a full exploration of this possibility, but here we attempt to probe which topological structures within the \B basis carry the discrimination power for the different tagging problems.
Since we have shown that the \Bs with $d\le 7$ have sufficient discrimination power to qualitatively match the performance of alternative tagging methods, we will restrict to this set of observables.

In \Fig{fig:Wefpnsweep:a}, we vary the maximum number of vertices in the \B graphs, where the maximum $N$ is $14$ for $d \le 7$, finding that the performance roughly saturates at $N=9$, highlighting the importance of higher $N$ \Bs.
The algorithmic advances described in \Sec{sec:complexity} allow for the efficient computation of these higher $N$ \Bs, which have complexities as intractable as $\mathcal O(M^9)$ with the naive algorithm.
Additionally, note that nearly every \B (all except those corresponding to complete graphs) has a non-vanishing angular weighting function, which is a new feature compared to the ECFs and ECFGs (see \Sec{subsec:goingbeyond}).
In \Fig{fig:Wefpnsweep:b}, we vary the maximum computational complexity $\chi$ of the \B graphs, where the maximum $\chi$ is $4$ for $d \le 7$.
Remarkably, the full performance of linear classification with the $d\le7$ \Bs can be obtained with merely those observables calculable in $\mathcal O(M^2)$ with VE.
Thus, fortuitously for the purposes of jet tagging, it seems that restricting to the most efficiently computable \Bs (in the VE paradigm) is sufficient for extracting the near-optimal IRC-safe observable for jet discrimination.
Similar results hold for quark/gluon discrimination and top tagging, shown in \Fig{fig:appnchisweep}.

\section{Conclusions}
\label{sec:conclusion}

In this paper, we have introduced the \Bs, which linearly span the space of IRC-safe observables.\footnote{In the course of this research, we encountered a more descriptive acronym than ``\Bs'' albeit with an unintended biblical reference:  Polynomials of Energies and Angles Result in a Linear Spanning Basis for Energy Flow Observables Relevant for Extracting Substructure With Improved Nuance and Efficiency.}
The core argument, presented in \Sec{sec:basis}, is that one can systematically expand an arbitrary IRC-safe observable in terms of energies and angles and read off the unique resulting analytic structures.
This expansion yields a new way to understand the importance of $C$-correlators \cite{Tkachov:1995kk,Sveshnikov:1995vi,Cherzor:1997ak,Tkachov:1999py} for IRC safety, and it enables a powerful graph-theoretic representation of the various angular structures.
The multigraph correspondence makes manifest a more efficient algorithm than the naive $\mathcal O(M^N)$ one for computing \Bs, overcoming a primary obstacle to exploring higher-$N$ multiparticle correlators for jet substructure.

To demonstrate the power of the energy flow basis, we performed a variety of representative regression and classification tasks for jet substructure.
Crucially, linear methods were sufficient to achieve good performance with the \Bs.
As a not-quite apples-to-apples comparison in three representative jet tagging applications, linear classification with 1000 \Bs achieved comparable performance to a CNN acting on a jet image with $33 \times 33 = 1089$ pixels.
Because of the wide variety of linear learning methods available~\cite{bishop2006pattern}, we expect that the \Bs will be a useful starting point to explore more  applications in jet substructure and potentially elsewhere in collider physics.

There are many possible refinements and extensions to the energy flow basis.
In this paper, we truncated the \Bs at a fixed maximum degree $d$; alternatively, one could truncate the prime \Bs at a fixed $d$ and compute all composite \Bs up to a specified cutoff.
Since the EFPs yield an overcomplete basis, it could be valuable to cull the list of required multigraphs.
A similar problem of overcompleteness was solved for kinematic polynomial rings in \Ref{Henning:2017fpj}, and that strategy may be relevant for EFPs with a suitable choice of measure.
In the other direction, it may be valuable to make the energy flow basis even more redundant by including EFPs with multiple measures.
With a vastly overcomplete basis, one could use techniques like lasso regression~\cite{tibshirani1996regression} to zero out unnecessary terms.
While we have restricted our attention to IRC-safe observables, it would be straightforward to relax the restriction to just infrared safety.
In particular, the set of IR-safe (but C-unsafe) functions in \Eq{eq:Cexample} can be expanded into multigraphs that have an extra integer decoration on each vertex to indicate the energy scaling.
Finally, the EFPs are based on an expansion in pairwise angles, but one could explore alternative angular expansions in terms of single particle directions or multiparticle factors.

To gain some perspective, we find it useful to discuss the \Bs in the broader context of machine learning for jet substructure.
Over the past few years, there has been a surge of interest in using powerful tools from machine learning to learn useful observables from low-level or high-level representations of a jet~\cite{Cogan:2014oua,deOliveira:2015xxd,Komiske:2016rsd,Almeida:2015jua,Baldi:2016fql,Kasieczka:2017nvn,Pearkes:2017hku,Butter:2017cot,Aguilar-Saavedra:2017rzt,Guest:2016iqz,Louppe:2017ipp,Datta:2017rhs, Baldi:2014kfa,Baldi:2014pta,Gallicchio:2010dq,Gallicchio:2012ez,Gallicchio:2011xq}.
The power of these machine learning methods is formidable, and techniques like neural networks and boosted decision trees have shifted the focus away from single- or few-variable jet substructure taggers to multivariate methods.
On the other hand, multivariate methods can sometimes obscure the specific physics information that the model learns, leading to recent efforts to ``open the box'' of machine learning tools~\cite{deOliveira:2015xxd,Butter:2017cot,Chang:2017kvc,Komiske:2017ubm,Metodiev:2018ftz}.
Even with an open box, though, theoretical calculations of multivariate distributions are impractical (if not impossible).
Furthermore, training multivariate models is often difficult, requiring large datasets, hyperparameter tuning, and preprocessing of the data.

The EFPs represent both a continuation of and a break from these machine learning trends.
The EFPs continue the trend from multivariate to hypervariate representations for jet information, with $\mathcal{O}(100)$ elements needed for effective regression and classification.  
On the other hand, the linear-spanning nature of the EFPs make it feasible to move away from ``black box'' nonlinear algorithms and return to simpler linear methods (explored previously for jet substructure in e.g.\ \cite{Thaler:2011gf,Cogan:2014oua}) without loss of generality.
Armed with the energy flow basis, there is a suite of powerful tools and ideas from linear regression and classification which can now be fully utilized for jet substructure applications, with simpler training processes compared to DNNs and stronger guarantees of optimal training convergence. 
Multivariate methods would ideally be trained directly on data to avoid relying on imperfect simulations, as discussed in \Ref{Komiske:2018oaa}.
The energy flow basis may be compelling for recent data-driven learning approaches~\cite{Komiske:2018oaa,Dery:2017fap, Metodiev:2017vrx} due to its completeness, the simplicity of linear learning algorithms, and a potentially lessened requirement on the size of training samples.

As with any jet observable, the impact of non-perturbative effects on the \Bs is important to understand.
Even with IRC safety, hadronization modifies the distributions predicted by pQCD and therefore complicates first-principles calculations. 
It would be interesting to see if the shape function formalism~\cite{Korchemsky:1999kt,Korchemsky:2000kp} could be used to predict the impact of non-perturbative contributions to \B distributions.
Alternatively, one standard tool that is used to mitigate non-perturbative effects is jet grooming~\cite{Butterworth:2008iy,Ellis:2009su,Ellis:2009me,Krohn:2009th,Dasgupta:2013ihk,Larkoski:2014wba}, which also simplifies first-principles calculations and allows for ``quark'' and ``gluon'' jets to be theoretically well-defined~\cite{Frye:2016aiz}.
We leave a detailed study of the effects of non-perturbative contributions and jet grooming on \Bs to future work.

Eventually, one hopes that the EFPs will be amenable to precision theoretical calculations of jet substructure (see e.g.\ \Refs{Feige:2012vc, Dasgupta:2013ihk,Dasgupta:2013via, Larkoski:2014tva, Procura:2014cba,Larkoski:2015kga,Dasgupta:2015lxh,Frye:2016aiz,Dasgupta:2016ktv,Marzani:2017mva,Larkoski:2017cqq}).
This is by no means obvious, since generic EFPs have different power-counting structures from the ECFs~\cite{Larkoski:2013eya} or ECFGs~\cite{Moult:2016cvt}.
That said, phrasing jet substructure entirely in the language of energy flow observables and energy correlations may provide interesting new theoretical avenues to probe QCD, realizing the $C$-correlator vision of \Refs{Tkachov:1995kk,Sveshnikov:1995vi,Cherzor:1997ak,Tkachov:1999py}.
Most IRC-safe jet observables rely on particle-level definitions and calculations, but there has been theoretical interest in directly analyzing the correlations of energy flow in specific angular directions~\cite{basham1978energy,basham1979energy,Belitsky:2001ij}, particularly in the context of conformal field theory~\cite{Hofman:2008ar,Engelund:2012re,Zhiboedov:2013opa,Belitsky:2013xxa,Belitsky:2013bja}.
The energy flow basis is a step towards connecting the particle-level and energy-correlation pictures, and one could even imagine that the energy flow logic could be applied directly at the path integral level.
Ultimately, the structure of the EFPs is a direct consequence of IRC safety, resulting in a practical tool for jet substructure at colliders as well as a new way of thinking about the space of observables more generally.

\acknowledgments
 
We thank J.J. Carrasco, Yang-Ting Chien, Timothy Cohen, Kyle Cranmer, Matthew Dolan, Fr\'ed\'eric Dreyer, Marat Freytsis, Brian Henning, Gregor Kasieczka, Andrew Larkoski, Alex Lombardi, Ian Moult, Benjamin Nachman, Lina Necib, Miruna Oprescu, Bryan Ostdiek, Gilad Perez, Matthew Schwartz, and Jon Walsh for illuminating discussions.
This work was supported by the Office of Nuclear Physics of the U.S. Department of Energy (DOE) under grant DE-SC-0011090 and the DOE Office of High Energy Physics under grant DE-SC-0012567.
Cloud computing resources were provided through a Microsoft Azure for Research award.

\appendix

\section{Energy flow and the stress-energy tensor}
\label{sec:stressenergy}

\begin{figure}[t]
\centering
\includegraphics[scale=0.8]{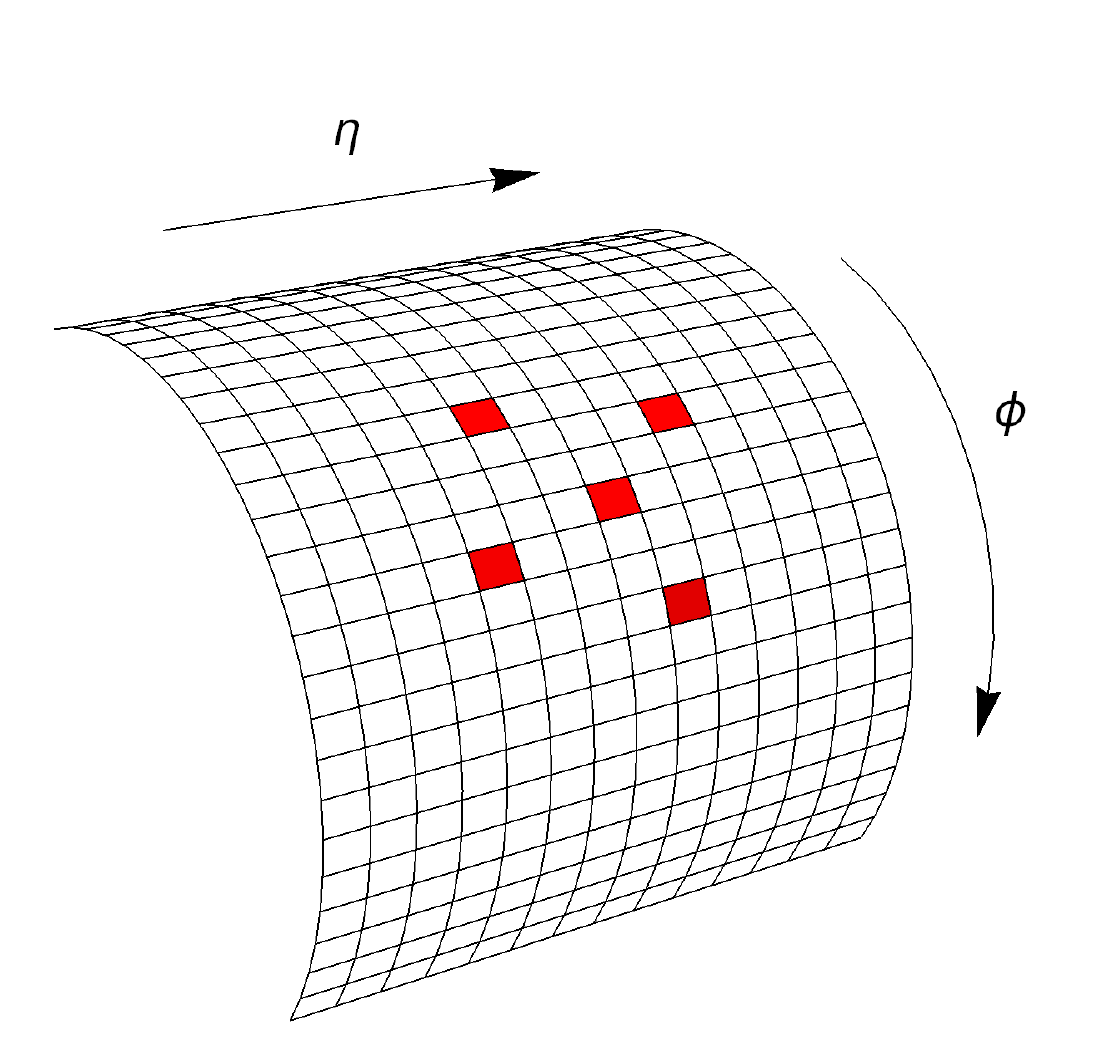}
\caption{An example calorimeter cell configuration to measure a 5-point energy correlator. The red regions indicate the five calorimeter cells chosen to measure the energy infinitely far from the interaction. For each event, the values of the five energy deposits are multiplied together to obtain the value of the observable in \Eq{eq:npointcorr}.}
\label{fig:5correxample}
\end{figure}

In this appendix, we review the connection between the energy flow of an event, as described by the stress-energy tensor, to multiparticle energy correlators~\cite{Tkachov:1995kk,Sveshnikov:1995vi,Cherzor:1997ak,Tkachov:1999py}.

Consider an idealized hadronic calorimeter cell at position $\hat n$ in pseudorapidity-azimuth $(\eta,\phi)$-space, spanning a patch of size $d\eta\,d\phi$. The \emph{energy flow operator} $\mathcal E_T(\hat n)$ corresponding to the total transverse momentum density flowing into the calorimeter cell can be written in terms of the stress-energy tensor $T_{\mu\nu}$~\cite{Sveshnikov:1995vi,Korchemsky:1997sy,Lee:2006nr,Bauer:2008dt,Mateu:2012nk} as:
\begin{equation}\label{eq:energyflow}
\mathcal E_T(\hat n) = \frac{1}{\cosh^3  \eta} \lim_{R\to\infty} R^2\int_0^\infty dt\, \hat n_i\, T^{0i}(t, R\hat n),
\end{equation}
with its action on a state $\ket X$ of $M$ massless particles given by:
\begin{equation}\label{eq:transcal}
\mathcal E_T(\hat n) \ket X = \sum_{i\in X} p_{T,i} \delta(\eta -  \eta_i)\delta(\phi - \phi_i) \ket X.
\end{equation}

Next, consider $N$ calorimeter cells at positions $(\hat n_1, \cdots, \hat n_N)$. 
An illustration of an example calorimeter cell configuration is shown in \Fig{fig:5correxample}. 
For an event $X$, multiply together the measured energy deposits in each of these $N$ cells. 
The corresponding observable is then the energy $N$-point correlator as defined in \Refs{basham1978energy,basham1979energy}:
\begin{equation}\label{eq:npointcorr}
\mathcal E_T(\hat n_1) \cdots \mathcal E_T(\hat n_N)\ket X = \sum_{i_1 \in X} \cdots \sum_{i_N \in X} \left[\prod_{j=1}^Np_{T,i_j}\delta^2(\hat n_j-\hat p_{i_j})\right]\ket X,
\end{equation}
where $\hat n_a=( \eta_a,\phi_a)$ for the calorimeters cells and $\hat p_a=( \eta_a,\phi_a)$ for the particles in the event.

We can define a new set of observables in terms of the $N$-point correlators in \Eq{eq:npointcorr}. 
Consider averaging \Eq{eq:npointcorr} over all calorimeter cells with an arbitrary angular weighting function $f_N(\hat n_1,\ldots,\hat n_N)$. 
The resulting observables are then of the form:
\begin{align}\label{eq:corravg}
\mathcal C_N^{f_N} \ket X &= \int d^2\hat n_1 \cdots d^2 \hat n_N \, f_N(\hat n_1,\ldots, \hat n_N)\mathcal E_T(\hat n_1) \cdots \mathcal E_T(\hat n_N) \ket X
\\& = \sum_{i_1\in X} \cdots \sum_{i_N \in X} p_{T,i_1} \cdots p_{T,i_N} f_N(\hat p_{i_1},\ldots, \hat p_{i_N})\ket X,\label{eq:ccorr}
\end{align}
namely, these observables $\mathcal C_N^{f_N}$ written in the form of \Eq{eq:ccorr} are exactly the $C$-correlators defined in \Eq{eq:genccorr}. 
Thus the averaging process of \Eq{eq:corravg} relates the particle-level $C$-correlators of \Eq{eq:genccorr} to the energy flow of the stress-energy tensor $T_{\mu\nu}$.

\section{Quark/gluon discrimination and top tagging results}
\label{app:moretagging}

\begin{figure}[t]
\centering
\subfloat[]{\includegraphics[scale=.76]{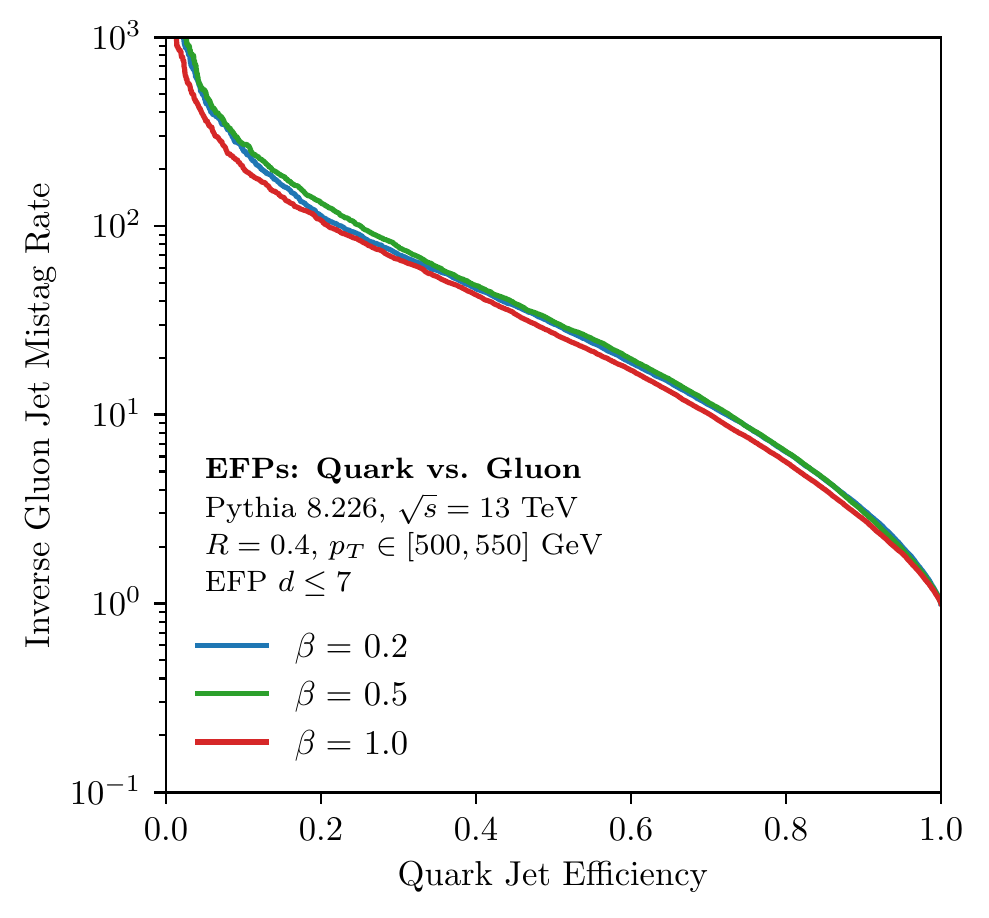}}
\subfloat[]{\includegraphics[scale=.76]{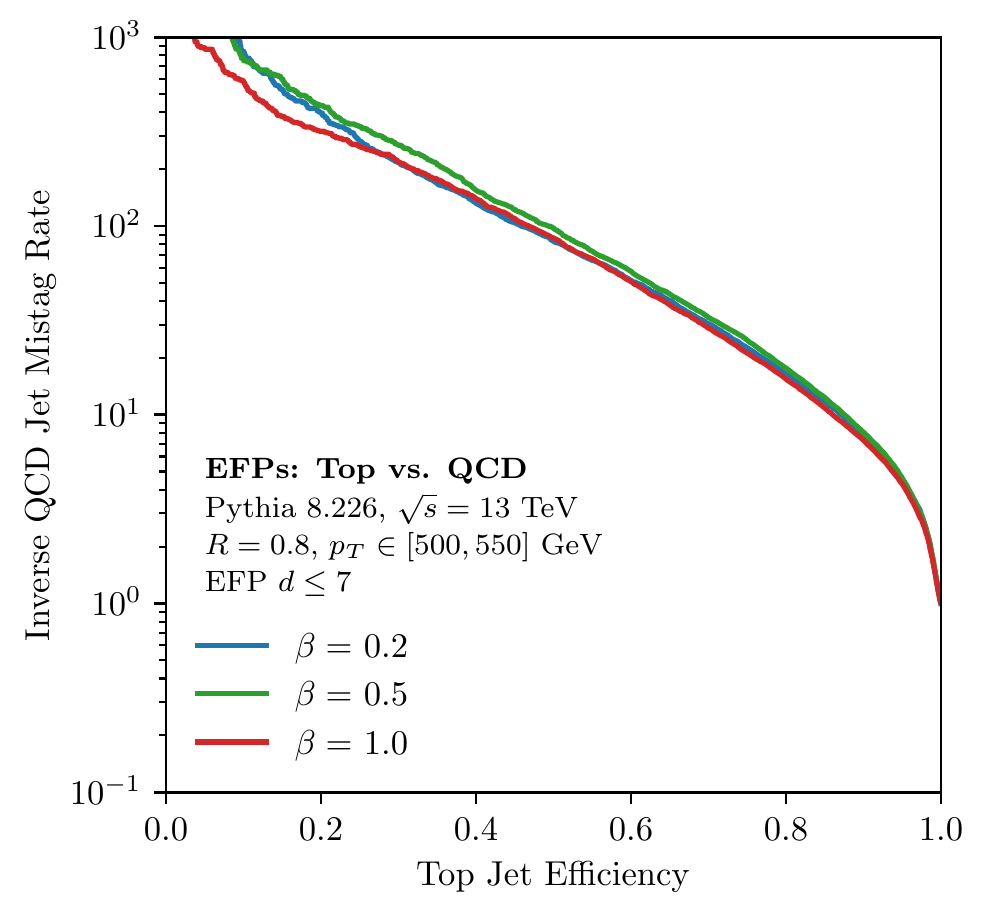}}
\caption{Same as \Fig{fig:Wbetasweep}, but for (a) quark/gluon discrimination and (b) top tagging.  Similar to the $W$ tagging case, the $\beta = 0.5$ choice has the best performance (marginally) for both tagging problems.}
\label{fig:appbetasweep}
\end{figure}

In this appendix, we supplement the $W$ tagging results of \Sec{sec:linclass} with the corresponding results for quark/gluon discrimination and top tagging.
The details of the event generation are given in \Sec{sec:eventgen}.

We compare the \B linear classification performance with $\beta = 0.2$, $\beta = 0.5$, and $\beta = 1$ in \Fig{fig:appbetasweep}.  
Consistent with the $W$ tagging case in \Fig{fig:Wbetasweep}, we find that the optimal performance is achieved with $\beta = 0.5$.  We therefore use $\beta = 0.5$ for the remainder of this study.

\begin{figure}[t]
\centering
\subfloat[]{\includegraphics[scale=.76]{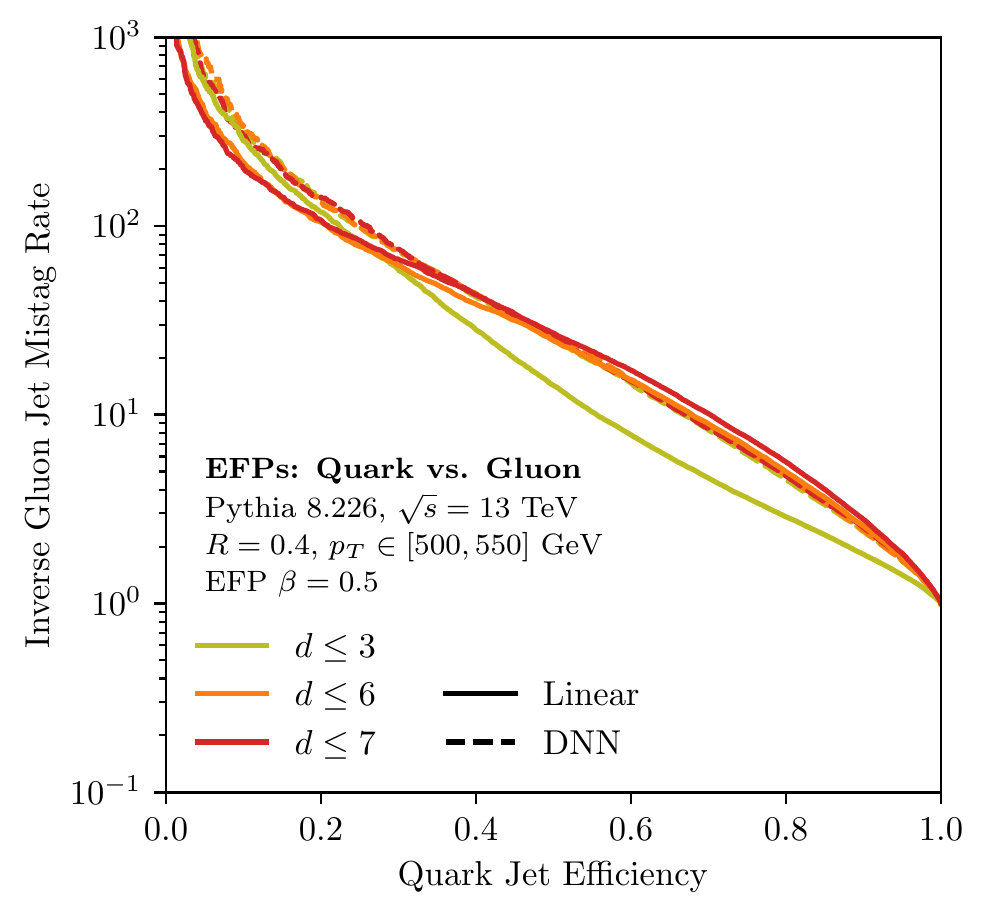}}
\subfloat[]{\includegraphics[scale=.76]{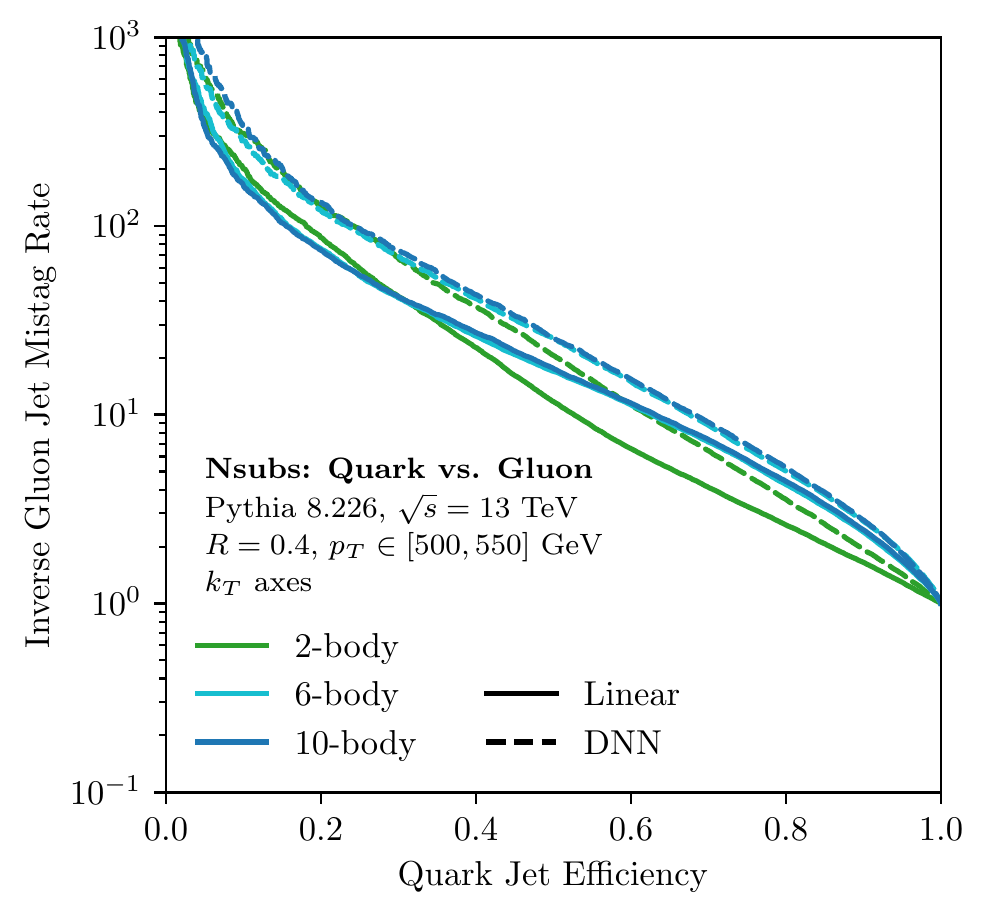}}

\subfloat[]{\includegraphics[scale=.76]{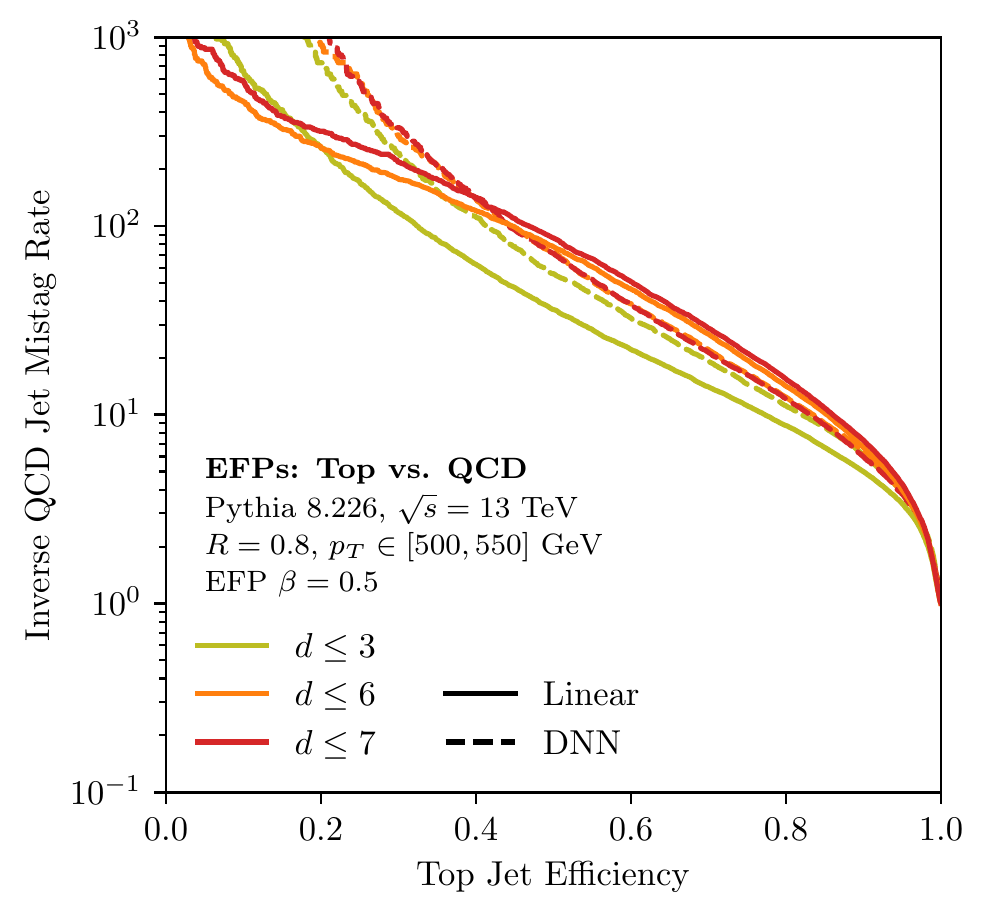}}
\subfloat[]{\includegraphics[scale=.76]{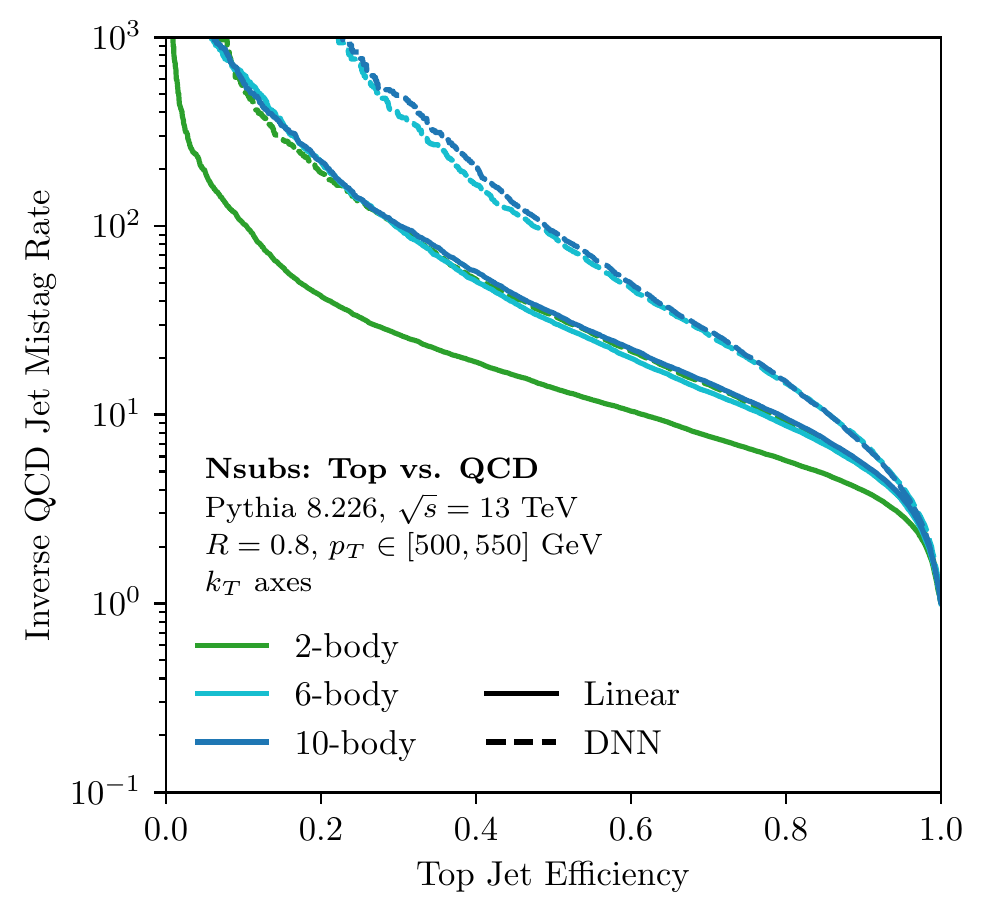}}
\caption{Same as \Fig{fig:Wefpnsubsweep}, but for quark/gluon discrimination (top) and top tagging (bottom).  As in the $W$ tagging case, the linear combinations of \Bs can be seen to approach (or even exceed) the nonlinear combinations, particularly for higher signal efficiencies.
\label{fig:appefpnsubsweep}}
\end{figure}

In \Fig{fig:appefpnsubsweep} we compare the linear and nonlinear performances of the energy flow basis and the $N$-subjettiness basis.
There is a clear gap between the linear and nonlinear $N$-subjettiness classifiers, whereas no such gap exists for the \Bs.  Interestingly, the linear classifier of \Bs tends to outperform the DNN at medium and high signal efficiencies, indicating the difficulty of training high-dimensional neural networks.
This behavior was not seen in \Fig{fig:Wefpnsubsweep}, most likely because the achievable efficiency is overall higher in the $W$ tagging case.

\begin{figure}[t]
\centering
\subfloat[]{\includegraphics[scale=.76]{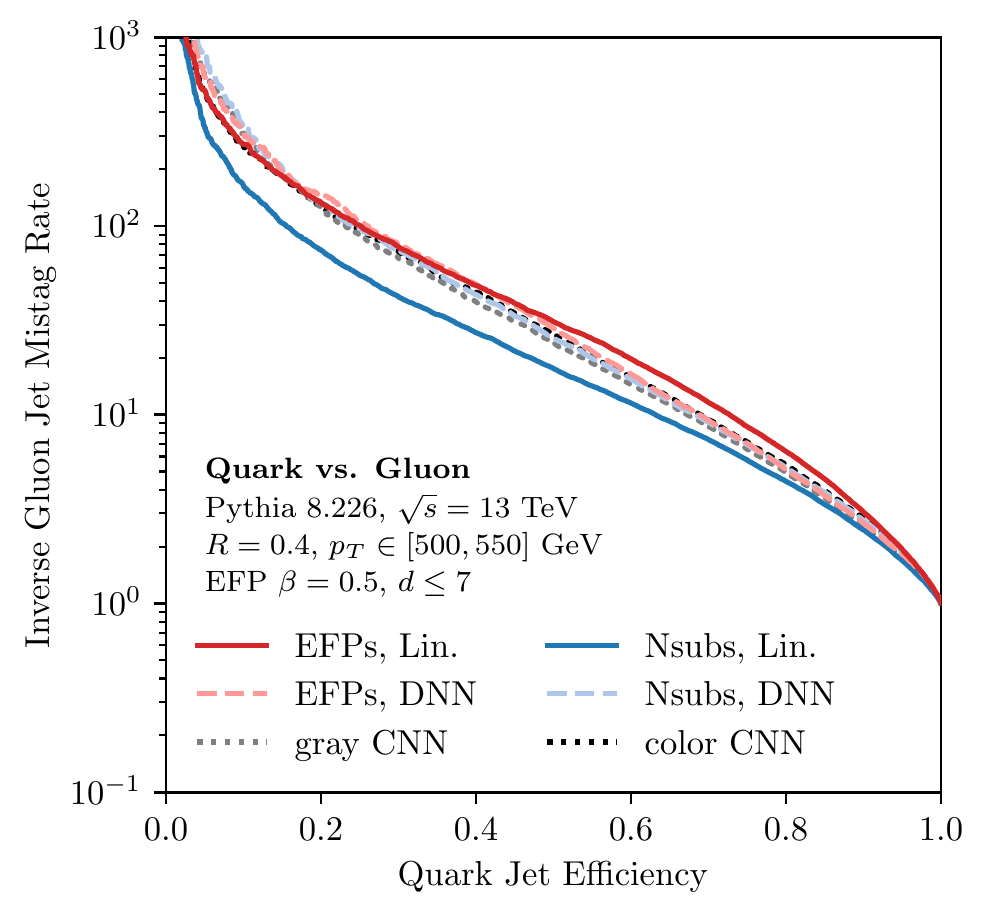}}
\subfloat[]{\includegraphics[scale=.76]{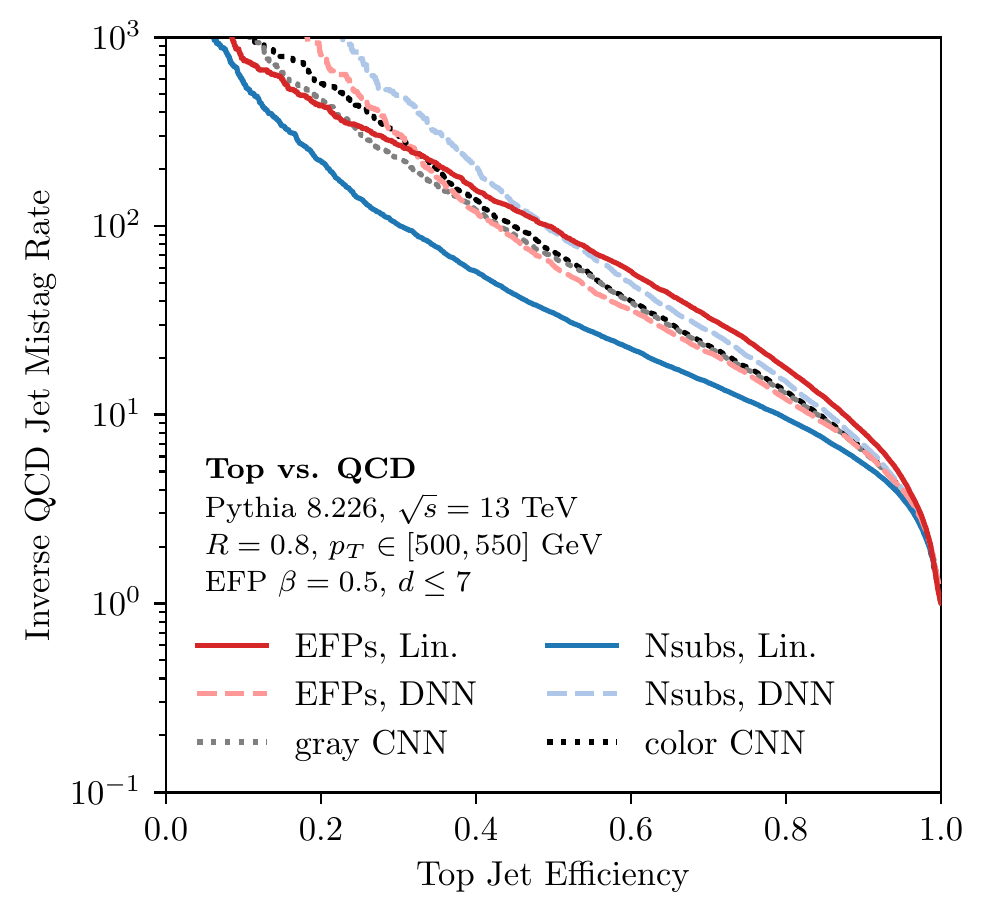}}
\caption{Same as \Fig{fig:Wtagcomp} for (a) quark/gluon discrimination and (b) top tagging.
As in the $W$ tagging case, the linear classification with \Bs can match (or even outperform) the other methods at high signal efficiencies.}
\label{fig:apptagcomp}
\end{figure}

A summary of the six tagging methods is shown in \Fig{fig:apptagcomp}, comparing linear and nonlinear combinations of the energy flow basis and $N$-subjettiness basis to grayscale and color jet images. 
As in \Fig{fig:Wtagcomp}, linear combinations of \B tend to match or outperform the other methods, especially at high signal efficiencies.

\begin{figure}[t]
\centering
\subfloat[]{\includegraphics[scale=.76]{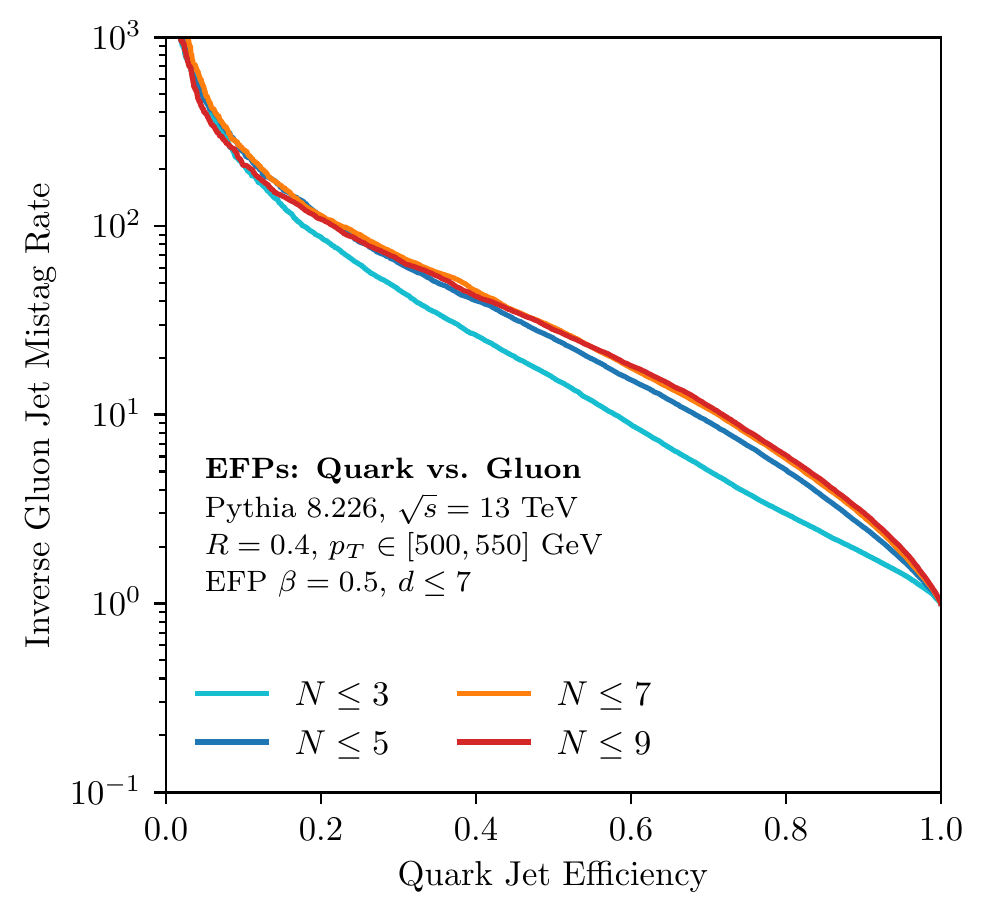}}
\subfloat[]{\includegraphics[scale=.76]{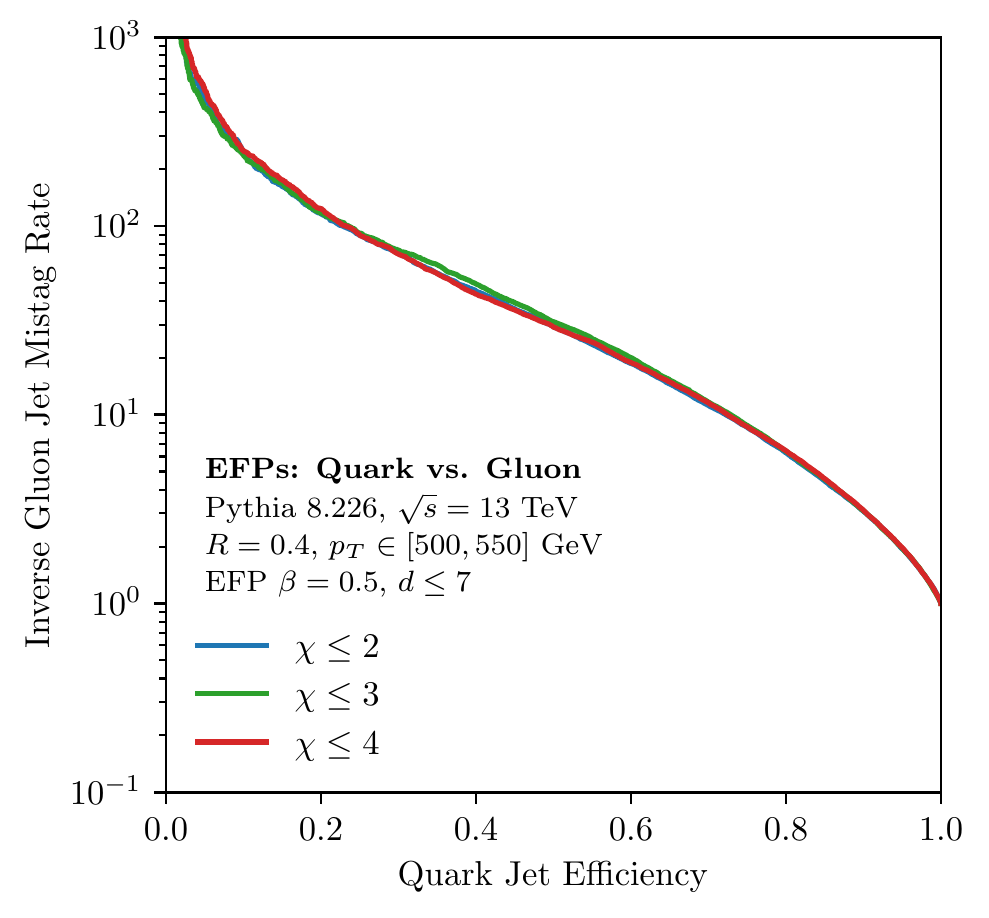}}

\subfloat[]{\includegraphics[scale=.76]{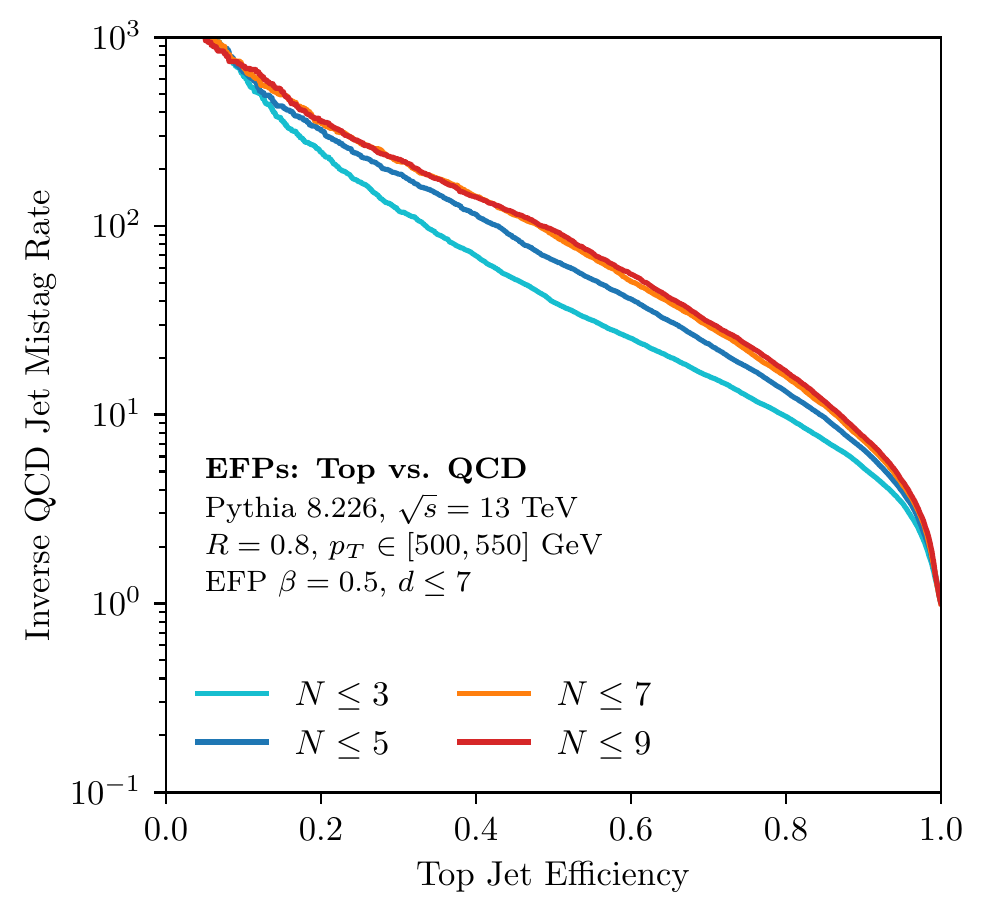}}
\subfloat[]{\includegraphics[scale=.76]{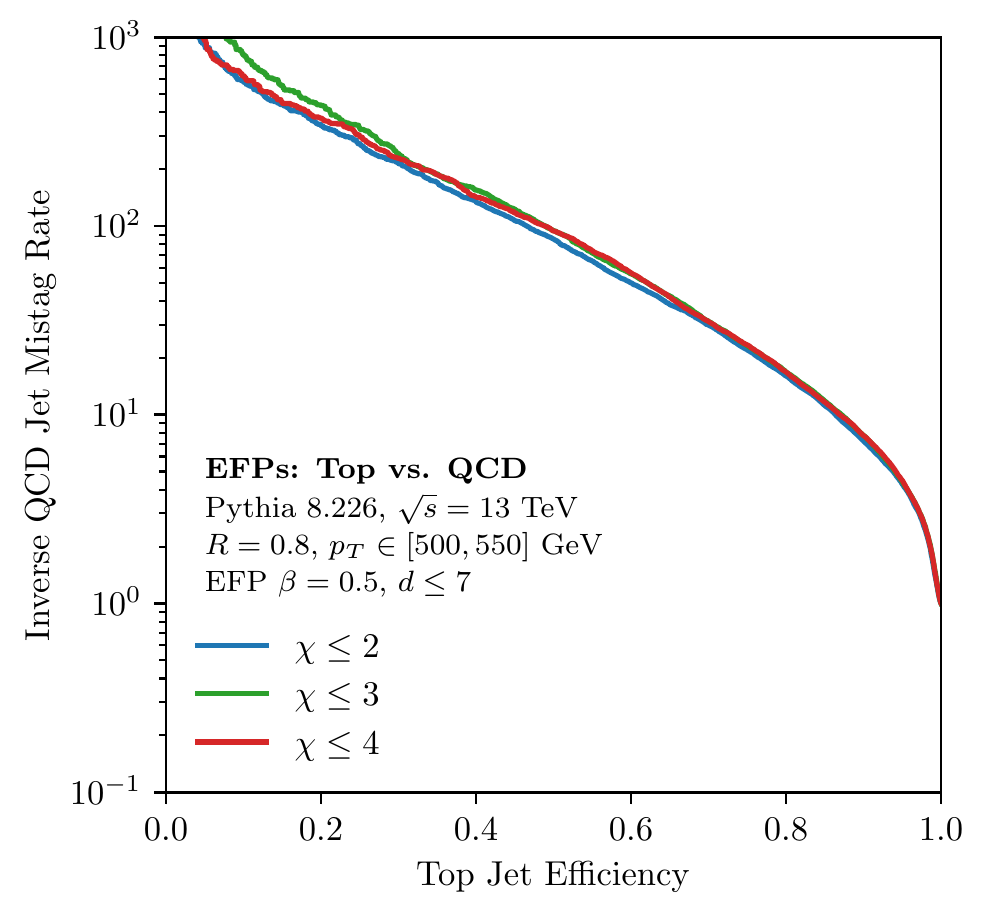}}
\caption{Same as \Fig{fig:efpnsweep} but for (top) quark/gluon discrimination and (bottom) top tagging.}
\label{fig:appnchisweep}
\end{figure}
\afterpage{\clearpage}

Finally, we truncate the set of \Bs with $d\le 7$ by the number of vertices $N$ and by the VE computational complexity $\chi$ in \Fig{fig:appnchisweep}.  
As in \Fig{fig:efpnsweep}, the higher $N$-particle correlators contribute to the classification performance up to at least $N=7$, whereas the higher-complexity \Bs beyond $\chi = 2$ do not significantly contribute to the classification performance.

\newpage

\bibliography{energyflow}

\end{document}